\documentclass[graybox]{svmult}

\usepackage{type1cm}        
\usepackage{makeidx}         
\usepackage{graphicx}        
\usepackage{multicol}        
\usepackage[bottom]{footmisc}
\usepackage{algpseudocode}
\usepackage{hyperref}

\usepackage{newtxtext}       %
\usepackage[varvw]{newtxmath}       
\usepackage{tikz}        
\usepackage{caption}     

\usepackage{etoolbox}
\makeatletter
\patchcmd{\ALG@doentity}{\item[]}{\item[\relax]}{}{}
\makeatother

\makeindex             

\usepackage{soul,xcolor,hyperref,cleveref,doi,todonotes,listings}
\usepackage{algorithm}
\usepackage{algpseudocode}
\usepackage{threeparttable,makecell,multirow}
\usepackage{subcaption}

\newcommand{\bigo}[1]{\mathcal{O}\left(#1 \right) }
\newcommand{\ra}[1]{\renewcommand{\arraystretch}{#1}}
\newcommand{\extadd}{\mathbin{\hspace{0.65em}\text{\makebox[0pt]{\resizebox{0.3em}{0.85em}{\(\updownarrow\)}}\raisebox{0.15em}{\makebox[0pt]{\resizebox{1.3em}{0.2em}{\(\leftrightarrow\)}}}}\hspace{0.65em}}}
\newcommand{\ignore}[1]{}
\newcommand{\rev}[1]{{\color{black}{#1}}}

\newlength{\doublecolumnimgwidth}
\usepackage[utf8]{inputenc}
\begin{document}

\title*{Parallel Sparse and Data-Sparse Factorization-Based Linear Solvers}
\author{Xiaoye Sherry Li\orcidID{0000-0002-0747-698X} and\\ Yang Liu\orcidID{0000-0003-3750-1178}}
\institute{Lawrence Berkeley National Laboratory, \email{xsli@lbl.gov} \\
Lawrence Berkeley National Laboratory,
\email{liuyangzhuan@lbl.gov} }
%
\maketitle


\abstract{ 
Efficient solutions of large-scale, ill-conditioned and indefinite
algebraic equations are ubiquitously needed in numerous computational fields,
including multiphysics simulations, machine learning, and data science.
Because of their robustness and accuracy, direct solvers
are crucial components in building a scalable solver toolchain.
In this chapter, we will review recent advances of sparse direct solvers
along two axes: 1) reducing communication and latency costs in both
task- and data-parallel settings, and 2) reducing computational complexity
via low-rank and other compression techniques such as hierarchical matrix algebra.
In addition to algorithmic principles, we also illustrate the key parallelization
challenges and best practices to deliver high speed and reliability
on modern heterogeneous parallel machines.
}

\ignore{
2018 conference: \\
\url{https://wsc.project.cwi.nl/woudschoten-conferences/2018-woudschoten-conference}

Chapter page limit: ~40 (including refs)
}

\section{Introduction}\label{sec:intro}
\ignore{
\hl{
based on the 2 lectures: 
Lecture 1: Exploiting Parallelism in Sparse Matrix Computations.
In this lecture, we will first briefly survey the fundamentals of high
performance computing, including computer architectures, parallel
programming models \& languages, and analysis of parallel algorithms
(e.g., performance upper bounds models: Amdahl's law, roofline model, 
 latency-bandwidth model.)
We will then study the dataflow organizations of different algorithms, 
such as left-looking, right-looking, and multifrontal algorithms.
We will present various ways of organizing parallelism using the fundamental
data structures like elimination tree or directed acyclic graph.

Lecture 2:
Distributed-memory Sparse Direct and Hybrid Solvers.
In this lecture, we will present some implementation techniques used in
developing scalable parallel factorization based sparse solvers for
distributed-memory parallel machines. We will use the software libraries
SuperLU, PDSLin and STRUMPACK to illustrate the design choices and the
actual performance on modern parallel architectures. Particularly, we will
demonstrate how to reduce inter-process communication and synchronization.
} 
}
Efficient solutions of large-scale, ill-conditioned and indefinite
algebraic equations are ubiquitously needed in numerous computational fields,
including multiphysics modeling and simulations, machine learning, and data science.
Because of their robustness and accuracy, direct solvers
are crucial components in building a scalable solver toolchain.
\rev{The field of sparse factorization-based solvers is vast and has been active since the 1970s.
It evolved from classical Gaussian elimination to combinatorial techniques that control fill-in and memory use. Early ideas focused on ordering and factorization strategies to minimize fill (notably nested dissection and minimum-degree orderings) and on scalable factorization forms (LU, Cholesky, multifrontal and, later, supernodal approaches). These ideas were embodied in portable software packages, most notably pioneered in SPARSPACK~\cite{sparspak1981} and Harwell Subroutine Libraries~\cite{scott2023}.
Since then, a vast amount of research has been conducted and many more portable and robust  software packages were developed, see a partial list in~\cref{sec:software}.  
 For orientation, classic references include~\cite{georgeliu81,davisbook2006,dufferismanreid2017}.
}
In this chapter, \rev{we will focus mostly on the developments in the distributed memory and GPU programming environments.} We will review recent advances of sparse direct solvers
along two axes: 1) reducing communication and latency costs in both
task- and data-parallel settings, and 2) reducing computational complexity
via low-rank and other compression techniques such as hierarchical matrix algebra.
In addition to algorithmic principles, we also illustrate the key parallelization
challenges and best practices to deliver high speed and reliability
on modern heterogeneous parallel machines.

\subsection{Problem statement}
\label{sec:problem-intro}
A linear system is defined as $Ax = B$, where matrix $A$ is of size $n\times n$ and the right-hand side $B$ is of size $n\times n_{rhs}$. 
This chapter focuses on sparse linear systems in two scenarios. In the first case $A$ is \emph{structure-sparse} with many explicit zeros.
This type of sparse systems typically arises from discretization of partial differential equations.
In the second case $A$ is dense but admits \emph{data-sparse} representations, for example, it is numerically low-rank or has numerically low-rank blocks.
 This type of sparse systems often arises from integral equations, boundary value problems and many kernel matrices used in machine learning.
Even though the solution techniques for the two types of sparse systems are distinct, we will illustrate that a hybrid method combining both types of  sparsity-exploiting techniques is even more efficient than a traditional sparse direct solver. For the right-hand side, we focus on the most common case with dense $B$ and $n_{rhs} =1$.

In this chapter, by ``direct solver" we mean more broadly the solvers or preconditioners that are based on certain matrix factorizations. That is, the core operations are factorizations and solutions using the factored matrices. 
In data-sparse cases, the factorizations are usually approximate, therefore they may not be used simply in direct solvers, instead, they can be used as effective preconditioners to accelerate convergence of a Krylov iterative solver. We note that the low-rank approximate factorization is very different from the conventional ILU-type approximate factorization, which will be clear in ~\cref{sec:data-sparse}.

We will focus on two main algorithmic advances in recent years:
1) asymptotic reduction of data movement on parallel machines, and 
2) asymptotic reduction of arithmetic operations and memory footprint.
Here, by asymptotic reduction we mean lowering the exponent in ``big O" notation in analytical formulae.
For example, in the first case, we will show orders of magnitude lower communication bandwidth and latency cost for some algorithms. 
In the second case, we will show almost linear complexity for arithmetic and memory for some algorithms.
The complexity formulae will be summarized in ~\cref{tab:asympt2d,tab:asympt3d,tab:SpTRS-comm-complexity,tab:hierarchical_matrix}.
Guided by these algorithmic principles, a large number of implementation techniques were developed. We will focus on the distributed-memory algorithms for both CPUs and GPUs, targeting exascale.

For most practical problems, we first {\em transform} the original linear system into a more favorable form that can be solved more efficiently.
A general description of the algorithm for direct solution of linear equations $AX=B$ is as follows:
\begin{enumerate}
\item Compute a {\em factorization}, for example, LU decomposition: 
\begin{equation} \label{eq:factorization}
P_c P_r D_r A D_c P_c^T = L U.
\end{equation}
  Here $D_r$ and $D_c$ are diagonal matrices to equilibrate the system, which can often reduce the condition number. 
  $P_r$ and $P_c$ are {\em permutation matrices} that reorder the rows and the columns of $A$. They serve the purpose of maintaining numerical stability and preserving sparsity in the sparse factored matrices.
\item Compute the {\em solution} of the transformed system:
$ (P_c P_r D_r A D_c P_c^T) P_c D_c^{-1} X = P_c P_r D_r B $. 
Thus, $X$ can be evaluated as:
\begin{equation} \label{eq:solution}
  X = A^{-1}B                              
   = D_c P_c^T U^{-1}L^{-1} (P_c P_r D_r B ).
\end{equation}
  \rev{The application of $L^{-1}$ (or $U^{-1}$) in the case of LU decomposition is carried out as triangular solution}. Or, it could mean some other solution methods for other types of factorizations, e.g., see \cref{sec:data-sparse}.
\end{enumerate}

\noindent\emph{Remark.} 
Computing $D_r, D_c, P_r$ and $P_c$ are preprocessing steps, which by themselves have large bodies of literature.
Although they are not the focus of this chapter, they are critical components for a robust and efficient solver and we will give appropriate references throughout the chapter,
see, for example,~\cite{li2015}.

\ignore{
\begin{tikzpicture}
    \draw[->] (-2,0) -- (2,0) node[midway, above] {Left-Right Arrow};

    \draw[->] (0,-2) -- (0,2) node[midway, right] {Vertical Arrow};
\end{tikzpicture}
}

\vspace{4pt}
Most of the algorithms discussed in this chapter are based on Gaussian elimination (GE) and its sparse variants. 
In the GE algorithm the Schur complement updates can be performed in different orders, leading to different variants such as 
left-looking (fan-in) or right-looking (fan-out). These variants
are mathematically equivalent under the assumption that the 
operations are associative, although this is not entirely true in floating point arithmetic. However, different variants  have very
different memory access and communication patterns. 
We mainly use the right-looking algorithm and its multifrontal variant to illustrate parallelism. \Cref{fig:GE-rl-mf} gives a high-level sketch of the two algorithms and compares them side-by-side. In the algorithms a node usually refers to a \emph{supernode}. 
Loosely speaking, a supernode consists of several consecutive rows/columns having the same nonzero pattern.
Supernodes help lift most of the submatrix operations to use Level-3 BLAS, leading to higher arithmetic intensity.

\vspace{4pt}
\noindent\emph{Remark.}
The multifrontal algorithm presented here works only for sparse matrices with \emph{symmetric patterns}, hence, the computation can be organized in a tree structure, called assembly tree or elimination tree (etree)~\cite{liu90}. 
For nonsymmetric-pattern sparse matrices, the computation graph is a directed acyclic graph (DAG), the same as the right-looking or left-looking algorithm. See more explanation of a DAG in \cref{sec:sptrsv}.
A good example of nonsymmetric-pattern multifrontal algorithm and code is UMFPACK~\cite{davis2004algorithm}.

\vspace{4pt}
In the algorithm description, we use the multifrontal term \emph{extend-add}, with symbol $\extadd$, to denote the sparse operation of
\emph{scattering-and-adding} an element from the source array into an element of the destination array.
The nonzero structure of the destination is a superset of the nonzero structure of the source, hence the need for ``scatter". 
Note that in the context of Schur complement update, the ``add" operation should really be interpreted as ``subtract".
For notational convenience, we define two index sets $R \subseteq \{K\!\! :\!\! N\}$ referring to the remaining part of the matrix at the $K$th elimination step, and $R_1=R \setminus \{K\}$.
These two index sets change after each step of GE, depending on sparsity.
In the supernodal right-looking algorithm (Listing (1) in~\cref{fig:GE-rl-mf}), steps 1 and 2 involve dense matrix operations. Step 3 involves sparse scattering.
In the multifrontal algorithm (Listing (2) in~\cref{fig:GE-rl-mf}), steps 1 and 2 are the same as the right-looking algorithm. The difference is in the treatment of Schur complement update. In right-looking, the Schur complement update is performed right away \emph{in-place}, whereas in multifrontal, multiple partial Schur complement update matrices (denoted as $T$) are passed along the etree. The scatter operation occurs at step 0.

In~\cite{amestoy-li2001}, Amestoy et al. provided 
a comprehensive study and comparison of these two algorithms that were implemented in the distributed memory SuperLU\_DIST~\cite{superlu_web} and MUMPS~\cite{amestoy2000mumps}, respectively.
Empirically, using a 512-processor Cray T3E and a set of matrices from real applications, the paper compared the performance characteristics of the two solvers with respect to uniprocessor speed, interprocessor communication, the memory requirement, and scalability.

The GE algorithms involving hierarchically low-rank matrices will be treated in~\cref{sec:data-sparse}.

\begin{figure}[t]
\begin{subfigure}[t]{.5\textwidth}\label{fig:rl-lu}
\begin{lstlisting}[caption={right-looking LU}, label={alg:rl-lu},basicstyle=\footnotesize\ttfamily]]

For each node |$K$| in DAG from sources to sink
  1. Factorize diagonal |$A(K,K)=L(K,K)U(K,K)$|
     (may involve pivoting) 
  2. Compute off-diag panels |$L(R_1, K)$|, |$U(K, R_1)$| 
     (via triangular solves TRSM) 
  3. In-place Schur complement update: 
     |$A(R_1, R_1)\leftarrow A(R_1, R_1) \extadd (-L(R_1, K) \cdot U(K, R_1))$|
     (via matrix multiplication GEMM)
Endfor
\end{lstlisting}
\end{subfigure}
  \begin{subfigure}[t]{.5\textwidth}
  \begin{lstlisting}[caption={symmetric-pattern multifrontal LU}, label={alg:mf-lu}, basicstyle=\footnotesize\ttfamily]]
for each node |$K$| in etree from leaves to root
  0. Assemble children's |$T$|s to frontal matrix:
     |$F \leftarrow \begin{bmatrix} A(K,K) &A(K,R_1)\\
		A(R_1,K) & 0 \end{bmatrix}\extadd T_1 \extadd T_2 \extadd \ldots$|
  1. Factorize diagonal |$F(K, K) = L(K,K) U(K,K)$| 
     (may involve pivoting) 
  2. Compute off-diag panels |$L(R_1, K)$|, |$U(K, R_1)$|
     (via triangular solves TRSM) 
  3. Compute and pass update |$T_K$| to parent:
     |$T_K = F(R_1,R_1)-L(R_1, K) \cdot U(K, R_1)$|
     (via matrix multiplication GEMM)
Endfor
\end{lstlisting}
  \end{subfigure}
  \caption{Sketch of the right-looking and multifrontal algorithms. We define $R \subseteq \{K\!\!:\!\!N\}$ consisting of indices corresponding to nonzeros of LU, and $R_1=R \setminus \{K\}$. In Listing (2), $T_i$ represents node $i$'s update matrix corresponding to the partial Schur complement update. The two index sets $R$ and $R_1$ change after each step of GE, reflecting the  Schur complement sparsity after each elimination step.}
  \label{fig:GE-rl-mf}
\end{figure}

\subsection{Challenges and motivations}
\label{sec:challenges}

Parallel computing has become an increasingly indispensable tool in various
computing disciplines, particularly in modeling physical phenomena in science
and engineering simulations. Developing parallel algorithms and software has become a central task in the scientific community.
Despite a long history of advances in parallelizing sparse linear algebra algorithms, there still exist many challenges, either architecture-related or application-related, which are summarized below.
\begin{enumerate}
\item Inter-process communication cost and load
imbalance become an increasing scalability hindrance. This is due to the growing gap in hardware speed of data transfer and the speed of floating point operations. Therefore, it is essential to redesign algorithms to reduce communication and synchronization. We address this challenge in~\cref{sec:reduce-comm}.
\item Modern HPC architectures become increasingly  heterogeneous. For the three exascale machines in the United States~\cite{top500-web},
  more than 90\% of the computing power comes from GPUs. 
Existing sparse direct solvers usually exploit only task parallelism, e.g., through OpenMP~\cite{openmp-web}
to use on-node CPUs 
and MPI~\cite{mpiforum-web}
to communicate across nodes. Novel methods are needed to make best use of massive data parallelism provided by GPUs. 
Worse yet, no common programming standard is developed for GPUs from different computer vendors, e.g., AMD, NVIDIA, Intel. 
We address this challenge in~\cref{sec:gpu}.

\item Direct solvers based on classical  LU decomposition and triangular solution are inherently
  not scalable from the viewpoint of algorithmic complexity, because the number of floating point operations and memory storage units grow super-linearly with respect to problem size. The growth rate is even higher for discretized PDEs in 3D geometry than in 2D.
Hence, the number of computing nodes must increase more than the increase of problem size in order to be able to store the factored matrices in the aggregate memory of all nodes.
This limits the size of the problems that are solvable given fixed hardware resources and time budget.
A truly scalable algorithm should have a nearly linear time and memory complexity with respect to problem size. 
We address this algorithmic challenge in~\cref{sec:data-sparse}.
\end{enumerate}



\subsection{Sources of parallelism}

\paragraph{{\bf Parallel architectures}}
Parallelism occurs at many levels of the hierarchy of modern processor architectures with varying granularity: bit level, instruction level, data, and task parallelism~\cite{hennessy2025computer}.

The vector machines with SIMD (Single
Instruction, Multiple Data) instructions were the earliest parallel machines which are based on the pipeline principle. 
They provide high-level operations on arrays of elements. SIMD instructions perform
exactly the same operations on multiple data objects, thus produce multiple results at the same time. 
For example, in the late 1990s, Intel introduced the SSE (Streaming SIMD
Extensions) (and SSE2 extension for double precision) in the x86 instruction sets, which were later expanded to MMX, AVX, and AVX512.
For the SIMD architectures, parallelization is usually achieved by a compiler which analyzes the program and generates the vector instructions.
However, the SIMD instructions are not well suited for operations with irregular control structures. 

The growth of multi-core CPUs with independent cores that can handle distinct threads of execution serves a much broader spectrum of applications that can take advantage of parallel computing. These so-called Multiple Instruction Multiple Data (MIMD) machines are suitable for general-purpose computing tasks, including servers, and high-performance computing (HPC) systems.
In the MIMD architectures, shared memory machines are widely used now on desktops and laptops. Multiple processors access a common memory space and communicate through this shared memory. Programming such systems is made easier by using the standard OpenMP API that is portable to many platforms and programming languages. 
Distributed memory machines are needed to provide true scalability, where each processor has its own local memory. Processors communicate via an interconnect network. Programming such systems is significantly harder than a shared memory machine.
The MPI standard brought a portable solution for message passing which quickly became the de facto standard. 
Nowadays, most HPC systems have a hybrid architecture which combines elements of both shared and distributed memory systems. Additionally, each node often has a heterogeneous organization with both multicore CPUs and GPU accelerators. It is insufficient to use MPI alone to program such systems.
We often use a hybrid programming model such as MPI+X, where MPI is used to generate multiple processes across multiple nodes, and X is used within the address space of each MPI process. Here, X can be OpenMP and/or GPU-specific languages.

\paragraph{{\bf Parallel algorithms in matrix computation}}
Parallelization is to divide the whole computational task into multiple subtasks and map them to the underlying architecture. 
The art of designing an efficient parallel algorithm follows several key guiding principles: 1) dividing the subtasks evenly to maintain balanced workload across all the processes; 
2) making subtasks as independent as possible to avoid information sharing and hence the waiting time;
3) increasing arithmetic intensity relative to memory access and interprocess communication; and
4) using high-level abstraction.

Matrix computation is one of the prime fields that could effectively utilize parallel computers at the dawn of parallel computing. Even though it is a mature field in parallel algorithms design, there are still ample opportunities for algorithm advances, which is mainly due to the frequent innovations in HPC architectures and novel applications that rely on numerical linear algebra. 
Over the years, the single most important abstraction is the Basic Linear Algebra Subprograms (BLAS) library~\cite{netlib-blas-web}
which contains low-level routines that perform basic vector and matrix operations, providing essential building blocks for linear algebra applications.
Various optimized BLAS libraries are widely accessible, such as OpenBLAS, ATLAS and Intel MKL for CPUs, and NVIDIA cuBLAS and AMD rocBLAS for GPUs. 
The BLAS abstraction hides a lot of low level architectural optimization from the algorithm developers, such as vectorization or threads/warps on GPUs.

In the sparse matrix landscape, many algorithms are also restructured to express the operations in terms of BLAS. However, the dense blocks in sparse matrices are usually small, making an algorithm far from performing at the peak speed if it depends only on parallelizing BLAS. 
Therefore, in addition to data parallelism encompassed in BLAS, it is essential to exploit task parallelism as well. In sparse factorization-based solvers, two important data structures can greatly help design a good matrix-to-process mapping  scheme and a parallel execution strategy. One data structure is a DAG as is used in the right-looking sparse LU. Another data structure is a tree as is used in the symmetric-pattern multifontal LU and the hierarchically rank-structured algorithms in~\cref{sec:data-sparse}. Both DAGs and trees encode the data and task dependencies among submatrix operations. There is a rich body of literature about both static and dynamic scheduling of a task graph described by a DAG or tree on a distributed memory machine to maintain load balance and to minimize communication.
The parallel algorithms presented in this chapter are mostly based on static scheduling strategies. The challenge with dynamic scheduling is the overhead associated with the scheduler itself. It is cost effective only when the computational task has large granularity.

\subsection{Analysis of parallel algorithms} 
We emphasize established performance bound models. 
\begin{itemize}
\item \emph{Amdahl’s law} provides an upper bound on achievable speedup based on the fraction of sequential work, highlighting the diminishing returns of parallelization when sequential portions remain. 
\item The \emph{roofline model} helps relate algorithmic arithmetic intensity to architectural bandwidth and peak performance, serving as a visual performance diagnostic. 
\item The \emph{latency–bandwidth model} is discussed to quantify the per-message startup overheads and sustained transfer rates, guiding optimizations in communication patterns.
\end{itemize}

The rest of the chapter is organized as follows. \Cref{sec:reduce-comm} describes recent algorithms for  mitigating communication cost for sparse LU factorization (SpLU) and sparse triangular solution (SpTRSV). This section also surveys analyses of the communication lower bounds for these operations.
\Cref{sec:gpu} describes various novel algorithms to exploit the massive fine-grain parallelism provided by the GPUs. 
\Cref{sec:data-sparse} describes various data-sparse compression techniques to reduce both floating point and memory complexity. In addition to the parallel construction  algorithms for compression, the section also illustrates how the GE algorithm works with the compressed formats and how they can be efficiently parallelized.
\Cref{sec:data-and-structure} describes how the data-sparse formats can be embedded in the structure-sparse algorithm, leading to a nearly linear complexity fast solver.
\Cref{sec:software} gives a brief survey of existing parallel sparse solvers, using one or multiple algorithms discussed in this chapter.
Finally, \cref{sec:open-problems} lists a number of open problems. 

\section{Mitigating Communication and Synchronization Costs}
\label{sec:reduce-comm}
Communication cost occurs when transferring data between fast and slow memory or between two nodes across the network. This chapter focuses more on the latter.
A commonly used cost model for communication performance is the
{\em $\alpha$-$\beta$ model},
where $\alpha$ refers to the latency and $\beta$ is the inverse of the bandwidth
between two processes. The time to send a message of length $n$ is roughly:
\[
\text{Time} = \text{latency} + n / \text{bandwidth} = \alpha + n\times \beta\;.
\]
This is a simplified and ideal model, without taking into account such practicalities
as network congestion etc. For most parallel machines
$\alpha \gg \beta \gg \text{ time\_per\_flop}$. For example, on the current Perlmutter node at NERSC~\cite{perlmutter-web}, 
typical sparse matrix algorithms run at about 4\% of the peak, i.e., the realistic time per flop is $~10^{-5} \mu s$. According to the measurement by Lin~\cite{lin2026} of typical message sizes, 
$\alpha = 1.7 \mu s \approx 170,000 \text{ flops}$,
and $\beta = 4\cdot 10^{-5} \mu s/\text{byte} \approx 30 \text{ flops\_per\_double\_word}$.
Therefore, the fundamental principle is to organize the parallel algorithm
so that it maintains high data locality and sends fewer long messages rather
than sending many short messages. In the following subsections we will describe various algorithmic and implementation techniques to mitigate communication cost for SpLU and SpTRSV.

\ignore{
\subsection{Overlapping communication and computation}
We developed a pipelining algorithm for sparse LU to further overlap communication and computation~\cite{}.
With the above two main techniques, the strong scalability of the sparse direct solvers was greatly improved. 
}
\subsection{Synchronization-avoiding via one-sided communication}
\label{sec:sptrsv}

Sparse computations in general have lower arithmetic intensity than the dense counterparts.
The problem is exacerbated by a high degree of data and task dependency. For example, sparse triangular solution (SpTRSV) is notoriously difficult to parallelize -- its arithmetic intensity is of $\bigo{1}$, similar to sparse matrix-vector multiplication (SpMV). Yet, it has a higher degree of dependency than SpMV, which imposes more sequentiality than SpMV. 

The execution of factorization and triangular solution can be concisely prescribed by the DAG  data structure,
in which a vertex $i$ represents a certain computation task $T_i$ and a directed edge from vertex $T_i$ to vertex $T_j$ encodes the dependency of $T_j$ on $T_i$, e.g., $T_j$ needs some data produced by $T_i$. 
The vertex weight represents the amount of computation and the edge weight represents the amount of data to be  transferred between two tasks.
If an edge crosses the boundary between two MPI processes, inter-process communication needs to happen for data transfer.

The DAGs for sparse computations can be highly irregular in that both the vertex weights and the edge weights vary a lot. The communication links between pairs of processes also vary.
It is particularly challenging to achieve good parallel efficiency for SpTRSV, which typically
has limited task-level parallelism and relies on fine-grain synchronization to exploit this parallelism. 

\begin{figure}[h!]
\centering
\subfloat[t][nonzero pattern of $L$ \label{fig:levelset-matrix}]{\includegraphics[width=0.35\textwidth]{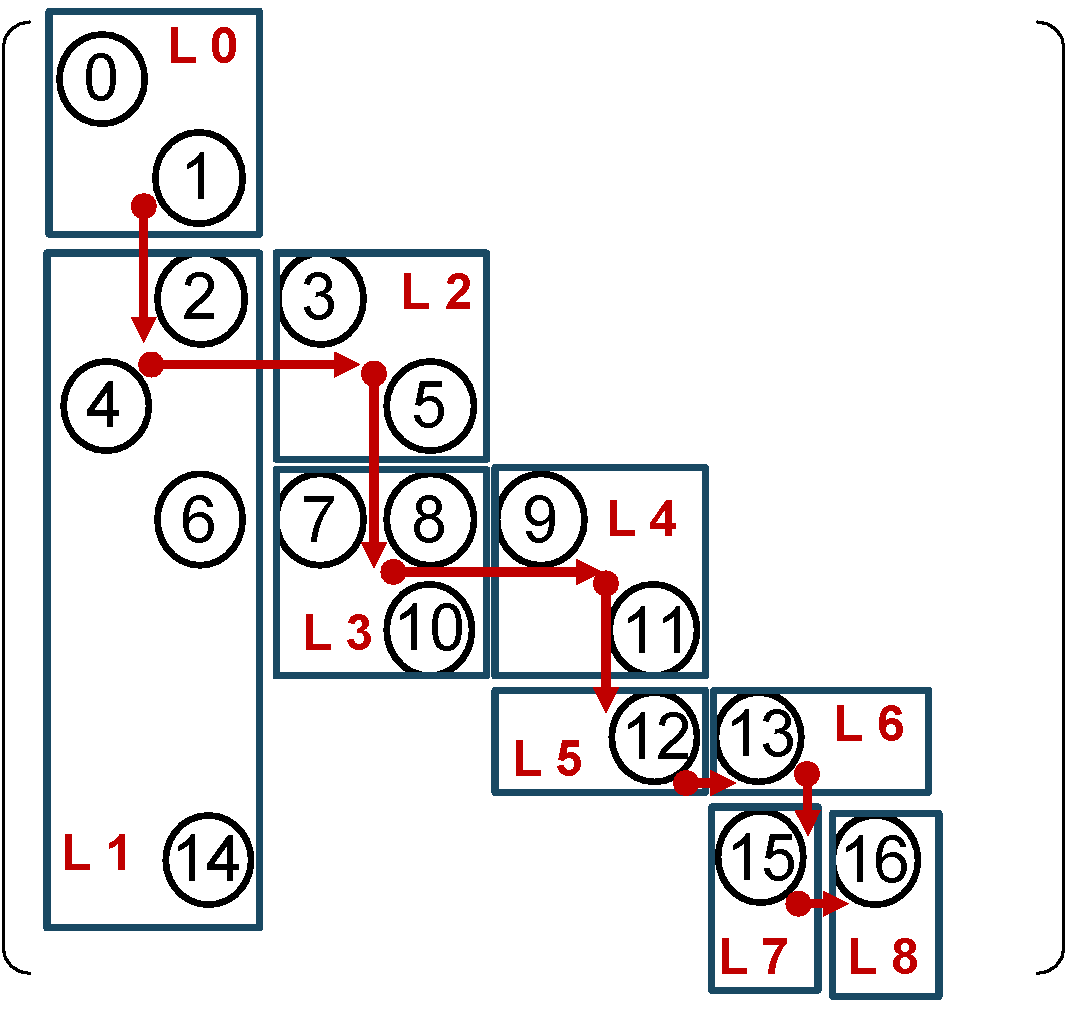} } \qquad\qquad
\subfloat[b][Nine level sets of task DAG \label{fig:levelset-dag}]
{\includegraphics[width=0.3\textwidth]
{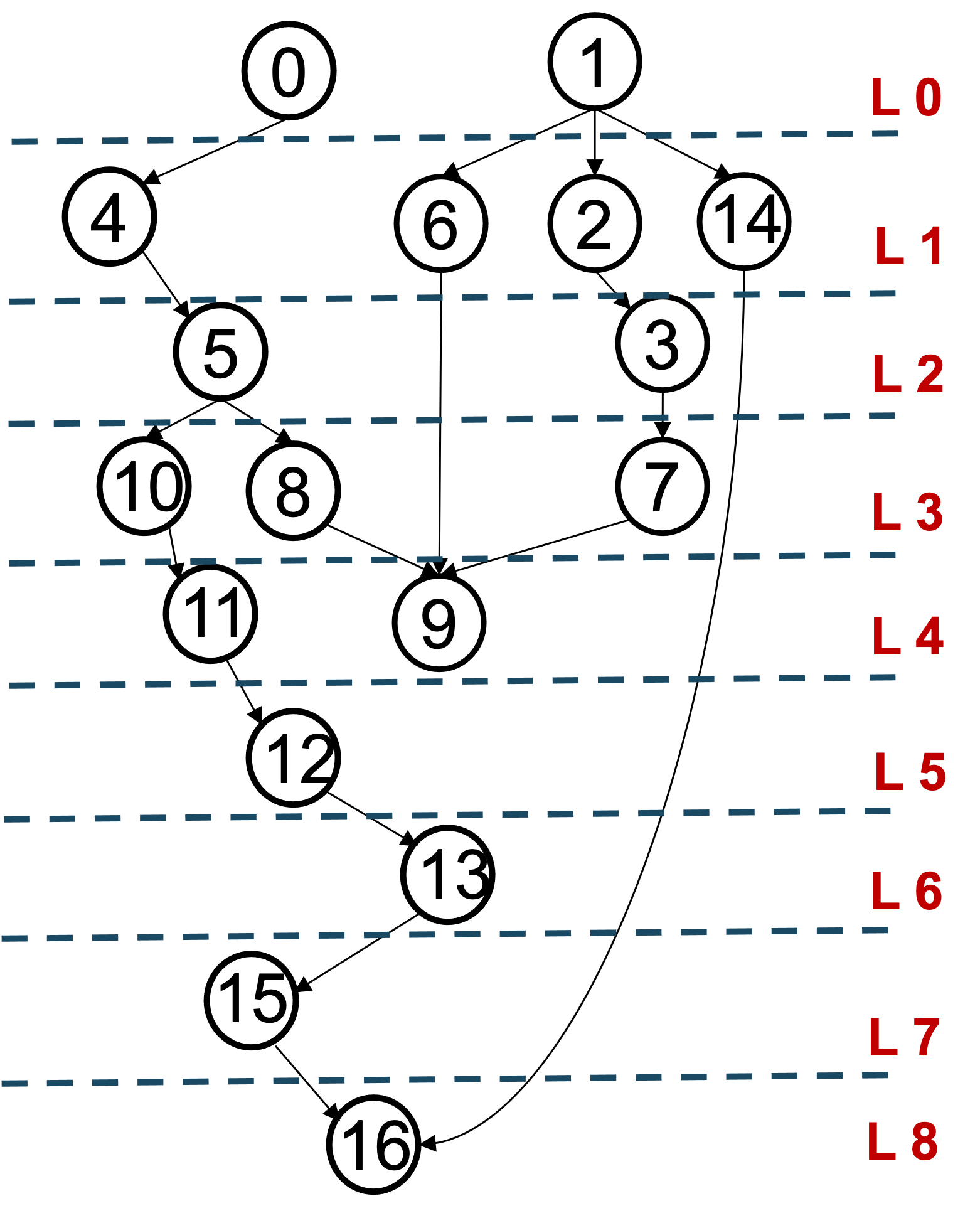}}
\caption{Illustration of level set in a lower triangular SpTRSV, $L x = b$.}
\label{fig:sptrsv-levels}
\end{figure}

For SpTRSV on-node parallelism, extensive studies have led to efficient algorithms with satisfactory parallel efficiency. All these algorithms aim to schedule ready tasks as soon as possible, including level-set based methods~\cite{sptrsv-level-1989,weifeng-sptrsv2016}, level tile method~\cite{swSpTRSV}, dynamic scheduling~\cite{bradley2016}, \rev{and scheduling via the standard OpenMP-barrier type of  loop~\cite{Yzelman2025}.}
\Cref{fig:sptrsv-levels} shows an example of a sparse lower triangular matrix $L$ and its associated task graph. Assuming a supernodal blocked algorithm is used, then each diagonal block $L_{ii}$ defines a dense TRSV operation $x_i = L_{ii}^{-1} \bar{b}_i$ and each off-diagonal nonzero block $L_{ij}$ defines a dense GEMV operation $L_{ij} x_j$.
In a level-set algorithm, multiple processes perform parallel operations within a level, then synchronize globally before moving to the next level. Park et al.~\cite{park-sptrsv-2014} developed a scheme to reduce the amount of synchronization. The algorithm uses local point-to-point synchronization for the two dependent tasks across the level sets if they are assigned to different processes. Dependent tasks assigned to the same process do not incur synchronization. They show that this reduced synchronization leads to a 1.6$\times$ speedup over the strictly level-based algorithm on a 12-core Intel Xeon processor.

Implementing these techniques in a distributed manner is highly nontrivial, since the algorithms do not  map well to MPI's two-sided handshaking paradigm, where both sender and receiver actively engage in the data exchange. 
In contrast, one-sided communication is an appealing alternative communication model, which was already established in the MPI standard since MPI-2~\cite{mpi-2}.
\rev{This model was inspired by the Bulk Synchronous Parallel (BSP) model, which separates communication from computation. The BSP model has been implemented in the BSPlib API~\cite{BSPlib1998}.
}
In the one-sided communication paradigm, communication between processes does not require explicit synchronization between the sender and receiver.
Instead, one process can perform operations on the memory of another process without the receiving process’s explicit participation. 
MPI-2 introduced the MPI\_WIN and Remote Memory Access (RMA) operations to take advantage of the remote direct memory access (RDMA) features offered by modern interconnects.
RDMA enables a process to directly access memory on
remote processes without involvement of the activities
at the remote side.
The light-weight asynchronous one-sided communication primitives provide a new mechanism for efficient DAG execution at scale.

The first distributed-memory SpTRSV algorithm in SuperLU\_DIST uses a \emph{message-driven} approach~\cite{li-sc98}.
Every process runs an asynchronous self-scheduling loop and triggers appropriate computations depending on one of the two types of messages received. This algorithm overcomes some synchronization overhead, but still uses the MPI two-sided Isend/Recv communication protocol. This algorithm can hardly scale to 100 processes. The algorithm was improved by Liu et al.~\cite{LiuTriSolve2018} by using asynchronous communication trees for collective communication among subsets of processes. 

In~\cite{Sptrsv-oneside}, Ding et al. developed the first SpTRSV algorithm exploiting the one-sided communication paradigm. Firstly, they benchmarked a couple of one-sided MPI implementations on a Cray XC40 parallel machine with Intel Xeon E5-2698v3 processors, one is Cray MPI~\cite{kandalla2015optimizing} and the other is foMPI~\cite{belli2015notified}. 
The latter is a fast MPI-3.0 RDMA library on the Cray system. They found that on 16 nodes, the foMPI library achieved up to 3$\times$ more bandwidth than the Cray MPI two-side primitives for messages of size 64 to 256 bytes, which is the typical size in SpTRSV.
Secondly, they profiled the earlier distributed SpTRSV code in SuperLU\_DIST with communication tree optimization~\cite{LiuTriSolve2018} and found that with 1000 MPI processes, as much as 90\% of the parallel runtime is spent in communication.

Ding's algorithm uses a \emph{passive target mode}
in one-sided communication to implement two synchronization-free task queues to manage the messaging between producer-consumer pairs. 
One such task queue is dedicated to the solution components $x_i$ after they are computed at the diagonal blocks and are remotely enqueued, via \texttt{MPI\_Put}, to the column processes that need them for subsequent matvecs. 
Another task queue is dedicated to the partial sums of the matvecs at the diagonal blocks, via \texttt{MPI\_Accumulate}.
Ding et al. empirically showed that, using up to 4096  processes on the Cray XC40 system, their one-sided SpTRSV solver reduces communication time by
up to 2.5$\times$ and improves runtime 
up to 2.4$\times$ compared to the SuperLU\_DIST's two-sided MPI implementation.

Another contribution of Ding's work is the development of a critical path performance model for the DAG-based computational graph of SpTRSV. Ding showed that this performance model
can accurately assess the observed performance relative to computation and communication speeds and provide advice on what shape of the process grid is optimal~\cite{Sptrsv-oneside}. This model is generalizable to any parallel application with DAG as an underlying computation/communication graph.

Given the delicate task queue algorithm design and careful implementation, a natural question to ask is whether a generic DAG-based runtime system can achieve the same effect. 
In the last two decades, there has been tremendous progress in development and deployment of runtime frameworks for parallel algorithms portable to heterogeneous architectures. The two widely used systems in the linear algebra community are PaRSEC~\cite{parsec2012} and StarPU~\cite{augonnet:inria-00550877}. These two runtime systems were successfully used in parallel implementations of SpLU~\cite{lacoste-2014,Caracal2025}.
However, it remains an open problem whether such a generic runtime scheduler is cost-effective for SpTRSV, because the task sizes in SpTRSV are much smaller than those in SpLU. The only work we are aware of in this vein is by Totoni et al.~\cite{totoni2014}, where the message-passing parallel language and runtime system Charm++~\cite{Charm++2009} was used for load balancing and dynamic scheduling. They showed that the Charm++ assisted implementation can scale up to 512 cores of an IBM BlueGene/P machine.

To extend the one-sided communication paradigm to the GPU realm, Ding et al. leveraged the advantage of GPU-initiated
data transfers of NVSHMEM to implement and evaluate a
multi-GPU SpTRSV~\cite{Sptrsv-nvshmem}. 
Ding's SpTRSV algorithm is based on a novel producer-consumer paradigm to manage the computation and communication using two separate CUDA streams. 
On two nodes of NVIDIA Tesla V100 GPUs, the algorithm achieved a 3.7$\times$ speedup when using two nodes of twelve GPUs relative to the baseline on a single GPU, and
up to 6.1$\times$ compared to the cuSPARSE \texttt{csrsv2} solver over the range of one to eighteen GPUs. 


\subsection{Communication-avoiding 3D algorithm framework}
\label{sec:ca-3d}



As the speed discrepancy between communication and floating point operation becomes increasingly large, communication time often dominates the parallel runtime of a parallel algorithm, especially for a sparse matrix algorithm.
One primary innovation in parallel sparse direct solvers in the 1990s and 2000s was the 2D blocking algorithm for sparse LU decomposition and sparse triangular solution, such as those used in
SuperLU\_DIST~\cite{li-sc98,lidemmel03}, WSMP~\cite{gupta-toms-2002}.  
This was motivated from the dense matrix algorithms in the ScaLAPACK library~\cite{scalapackmanual}. 
In these 2D algorithms, the processes are organized in a two-dimensional grid, with communication restricted within process rows or process columns. 
The sparse L and U factors are partitioned into 2D blocks, which generally are of non-uniform size and are sparsity-induced. These 2D blocks are cyclically mapped to the 2D process grid.
\rev{Prior to these efforts, a more ﬁne-grained 2D approach was developed by van der Stappen et al. for the sparse LU on a mesh of transputers~\cite{Stappen93}, where they used the block size 1 by 1. This led to speedups of up to a factor 107 on 400 transputers.} 
However, \rev{despite the improved scalability}, when using thousands of processes, the 2D algorithm still spends more than 90\% of the time in communication~\cite{sao-JPDC-2019}.

In last two decades, there is a growing body of research on novel algorithms that require asymptotically lower amount of communication than the earlier parallel algorithms (e.g.,~\cite{Ballard_et_al_2014}). These algorithms are termed communication-avoiding algorithms, or CA algorithms for short.
One notable class of CA algorithms trade off extra memory for reduced communication. The rationale behind this is based on the analysis that the communication lower bounds decrease as the size of local memory increases~\cite[Section 2]{Ballard_et_al_2014}.
Based on this principle, the 3D and 2.5D CA algorithms were developed for dense matrix-multiplication and LU decomposition~\cite{irony2002trading,solomonik2011COLU}.%
\footnote{The 3D process configuration is a special case of the 2.5D configuration.}


Inspired by the success of 2.5D dense CA algorithms, Sao et al. developed the first 3D CA algorithm for SpLU~\cite{sao-JPDC-2019} and SpTRSV~\cite{sao2019}.
The 3D CA algorithm framework uses a 3D logical process grid, instead of the 2D process grid that was the state-of-the-art as of 2018.  The algorithm replicates some data
to reduce both the number of messages and the communication volume.
In addition, the 3D sparse algorithms can use the elimination tree to efficiently map the sparse matrix to the 3D process grid to further localize the communication among subsets of processes.
On the other hand, the 2.5D dense LU algorithm cannot exploit this feature.
As a result, the 3D sparse algorithm not only reduces communication but also
shortens the critical path of SpLU. For matrices with planar graph
structure (e.g., planar grids and meshes), the critical path length of Sao's 3D SpLU algorithm is $\bigo {n\,/\,\log n }$ whereas the critical path of a typical 2D algorithm is $\bigo{n}$.

\begin{figure}[t]
	\centering 
	\subfloat[]{\includegraphics[width=0.3\textwidth]{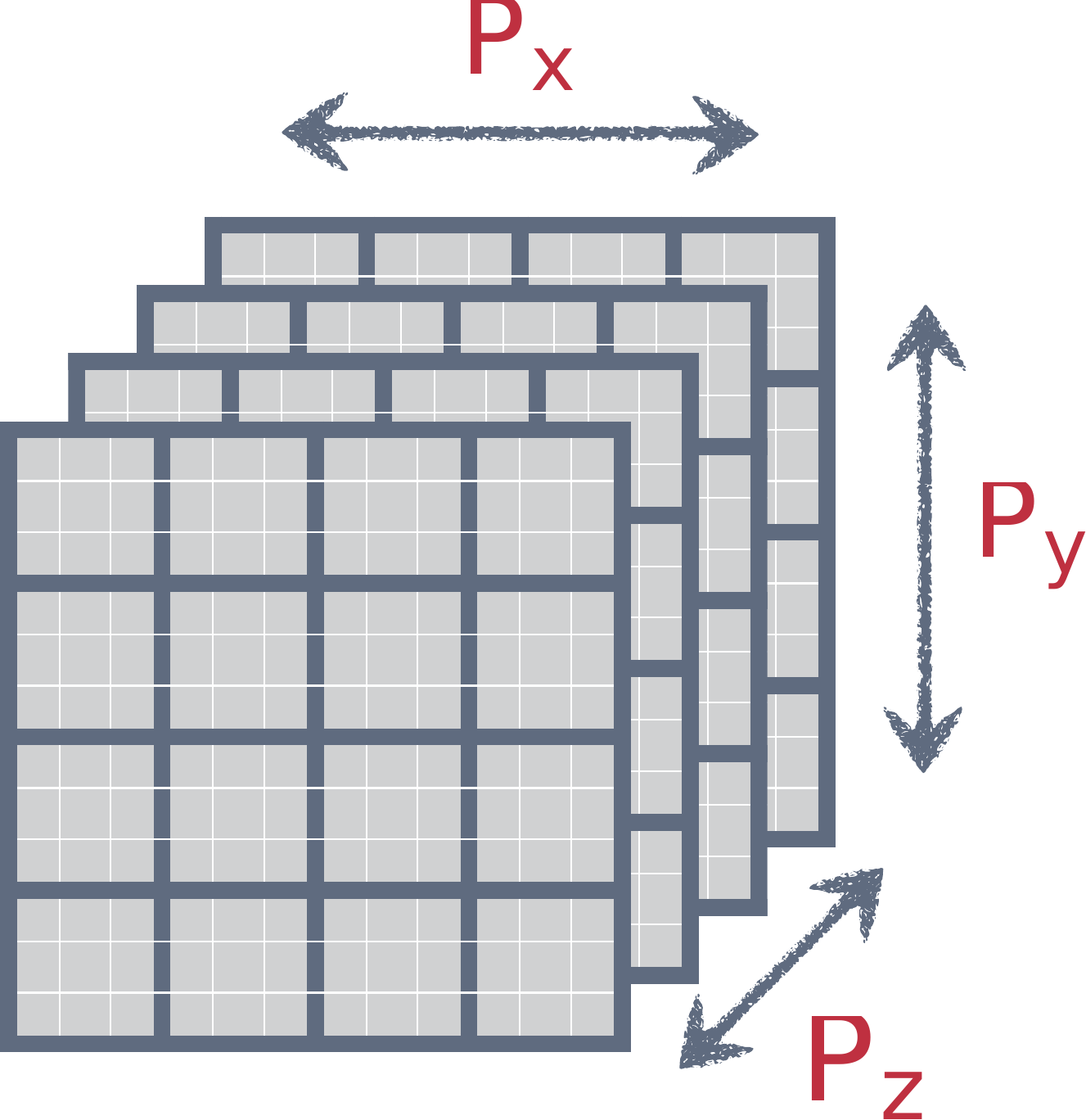}} \qquad
	\subfloat[]{\includegraphics[width=0.25\textwidth]{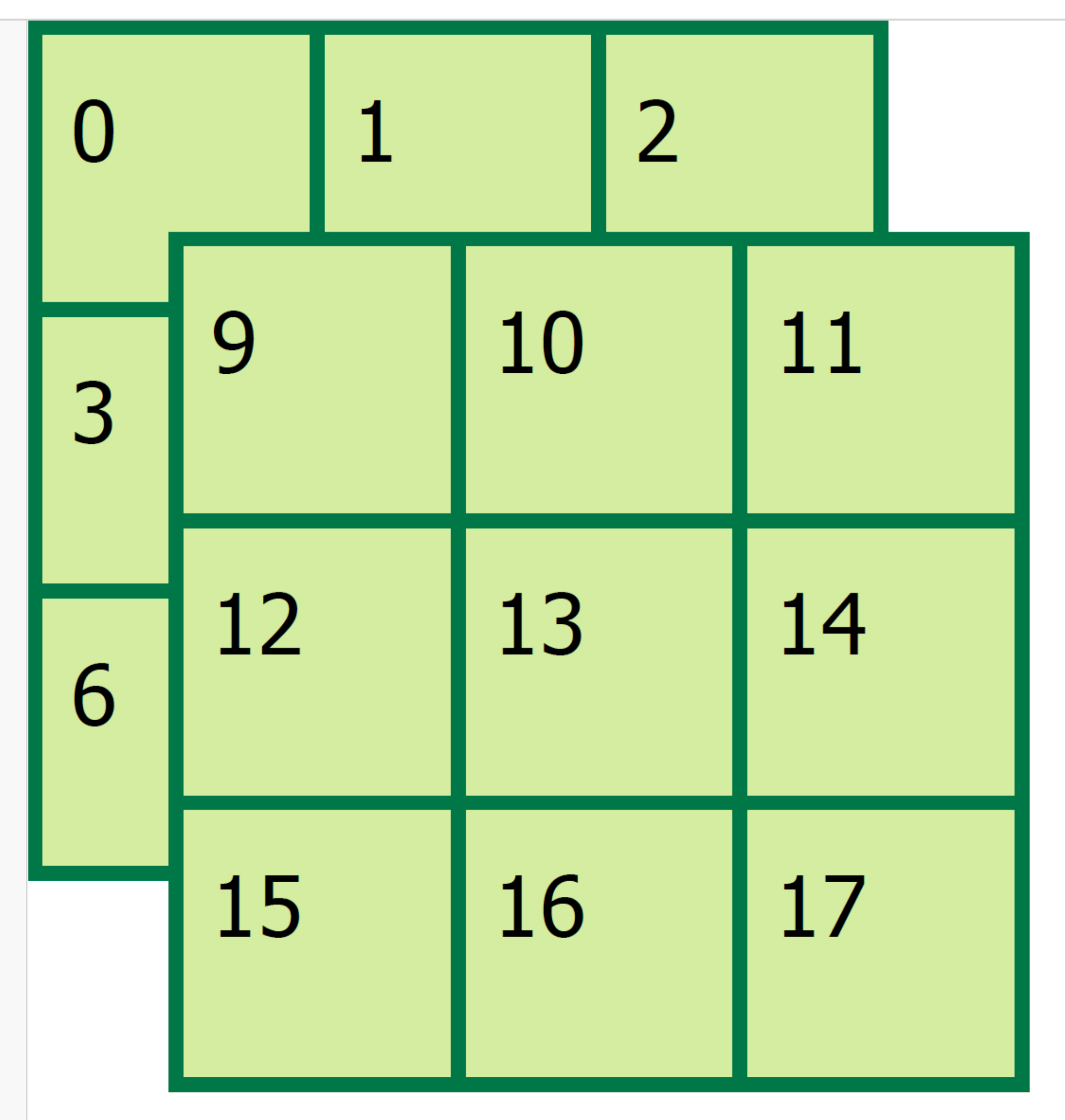}}
	\caption{The view of the logical 3D process grid and an example of 18 processes arranged as a $3\times3\times2$ process grid.}
	\label{fig:3dprocs}
\end{figure}

\begin{figure}[t]
\subfloat[t][ \label{fig:2level-a}]{\includegraphics[width=0.5\textwidth]{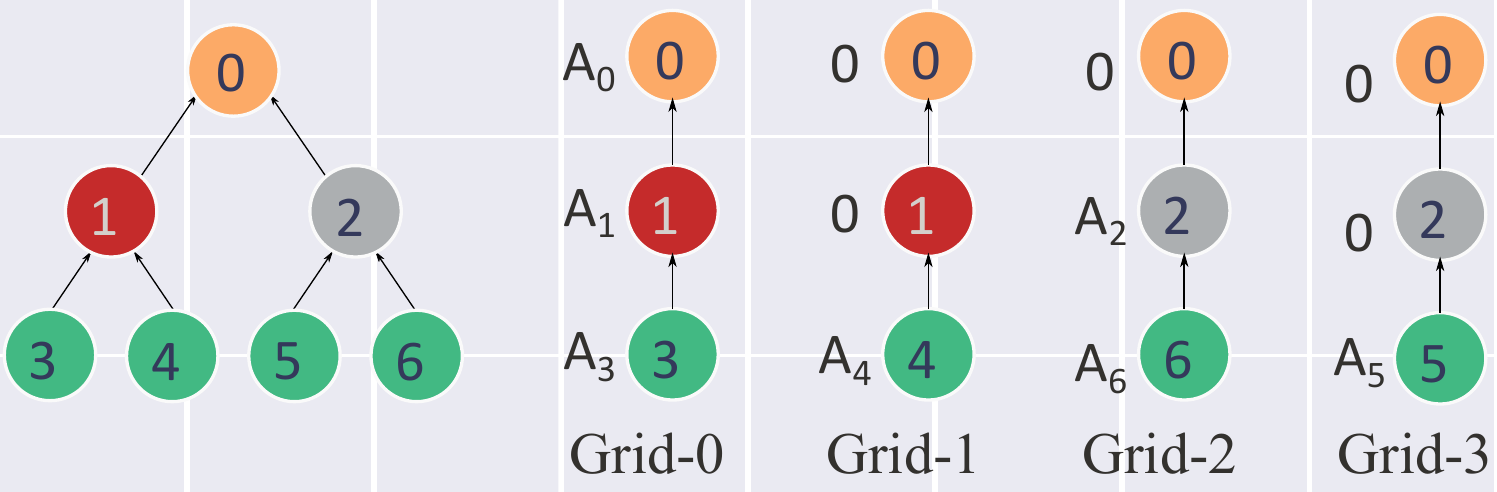} } \quad
\subfloat[b][]{\includegraphics[width=0.4\textwidth]{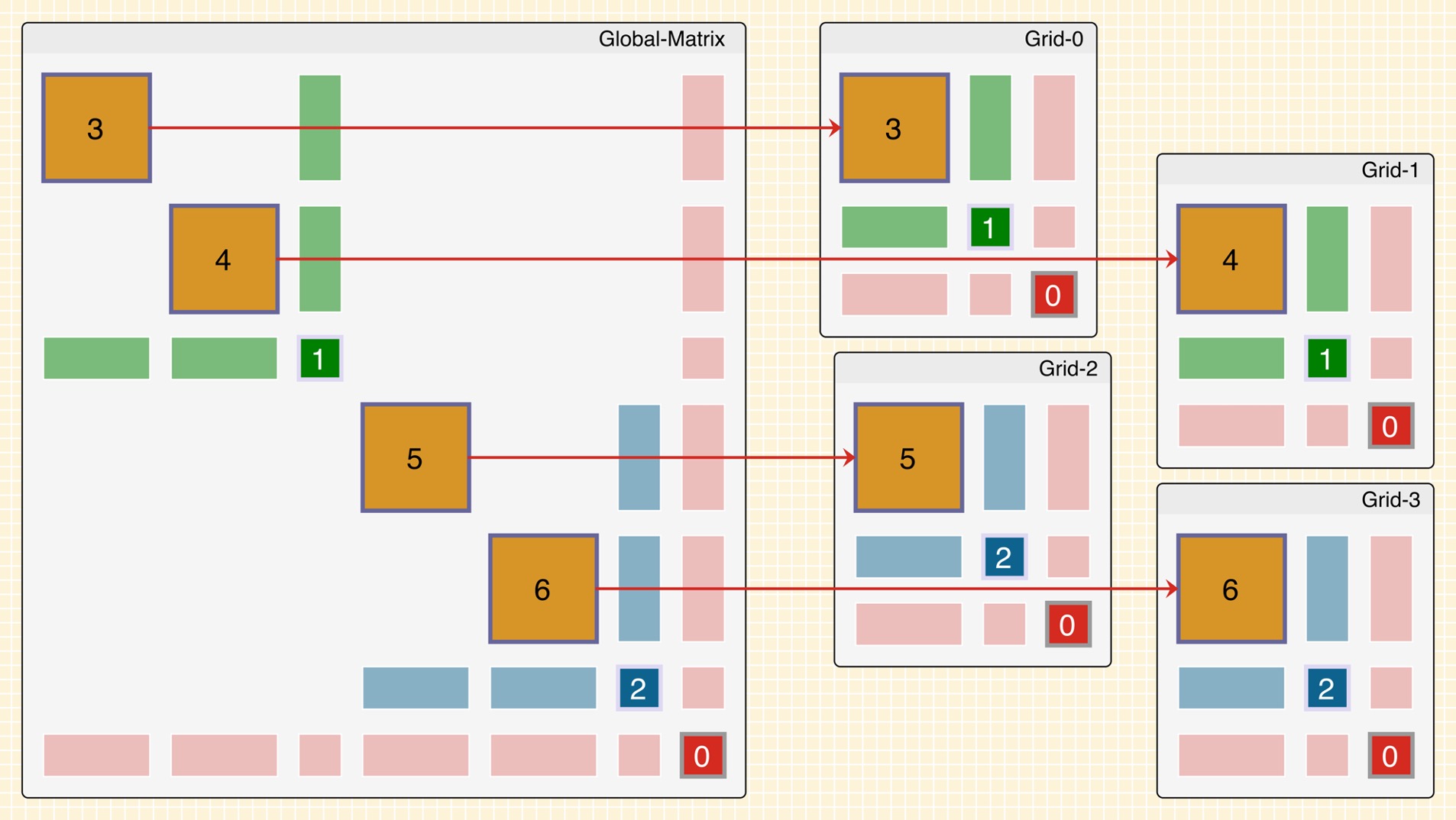}}\label{fig:2level-b}
\caption{Two-level etree partition and the matrix view of the submatrix mapping to four 2D process grids.}
\label{fig:2level-matrix}
\end{figure}

\Cref{fig:3dprocs} illustrates the 3D processes grid, configured as $P = P_x$  $\times$  $P_y$ $\times$  $P_z$.
It can be considered as $P_z$ layers of 2D processes.
The distribution of the sparse matrices is governed by the supernodal elimination tree: the standard etree is transformed into
an etree which is binary at the top $\log_2(P_z)$ levels and has $P_z$
subtrees at the bottom level. 
The transformation may make some subtrees disconnected, hence becoming forests. But for notational simplicity, we still use the term etree herein.
The matrices $A$, $L$, and $U$ corresponding to each
subtree are assigned to one 2D process layer. The 2D layers are referred
to as Grid-0, Grid-1, $\ldots$, for all $P_z$ grids.
\Cref{fig:2level-matrix} shows the submatrix mapping to the four 2D process
grids. 
The sparse LU on each 2D grid is performed using the same 2D algorithm locally within the 2D processes. Each 2D LU generates a Schur complement update to the common ancestor of the etree. The key feature of 3D LU is to replicate the Schur complement memory corresponding to the common ancestor tree among the $P_z$ 2D process layers, so that each 2D grid can accumulate the Schur complement (partial sum) in its local memory, in parallel with the other 2D grids.
Finally, the individual copies of partial sums are reduced (added) onto a single grid, where they are factorized in a 2D fashion.
The communication reduction is mainly due to the independent 2D subgrids working on smaller subproblems.
For example, in~\cref{fig:2level-a}, let $S_i^j$ denote the partial update of the Schur complement that is computed and residing in Grid-j but is destined to the etree node $i$.
At the leaf level, Grid-0 and Grid-1 first factorize nodes 3 and 4 in parallel.
Grid-0 produces the partial updates $S_1^0$ and $S_0^0$.
Grid-1 produces the partial updates $S_1^1$ and $S_0^1$.
Then the two grids reduce all the common ancestor nodes, namely nodes 1 and 0, to Grid-0. This amounts to subtracting $S_1^0 + S_1^1$ from $A_1$ and subtracting $S_0^0 + S_0^1$ from $A_0$. 
In parallel, similar operations occur for Grid-2 and Grid-3 along  the etree nodes 5, 6, 2 and 0.
At one level up, only Grid-0 and Grid-2 are active. 
They factorize nodes 1 and 2, producing $S_0^0$ and $S_0^2$, and reduce the updates on node 0 to Grid-0.
At the root level, only Grid-0 factors node 0.
The generalization of this process for any $P_{z}=2^{\ell}$ is given as Algorithm 1 in~\cite{sao-JPDC-2019}.
In practice, the 3D SpLU algorithm improves strong scaling from 1000 cores to 32,000 cores, leading to
speedups of up to 27$\times$ on planar and up to 3.3$\times$ on non-planar sparse matrices.

For a sparse matrix with arbitrary sparsity pattern, it is infeasible to quantify the communication reduction in ``big O" sense.  Fortunately, for a large class of discretized prototypical PDEs, there exist analytical closed forms of nonzero patterns in the LU factors. These include the finite difference discretization of PDEs defined on a 2D geometry giving rise to planar graphs, and PDEs defined on a 3D geometry giving rise to non-planar graphs. Sao et al. analyzed the memory-communication trade-off for these classes of problems.
They assume that the matrices are first reordered by the nested dissection ordering~\cite{george73} or recursive graph partitioning~\cite{Lipton79,gilbert87}, which is optimal for this class of problems~\cite{george73,Lipton79,gilbert87}. With this ordering, the nonzero counts (and hence the memory) in the LU factors are $\bigo{n \log n}$ for planar graphs and $n^{4/3}$ for non-planar graphs. The flop counts are $n^{3/2}$ for planar graphs and $n^2$ for non-planar graphs.
The etree corresponds to the \emph{separator tree} where each internal node is a separator at a certain middle level of the dissection.
The memory and communication analyses use the following two key facts:
\begin{itemize}
    \item Each separator node in the etree corresponds to a dense submatrix;
    \item Communication time is determined by the amount of communication occurring on the \emph{critical path} of the computational graph. 
    Here, the critical path is a leaf-to-root path in the separator tree.
\end{itemize}

\Cref{tab:asympt2d} summarizes the analysis results regarding the memory, bandwidth and latency costs, comparing the 2D and 3D sparse LU algorithms.
The analysis calculates the memory and communication in two separate phases --- first in the 2D algorithm for each subtree and second in the ancestor-reduction step. Thus, in the column ``3D-SparseLU", two separate terms contribute to memory and communication.
The calculation is only needed for Grid-0 as it is the only grid that participates in all levels and is on the critical path.
The last column suggests that setting $P_z =\bigo{\log n}$ would lead to the minimal memory and communication costs for the 3D algorithm, assuming $P\gg \log n$.
Similarly, \cref{tab:asympt3d} summarizes the memory, bandwidth and latency costs for the PDEs associated with the non-planar graph.


\begin{table}
\begin{centering}
\ra{1.2}
\begin{threeparttable}
\begin{tabular}{@{}llllll@{}} \hline
& &\multicolumn{3}{c} { \thead{{\bf 2D PDE}} } \\ \cline{3-5}
{Parameter} & & {2D-SparseLU } & { 3D-SparseLU } & 3D-SparseLU (min)  \\
  & & & & $P_{z}=\bigo{\log n}$  \\ \hline
\textbf{FLOP count } \\ 
  \quad per process ($F$) & & $\frac{\bigo{n^{3/2}}}{P}$ & $\frac{\bigo{n^{3/2}}}{P}$ & $\frac{\bigo{n^{3/2}}}{P}$ 
  \\
\textbf{Memory} \\ \quad per process ($M$) & &
{$\bigo{\frac{n}{P}\log n}$} & {$\bigo{\frac{n}{P}(\log (\frac{n}{P_{z}})+P_{z})}$} & {$\bigo{\frac{n}{P}\log n}$} \\
\textbf{Communication volume} \\ \quad per process ($W$)
& &
{$\bigo{\frac{n}{\sqrt{P}}\log n}$} &
{$\bigo{\frac{n}{\sqrt{P}}(2\sqrt{P_{z}}+\frac{\log n}{\sqrt{P_{z}}})+\frac{2nP_{z}\log P_{z}}{P}}$} &
$\bigo{\frac{n}{\sqrt{P}}\sqrt{\log n}}$ 
\\
{\textbf{Latency}}      & &
{$\bigo{n}$} & {$\bigo{\frac{n}{P_{z}}+\sqrt{n}}$} & {$\bigo{\frac{n}{\log n}}$} \\ \hline
\end{tabular}
\end{threeparttable}
\par\end{centering}{\tiny \par}
\caption{\label{tab:asympt2d}~\cite[Table 2]{sao-JPDC-2019} Asymptotic memory, communication, and latency costs for 2D and 3D Sparse LU algorithm for 2D model PDEs.}
\end{table}

\begin{table}
\begin{centering}
\ra{1.2}
\begin{threeparttable}
\begin{tabular}{@{}lllll@{}} \\ \hline
&  & \multicolumn{2}{c}{\thead{\bf 3D PDE} }\\
 \cline{3-4}
{Parameter} & & {2D-SparseLU } & { 3D-SparseLU} \\ \hline 
\textbf{FLOP count } \\ 
  \quad per process ($F$) & & $\bigo{\frac{n^2}{P}}$ & $\bigo{\frac{n^2}{P}}$ \\
\textbf{Memory} \\ \quad per process ($M$) & &
$\bigo{\frac{n^{\frac{4}{3}}  }{P}}$  & $\bigo{ M_{2D}\left (  P_{z} + \frac{1}{P^{4/3}_{z}} \right )}$\\
\textbf{Communication volume} \\ \quad per process ($W$) & &
$\bigo{\frac{n^{4/3}}{\sqrt{P}} }$
&$\bigo{W_{2D}\left(\kappa_{1}\sqrt{P_{z}}+\frac{1-\kappa_{1}}{{P_{z}}^{5/6}}\right) + W_{2D}\left(\kappa_{1}\frac{P_{z}\log P_{z}}{\sqrt{P}}\right)}$ \tnote{$\dagger$} \\             
{\textbf{Latency}}       &   & $\bigo{n}$                              & $\bigo{ L_{2D} \left ( \kappa_{2} n^{-1/3} +\frac{1}{P_{z}} \right ) }  $ \tnote{$\ddagger$}\\ \hline 
\end{tabular}
\begin{tablenotes}                         
\item[$\dagger$] The first and second term in the expression for $W_{3D}$ represent $W_{3D}^{\
xy}$ and $W_{3D}^{z}$, respectively. $\kappa_{1} = \frac{2(\alpha-1)}{\alpha^{4}-1}$, where $\alpha\
=2^{1/3}$.
\item[$\ddagger$] $\kappa_{2} = \frac{\alpha}{\alpha+1}$
\end{tablenotes}
\end{threeparttable}
\par\end{centering}{\tiny \par}
\caption{\label{tab:asympt3d}~\cite[Table 3]{sao-JPDC-2019} Asymptotic memory, communication, and latency costs for 2D and 3D Sparse LU algorithm for 3D model PDEs.}
\end{table}

\begin{figure}[t]
\subfloat[t][ \label{fig:pde2d}]{\includegraphics[width=0.45\textwidth]{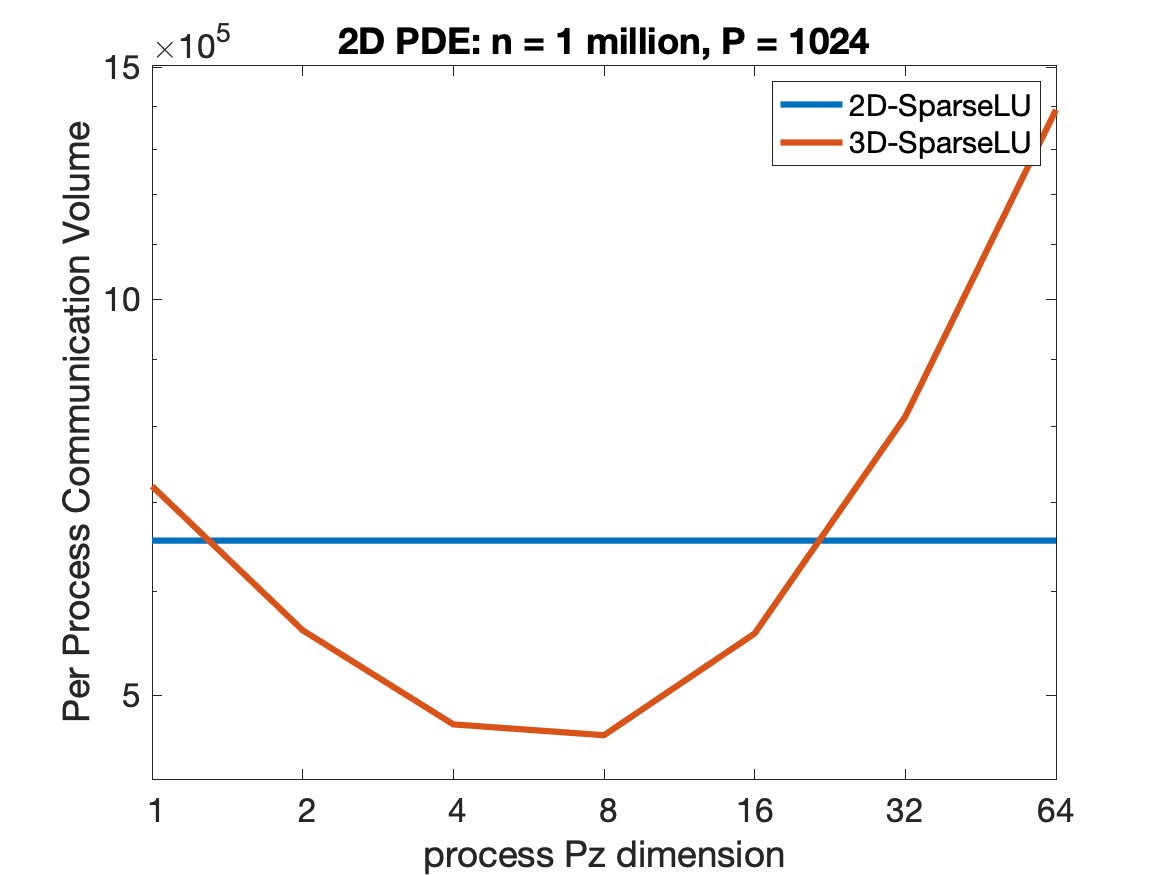} } \quad
\subfloat[b][ \label{fig:pde3d}]{\includegraphics[width=0.45\textwidth]{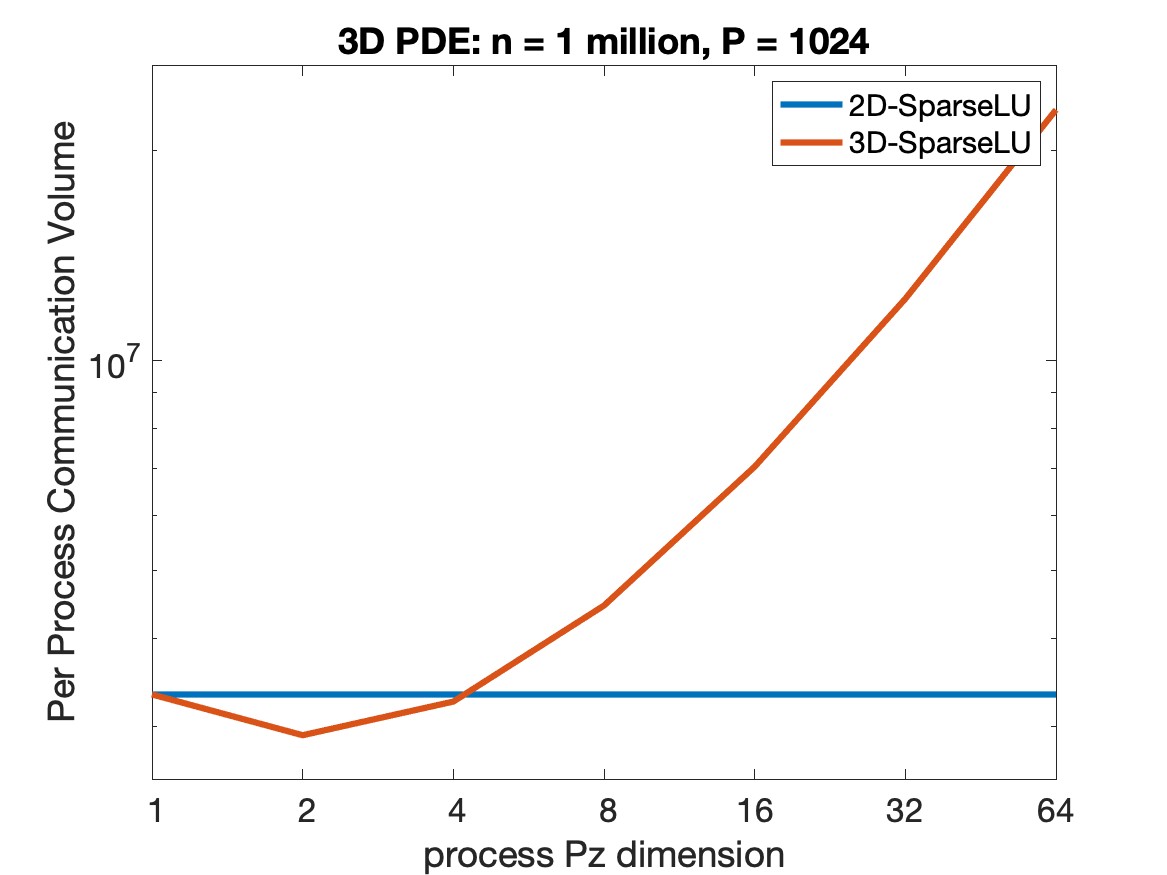}}
\caption{Asymptotic per process communication volumes given in~\cref{tab:asympt2d}
\rev{$\left(\frac{n}{\sqrt{P}}(2\sqrt{P_{z}}+\frac{\log n}{\sqrt{P_{z}}})+\frac{2nP_{z}\log P_{z}}{P}\right)$}
and~\cref{tab:asympt3d}
\rev{$\left(W_{2D}\left(\kappa_{1}\sqrt{P_{z}}+\frac{1-\kappa_{1}}{{P_{z}}^{5/6}}\right) + W_{2D}\left(\kappa_{1}\frac{P_{z}\log P_{z}}{\sqrt{P}}\right) \right)$}, respectively,
with different $P_z$ settings.}
\label{fig:volume-analysis}
\end{figure}

In both tables, the formulae for communication volumes in the 3D sparse LU are non-trivial functions depending on $P_z$.
To provide visual perspective of the two functions, we plot these formulae with increasing $P_z$ while fixing the matrix dimension $n$ to be 1 million and the total number of processes $P$ to be 1024, see~\cref{fig:volume-analysis}.
In both cases, there are some sweet spots for the $P_z$ values that would minimize the communication volume. It is not always true that the larger $P_z$ the less communication. Note that these formulae are just in ``big O" sense and do not contain the constant prefactors in each case. Therefore, the relative communication volumes between the 3D and 2D SpLU are just notional magnitudes and should not be considered to be the actual differences on a real machine.
In practice, Sao's paper showed that $P_z$ ranging from 8 to 16 is usually good for both planar and non-planar problems. 

\paragraph{\bf Communication-avoiding 3D SpTRSV}
Following the same 3D SpLU principle and the distribution of the LU factors, Sao et al. designed a new 3D SpTRSV algorithm~\cite{sao2019}. Since the diagonal block of the $L$ (or $U$) factor associated with a leaf subtree or an ancestor separator resides only on 2D process grids, the corresponding SpTRSV operations would still use the prior 2D algorithm.
Similar to 3D LU, there is some replicated data on the common ancestor nodes at the higher levels of etree.
Different from LU, where the replicated memory is for the partial Schur complement updates, here in SpTRSV, the replicated memory is for the partial updates to the right-hand side matrix $B$. 
Each 2D grid accumulates the local partial sum that is reduced to update a part of the right-hand side corresponding to the common ancestor nodes.

Consider solving the lower triangular system $L Y = B$, using the same example as in~\cref{fig:2level-matrix}. The right-hand side $B$ (and solution $Y$) is partitioned into 7 block rows,
$B_0, B_1, \ldots, B_6$, corresponding to the 7 etree nodes. Let $S_i^j$ denote the partial update that is computed and residing in Grid-j but is destined to the etree node $i$ for block $B_i$.
At the leaf level, Grid-0 and Grid-1 first solve nodes 3 and 4 in parallel, obtaining $Y_3 = L_3^{-1} B_3$ and $Y_4 = L_4^{-1} B_4$.
Grid-0 produces the partial updates $S_1^0 = - L_{1,3}\cdot Y_3$ and $S_0^0 = - L_{0,3}\cdot Y_3$ destined to $B_1$ and $B_0$, respectively.
Grid-1 produces the partial updates $S_1^1 = -L_{1,4}\cdot Y_4$ and $S_0^1 = -L_{0,4}\cdot Y_4$ destined to $B_1$ and $B_0$.
The two grids reduce the updates at nodes 1 and 0, forming the updated $B_1 = B_1 + S_1^0 + S_1^1$ and the pending update $S_0^0 = S_0^0 + S_0^1$ on Grid-0. Grid-0 proceeds to solve $Y_1 = L_1^{-1}\cdot B_1$ using the 2D algorithm.
In parallel, similar operations occur for Grid-2 and Grid-3 along the etree nodes 5, 6, 2 and 0. 
At one level up, only Grid-0 and Grid-2 are active. They solve for $Y_1 = L_1^{-1}\cdot B_1$ and $Y_2 = L_2^{-1}\cdot B_2$. They then reduce the merged updates 
$S_0^0 - L_{0,1}\cdot Y_1$ (Grid-0) and $S_0^2 - L_{0,2}\cdot Y_2$ (Grid-2) on node 0 to Grid-0. At the root of the etree, only Grid-0 solves for $Y_0$ using the 2D algorithm.
The generalization of this process for any $P_z = 2^\ell$ is given in~\cite[Algorithm 2]{sao2019}.
\rev{\Cref{tab:SpTRS-comm-complexity} summarizes the analysis results about the communication cost and per-process communication volume, comparing the 2D and
3D SpTRSV algorithms. The communication cost measures the number of words
sent along the critical path of the computation.}
Liu et al. \cite{Liu-SC23} further improved the 3D SpTRSV algorithm by allowing replicated computation on each grid by setting the corresponding RHS (Right-Hand Side) to 0. This greatly simplifies the 3D algorithm design, and removes the need for inter-grid communication and synchronization at each level, albeit with moderately increased inner-grid communication and computation. This approach achieved up to a 3.45$\times$ additional speedup compared with the above-mentioned algorithm \cite{sao2019} on up to 2048 CPU cores.   



\begin{table}
    \begin{centering}
    \ra{1.2}
    \begin{threeparttable}
        \begin{tabular}{@{}llllll@{}} \\ \hline
{Problem type}& {Communication } & & {2D SpTRSV } & {3D SpTRSV } \\ \hline
Planar    &Cost (W)  & &
    {$\bigo{ \frac{n}{\sqrt{P}}{+} \sqrt{n}}$} & {$\bigo{
        \frac{n}{\sqrt{P_{z}P}}{+}\sqrt{n}}$}  \\
(2D PDE)  & Average Volume  & &
    {$\bigo{ \frac{n}{\sqrt{P}}}$} &
    $\bigo{\frac{n}{\sqrt{P_{z}P}} }$ \\
    & Max Volume  & &
    {$\bigo{ \frac{n}{\sqrt{P}}}$} &
    {$\bigo{
        \frac{n}{\sqrt{P_{z}P}} {+}\frac{\sqrt{nP_{z}}}{\sqrt{P}}  }$} \\ \hline
Non-Planar    & Cost (W)  & &
    {$\bigo{ \frac{n}{\sqrt{P}}{+} n^{2/3}}$} & {$\bigo{
        \frac{n}{\sqrt{P_{z}P}}{+}n^{2/3}}$}  \\
(3D PDE)   &Average Volume  & &
    {$\bigo{ \frac{n}{\sqrt{P}}}$} &
    $\bigo{ \frac{n}{\sqrt{P_{z}P}}}$ \\
    & {Max Volume  } & &
    {$\bigo{ \frac{n}{\sqrt{P}}}$} &
    {$\bigo{\frac{n}{\sqrt{P_{z}P}} {+} n^{2/3}\frac{\sqrt{P_{z}}}{\sqrt{P}} }$} \\ \hline
    \end{tabular}        
    \end{threeparttable}
    \par\end{centering}{\tiny \par}
\caption{\label{tab:SpTRS-comm-complexity}
\cite[Table 2]{sao2019} Asymptotic communication
    cost and volume for 2D SpTRSV and 3D SpTRSV,
    on planar (2D PDE) and non-planar (3D-PDE) input problems}
    \end{table}



\subsection{Communication lower bounds}
Even though the 3D CA algorithm framework implemented in SuperLU\_DIST has low communication complexity among the widely used parallel sparse direct solvers, a natural question arises whether this communication complexity reaches the lower bound, that is, whether this is a communication optimal algorithm and whether we can continue to seek better algorithms to close the gap from optimum. 
The answer to this question is not simple. Firstly, the communication cost consists of two parts: bandwidth (volume) and latency (number of messages). Some algorithms may minimize one cost but not both, or may be able to trade-off one for another. Secondly, some algorithms may be able to trade-off floating point operations and/or memory for lower communication cost. 
Thirdly, for sparse matrices, the optimality of an algorithm may depend on the sparsity pattern.

Let $F$ denote the number of floating point operations per process, and let $M$ denote the local memory size. 
For a broad class of matrix-multiplication-like linear algebra algorithms, dense or sparse, it was established that the lower bounds on communication bandwidths and  communication latencies are:
\begin{equation}
W = \Omega\left(\frac{F}{\sqrt{M}}\right)\,, \qquad
S = \Omega\left(\frac{F}{M^{3/2}}\right)\,,
\label{eq:lower-bound}
\end{equation}
see~\cite{Ballard_et_al_2014} and the historical notes therein. These formulae suggest that when increasing the local memory size, the lower bounds for both $W$ and $S$ can be decreasing. This intuition prompted the development of some CA algorithms that replicate some data for reduced communication.
In~\cite{solomonik2011COLU}, Solomonik and Demmel presented a 2.5D dense LU algorithm that arranges $P$ processes as $c$ layers of 2D process grids. The dense matrix is duplicated on each layer, thus, the algorithm 
uses $M\approx c n^2/P$ local memory. The algorithm reduces the bandwidth by a factor of $\sqrt{c}$ in comparison with standard 2D LU algorithms (when $c=1$). However, it sends a factor of $\sqrt{c}$ more messages than the 2D algorithm. 

For SpLU, the two available open-source implementations with low communication complexity are: the multifrontal approach with subtree-to-subcube mapping proposed by Gupta et al.~\cite{gupta-kk97} and the right-looking 3D mapping approach proposed by Sao et al. (see ~\cref{sec:ca-3d} and~\cite{sao-JPDC-2019}). 
Both of these approaches fully exploit the etree parallelism and do not suffer the latency cost increase as does the 2.5D dense LU.
Gupta's algorithm  partitions the 2D process grid into smaller 2D process grids that are assigned to the frontal matrices at different levels. Sao's 3D mapping approach partitions
the 3D process grid into smaller 3D process grids.
The essence of both approaches is to use multiple copies of the matrix to perform multiple Schur-complement updates in parallel. 
The low communication complexity of the multifrontal algorithm is also attributed to trading additional memory for reduced communication. This additional memory stores the multiple partial update matrices, i.e., the $T$ matrices in~\cref{fig:GE-rl-mf}. 
For three-dimensional and higher-dimensional sparse matrices, \rev{e.g., those corresponding to PDEs on 2D and 3D spatial grids}, the sparse LU operation is dominated by the dense LU at the top level separator, therefore, the behavior of both Gupta's and Sao's algorithms is optimal
in terms of communication. But, for sparse matrices associated with planar graphs, 
both algorithms do not achieve the communication
lower bound specified by~\cref{eq:lower-bound}. Consider a model planar problem
that comes from a finite difference discretization on a 2D grid of size $n=k\times k$. For this, Gupta's algorithm has a gap of $\sqrt{\log n}$ from the bandwidth lower bound, and
Sao's algorithm has a gap of $\log n$ (see last column of~\cref{tab:asympt2d}, lower bound is $F/\sqrt{M} =\bigo{  n/(\sqrt{P}\sqrt{\log n})}$).

In~\cite{Sao-spaa23}, Sao et al. proposed a nearly-optimal algorithm in terms of communication, 
which combines the subtree-to-subcube mapping and the 3D mapping approach. Here, the lower level subtree of the
etree uses subtree-to-subcube mapping and subtrees at higher levels use the 3D mapping algorithm. 
Thus, for higher level subtrees, data replication is used up to a factor of $\log n$, without increasing the asymptotic memory cost. 
This hybrid algorithm achieves the communication lower bound within a factor of $\log \log n$, a considerable improvement over the $\sqrt{\log n}$ and $\log n$ factors of existing approaches. However, it remains an open problem to implement this last communication-optimal algorithm.
 
The software details of the 3D sparse direct solver SuperLU\_DIST are described in~\cite{TOMS_release_2023}.

\section{Uncovering Fine-grain Parallelism for GPU Acceleration}
\label{sec:gpu}
In the exascale computing era, GPUs have become mainstream computing engines. In the November 2025 edition of the TOP500 supercomputer list, all four exascale machines are using various  generations of GPUs from AMD, Intel and NVIDIA~\cite{top500-web}.
On those machines, more than 90\% of the computing power comes from GPUs. Despite their increasing adoption in scientific computing, GPU architectures present unique challenges for sparse computations. Modern GPUs can have thousands of cores, enabling them to run a large number of threads concurrently. This high thread count allows for simultaneous execution of many operations, greatly increasing throughput. This type of SIMD data-parallel model is well-suited to large dense matrix operations.
However, for sparse matrix operations, task parallelism is often more prevalent than data parallelism. 
Significant efforts have been made to expose fine-grained data parallelism. 

Programming on GPUs often requires explicit management of threads and thread blocks. Developers need to understand how to organize computations into many small kernels and manage synchronization between threads, which can be non-intuitive.
 While GPUs can handle many threads in parallel, their architectures are not designed for handling complex control flows. Therefore, most GPU-capable sparse direct solvers still use CPUs to handle control flows and use GPUs only in the \emph{offload mode}.

About fifteen years ago, several CPU-GPU SpLU algorithms were developed that mostly use GPU as a BLAS accelerator (i.e., dense LU, TRSM and GEMM in~\cref{fig:GE-rl-mf})~\cite{lucas10,Wang-Pierce11,george-gupta2011,vuduc2010}. All these earlier developments focused on single GPU or a few GPUs on a single node.
Starting 2014, Sao et al. initiated the first line of work on large-scale distributed memory GPU accelerated sparse direct solvers with implementations in SuperLU\_DIST. This required using the MPI+OpenMP+CUDA programming model with delicate coordinations of both inter-node communication, intra-node CPU-GPU communication and intra-node multithreading on each multicore node. 
In their first distributed memory GPU-enabled SpLU algorithm~\cite{sao2014}, the only part offloaded to GPUs is the GEMM computation that makes up the contribution to each Schur complement update. Sao et al. leveraged NVIDIA's cuBLAS library for the GEMM calls on the GPU side. To overcome the cost of data transfer between the CPU and GPU memories, they developed a sophisticated software pipelining for overlapping the GEMM operations on GPUs with the Scatter operations on CPU. This requires creating multiple CUDA streams to split a single large GEMM or Scatter operation into smaller pieces. 
This algorithm achieved up to 2$\times$ speedups on 8 nodes with 16 Tesla ``Fermi" GPUs. However, for truly sparse matrices, the performance gain is marginal.
Their second algorithm~\cite{sao2015} also offloaded the Schur complement update to GPU, including both GEMM and Scatter operations. The algorithm is called HALO for Highly Asynchronous Lazy Offload, referring to combining
asynchronous use of both CPU and GPU for each Schur complement update accounting for the GPU memory size and the Schur complement matrix size.
The asynchronous and lazy updates serve to hide
and reduce communication over the PCIe link between CPU and GPU and to overcome the small GPU memory capacity. This strategy delivered an additional 2.5$\times$ speedup over the GEMM-only-offload approach.
In their subsequent effort~\cite{sao-JPDC-2019}, on top of the HALO strategy, Sao et al. added additional GPU offloading for the block diagonal dense LU and the L and U panel factorizations, leveraging the cuBLAS TRSM routine and the cuSparse GETRF routine. 

In~\cite{ghysels2022high}, Ghysels and Synk developed the first distributed memory GPU accelerated symmetric-pattern multifrontal sparse direct solver with implementation in STRUMPACK. Their algorithm offloads both dense matrix operations and the sparse scatter–gather operations to GPUs.
They leveraged various vendor libraries for the dense operations, such as NVIDIA's cuBLAS and cuSOLVER and AMD's rocBLAS and rocSOLVE. For the smaller frontal matrices they developed custom CUDA and HIP kernels to reduce kernel launch overhead.  The multi-GPU setting uses SLATE~\cite{gates2019slate} as a modern GPU-aware replacement for ScaLAPACK. On 4 nodes of the Summit machine at Oak Ridge National Laboratory, the
code is $~10\times$ faster when using all 24 V100 GPUs compared to when it only uses the 168 POWER9 CPU cores.
On 8 Summit nodes, using 48 V100 GPUs, the sparse solver reaches over 50TFlop/s.

In a more recent effort in PanguLU~\cite{PanguLU}, Fu et al. developed a different SpLU algorithm which does not use any dense supernode data structure; instead, they use a uniform 2D block partition of the sparse matrix, in which each block is generally sparse and represented in the standard Compressed Sparse Column (CSC) format. In order to accommodate a vast number of different sparsity patterns, they developed 17 different sparse kernels for block diagonal LU, panel lower and upper triangular solution and Schur complement update. Some of them are for CPU and some others for GPU.
Based on the extensive microbenchmarks of various block patterns arising from different sparse matrices, they derived four decision trees to choose a specific kernel dynamically based on each block's sparsity pattern.
Accompanied with an efficient asynchronous scheduling strategy for parallel block operations, PanguLU achieved strong scaling of 47$\times$ and 74$\times$ on 128 NVIDIA A100 GPUs and 128 AMD MPI50 GPUs, respectively, compared to single GPUs.

The new generation of GPUs has much larger memory sizes than earlier generations. For example, an NVIDIA GH200 (2024) Grace Hopper GPU has nearly 100 GB memory whereas an NVIDIA Tesla M2090 (2011) GPU had only 6 GB memory. 
As such, many applications were ported to run entirely on GPUs and there is a demand for the algebraic solvers to run entirely on the GPUs. 
The recent work of the Caracal SpLU algorithm~\cite{Caracal2025} epitomizes the ``GPU-resident" paradigm. Two main techniques developed by Ren et al. greatly enhanced the performance of single GPU implementations: 1) a fine-grained static scheduling algorithm that utilizes multiple GPU streams to perform TRSM and GEMM on multiple sub-blocks simultaneously, and 2) a memory caching algorithm and clever use of the GPU memory pool to enhance data reuse and reduce the frequency of calling cudaMalloc/cudaFee that may cause GPU memory fragmentation. On a single NVIDIA A100 GPU, Caracal achieved up to 21\% of the A100’s theoretical peak performance, 7$\times$ speedup over SuperLU\_DIST, 94$\times$
speedup over PanguLU, and 16$\times$ speedup over PasTiX. 
Even though the Caracal approach eliminates the need for CPU-GPU data transfers during the LU factorization, the inter-GPU communication is still initiated on the CPU. That is, CPUs are not entirely out of the picture. From their limited scaling study using only 4 GPUs, their multi-GPU results are not superior to the other methods. Further improvement is needed for efficient use of large numbers of GPUs.

\paragraph{{\bf Batching}}
In the same vein as the GPU-resident solvers that intend to reduce the number of kernel launches, \emph{batching} also emerged as an efficient algorithm paradigm for GPUs. When executing many small tasks, the overhead of launching kernels and managing resources can dominate the computation time. Batching reduces this overhead by executing a large number of operations within a single kernel call, increasing overall throughput.
This leads to significant speedups compared to executing each task sequentially.
The batched algorithm is particularly suitable for applications that involve a large number of independent and \emph{small to medium sized} linear algebra problems. 
These problems arise from combustion modeling, bio-chemical modeling, fusion plasma modeling, and nuclear physics simulations, to name a few. The dimension of each matrix is of order hundreds to tens of thousands. Most of the GPU throughput would be wasted if we solve each of them individually.
Well-known libraries supporting dense batched matrix operations include MAGMA~\cite{Agullo_2009_jpcs,magma2024} and KBLAS~\cite{kblas-website}. 
The newer library Kokkos Kernels supports batched linear algebra operations for both dense and sparse matrices~\cite{kk-website}.

Batched dense operations are also critical components to enable the development of fast sparse direct solvers on GPUs.
For example, in the GPU version of STRUMPACK, each level of the assembly tree involves multiple independent diagonal LU, off-diagonal TRSM and GEMM for the update matrices. These are all dense operations and can be organized as three batched routines across all the nodes on one level of the tree. However, the sub-matrices within a level are not of the same size. Abdelfattah et al. expanded the batched MAGMA library to handle irregular patterns of matrices in a batch and added new routines irrLU, irrTRSM and irrGEMM~\cite{abdelfattah2022addressing}. They showed that the new MAGMA enabled over 6$\times$ speedup for STRUMPACK running on a single AMD MI100 GPU. 

The SuperLU team led the development of the batched sparse direct solver interface and the implementation in SuperLU\_DIST~\cite{batch_slu}. A salient feature of the new interface is to allow a heterogeneous batch structure such that each linear system can have different size, sparsity pattern and numerical property.
The interface allows each system in the batch to have different preprocessing: equilibration, and row and column reordering,
depending on the numerical values and sparsity pattern of each individual system. The results with some combustion and fusion applications demonstrate
that the batched sparse direct solver can be an order of magnitude faster than the batched banded solver or the batched dense solver in MAGMA.

There are several prominent efforts in developing batching sparse iterative solvers and preconditioners. The modern GPU-oriented Gingko library~\cite{ginkgo-toms-2022} supports batched iterative solvers for both symmetric and nonsymmetric sparse linear systems. The current version supports a fixed sparsity pattern across all the matrices in the batch. That is, their BatchCsr or BatchEll formats store only one copy of the meta-graph data structure, with different numerical values~\cite{ginkgo-batch}. In addition, Ginkgo supports batched Block–Jacobi preconditioners for which the inversion of the diagonal block is performed via a batched variable size Gauss–Jordan elimination. 
The Kokkos Kernels library also supports a number of batched iterative solvers including CG and GMRES. Liegeois et al. described in detail how they optimized the batched sparse matrix vector product (SpMV), which is the performance-dominant operation in most sparse iterative solvers~\cite{kk-batch}.


\section{Acceleration via Data-sparse Compression}
\label{sec:data-sparse} 
Large-scale numerical linear algebra (NLA) algorithms, even possibly exploring exact structural sparsity and communication-avoiding techniques, may remain computationally expensive in terms of flop counts, storage units, communication and data movement. Moreover, modern HPC architectures are increasingly heterogeneous and compute-intensive, but the relatively smaller improvement on memory bandwidth, communication latency and energy efficiency is still limiting the size of problems that can be solved. Data-sparse compression, as a competitive complement, refers to compact representations where the effective amount of information is much smaller than the ambient size of a dense matrix/tensor/array. Data-sparse compression reduces the amount of data that must be stored, moved and processed while maintaining a controllable accuracy level, and has emerged as a fundamental enabler in NLA and scientific computing. 

The data-sparse compression can be applied to the entire dense matrix of interest, its judiciously selected dense submatrices, or dense submatrices of a sparse matrix, leading to significant reduction in the asymptotic complexity and/or its large constant prefactors of the original exact NLA algorithm. Data-sparse compression can be categorized into two complementary families of techniques, i.e., low-rank compression and lossy floating-point compression, which are summarized as follows:  

\textit{Low-rank compression} exploits mathematical or algebraic structure in matrices to approximate them efficiently and accurately. Many matrices arising from PDEs \cite{engquist2018approximate, chandrasekaran2010numerical,faustmann2013new,FaustmannHFEM}, integral equations \cite{Bebendorf2000ACA,hackbusch2002h2}, kernel methods in machine learning \cite{rebrova2018study,chavez2020scalable,chenhan2016inv,chenhan2017n,ambikasaran2015fast,litvinenko2020hlibcov,minden2017fast,ambikasaran2013n} and optimization \cite{feliu2021approximate,alger2019data,ambartsumyan2020hierarchical} are not arbitrary dense matrices: they exhibit latent separability, long-range separability, smoothness, or decay, which implies that submatrices of the matrix have rapidly decaying singular values and can be approximated by products of small matrices of rank far less than the dimension of the matrix. Low-rank compression becomes essential in direct solvers and factorization-based preconditioners (compressing dense frontal matrices and Schur complements), iterative solvers (compressing operator evaluations and Krylov bases), fast transforms, and inverse problems (accelerating Jacobian/Hessian actions and PDE-constrained optimization). In the context of direct solution of large-scale sparse and dense linear systems, low-rank compression includes single low-rank compression \cite{xia2024nystrom,huang2025higp}, rank-structured matrix algorithms leveraging single-level low-rank compression \cite{shaeffer2008direct,al2020solving} or multilevel compression such as hierarchical matrices \cite{wolfgang2015hierarchical,chandrasekaran2006fast,l2016hierarchical,coulier2017inverse,minden2017recursive,ambikasaran2013n}, and their higher-order tensor extensions \cite{liu2025block,li2025hierarchical}. These methods can effectively reduce the $\bigo{n^3}$ computational complexity for factorizing a dense $n\times n$ matrix (or lower but still superlinear complexities for sparse matrices) to quasi-linear, linear or even sublinear complexities. Moreover, many modern low-rank compression methods can reduce the constant pre-factors, fully utilize parallel computers, and maintain low asymptotic complexity, making them capable of solving record-breaking large-scale dense and sparse linear systems.    

\textit{Lossy floating-point compression} exploits bit-level redundancy in numerical arrays. Rather than exploiting algebraic structure by low-rank compression, it reduces cost by quantizing or transforming data to a compressed representation with controlled errors. Matrices and multi-dimensional arrays, arising from PDE solutions and operators, machine learning features, fast transforms (e.g, FFT), and numerical algorithms (e.g., AMG) can exhibit local smoothness/correlation, and small dynamic range or low entropy, of which the bit-level representation can be efficiently transformed or quantized to exploit the redundancy. Existing lossy compressors such as ZFP \cite{lindstrom2025zfp}, SZ \cite{di2016fast}, and FPZIP \cite{lindstrom2006fast} can significantly reduce the storage cost of large scientific data and leverage shared-memory and GPU parallelism. To the best of our knowledge, lossy floating-point compression has been mostly used in reducing the memory requirement or communication volume, albeit incurring manageable computational overheads during compressing and decompressing the data.    

It is worth emphasizing that low-rank compression and lossy floating-point compression are two complimentary families of data-sparse compression algorithms: the former exploits mathematical structure (long-range smoothness and spectral decay) and the latter exploits bit-level redundancy (local smoothness and correlation). In practice, one can often attain optimal compression by hybridizing the two algorithms such as in \cite{kriemann2025hierarchical,claus2023sparse,nuca2025adaptive,kielstra2025linear} and we will see a few such examples in~\cref{sec:data-and-structure}. 

In the rest of this section, we will focus on the description of the fundamentals of rank-structured matrices for the direct solution of dense matrices.

\subsection{Overview of rank-structured matrix algorithms}\label{sec:overview_H}
Large-scale dense linear systems $AX=B$ arise from many computational science and engineering, machine learning and statistics problems, including matrices in boundary element methods and volume integral equations, Fourier integral operators, fractional Laplacian operators, kernel matrices and covariance matrices in machine learning, Hessian matrices in PDE-constrained optimization, and effectively dense matrices due to fill-in of a sparse direct factorization for PDEs. Exact arithmetic-based solvers for these large linear systems can be computationally very demanding even leveraging modern parallel HPC architectures. This is particularly the case when the system becomes ill-conditioned and/or involves many right-hand side vectors $B$, as in multi-physics, multi-scale simulations and machine learning pipelines. That said, these matrices often exhibit rich mathematical and physical structures that can be exploited to significantly reduce the solver complexities. We will see a few such examples in~ \cref{sec:app_example}.   

Rank-structured matrix algorithms \cite{martinsson2025fast} aim at solving these large $n\times n$ linear systems with less than the conventional methods' $\bigo{n^2}$ memory and $\bigo{n^3}$ computational requirements. Taking hierarchical matrices as an example, they leverage the ideas of multilevel matrix partitioning and low-rank compression. They can be viewed as the algebraic version of the fast multipole methods with additional support for matrix inversion. After a proper \textit{preprocessing} phase that reorders the matrix rows/columns and hierarchically clusters them into groups, certain off-diagonal blocks at each level $\ell$ of the reordered matrix reveal geometric structure or smooth interactions, leading to rapidly decaying singular values. This means these blocks can be approximated, up to a user-specified accuracy tolerance $\epsilon$, as low-rank matrices:
\begin{align}
A^\ell_{st}\approx U_{st}V^T_{st}.\label{eq:low-rank}
\end{align}
Here $A^\ell_{st}$ is the $n_s\times n_t$ block corresponding to row group $s$ and column group $t$, $U_{st}$ and $V^T_{st}$ are the low-rank factors, and the rank $r\ll\min(n_s,n_t)$ depends on the specified tolerance $\epsilon$ and the nature of the underlying kernel. As a result, one needs far less than $\bigo{n_sn_t}$ flop operations and storage units for this block. After compressing all off-diagonal blocks at each level of the hierarchical clustering (which is referred to as the \textit{construction} phase), one ends up with a compressed representation $H\approx A$ with almost $\bigo{nr}$ memory requirement. Here we assume all blocks have the same numerical rank $r$ for simplicity of explanation. 

Once the construction of $H$ is finished, many basic operations on $A$ can be approximated and accelerated using $H$ instead of $A$. These include \textit{factorization} of $H$ into $H^{-1}$ \cite{ambikasaran2013n,grasedyck2003construction} or its equivalents (e.g., ULV \cite{chandrasekaran2006fast,chandrasekaran2007fast}, LU \cite{grasedyck2003construction,shaeffer2008direct,Sushnikova2023}, QR \cite{kressner2018fast}, Cholesky \cite{xia2010fast,grasedyck2003construction} and telescoping factorization \cite{martinsson2005fast,l2016hierarchical}), \textit{application} of $H$ or $H^{-1}$ to given vectors (also called the \textit{solve} phase), recompression of the composition of $H$ and other operators into a new rank-structured matrix, etc.  

There is a wide variety of rank-structured matrices developed in applied mathematics, computer science and engineering (e.g., acoustics, seismics, electromagnetics and fluid dynamics) societies. Multilevel partitioning-based algorithms, i.e., hierarchical matrices, can be roughly classified into four categories depending on two criteria, i.e., the admissibility condition and the shared basis property. The admissibility condition refers to whether the low-rank compression of \cref{eq:low-rank} is applied to all off-diagonal blocks (dubbed weak admissibility) or only to those that are ``well-separated" from each other using certain distance metrics (dubbed strong admissibility); the shared basis property refers to whether the $U_{st}$ (and $V_{st}$) factors for two row (and column) groups in \cref{eq:low-rank} share common column (and row) bases with other groups. Weak-admissible and independent-basis algorithms typically refer to hierarchically off-diagonal low-rank (HODLR) matrices \cite{ambikasaran2013n}, and strong-admissible and independent-basis algorithms include $\mathcal{H}$ matrices \cite{wolfgang2015hierarchical} and their variants \cite{khan2024hodlrdd}. Weak-admissible and shared-basis algorithms include hierarchically semi-separable (HSS) matrices \cite{chandrasekaran2006fast,xia2010fast,xia2014n,gillman2012direct}, hierarchical interpolative factorization (HIF) \cite{l2016hierarchical}, and strong-admissible and shared-basis algorithms include $\mathcal{H}^2$ matrices \cite{wolfgang2015hierarchical,ma2017accuracy}, inverse fast multipole methods (IFMM) \cite{coulier2017inverse}, and strong recursive skeletonization factorization (RS-S) \cite{minden2017recursive}, etc. In general, strong admissibility, as opposed to weak admissibility, leads to smaller ranks for high-dimensional problems (e.g., solving PDEs on 3D manifolds or in 3D spaces, or performing kernel regression in high-dimensional feature spaces), but results in higher computational costs for low-dimensional problems; shared-basis based algorithms are more sophisticated but computationally less expensive than independent-basis based algorithms. A summary of the four major classes of hierarchical matrix algorithms is provided in~\cref{tab:hierarchical_matrix}. 

When dealing with matrices arising from oscillatory problems such as high-frequency wave equations and oscillatory integral transforms, the rank $r$ in 
\cref{eq:low-rank}
can still remain high even for strong-admissibility-based hierarchical matrix algorithms. In this regime, one can replace the low-rank compression $A^l_{st}$ in~\cref{eq:low-rank} with multilevel compression algorithms such as the directional fast multipole method \cite{engquist2007fast} or butterfly compression \cite{michielssen1996MLMDA,liu_butterfly:2020}, which leads to efficient extensions of hierarchical matrices capable of solving oscillatory problems with quasi-linear computational complexity. These algorithms include strong-admissible formats e.g., directional $\mathcal{H}^2$ matrices \cite{Steffen2017directionalh2} and hierarchical block LU-based solvers \cite{Han_2017_butterflyLUPEC,heldring2025fast}, as well as weak-admissible formats e.g., hierarchical off-diagonal butterfly matrices (HODBF) \cite{Liu_2017_HODBF,Sadeed2022VIE} and directional preconditioners \cite{ying2015directional}. 

As opposed to the hierarchical matrix algorithms, single-level partitioning based algorithms, named block low-rank (BLR) matrices or tile low-rank matrices \cite{shaeffer2008direct,al2020solving}, only compress off-diagonal blocks of $A^l_{st}$ in~\cref{eq:low-rank} at a fixed level $\ell$. These algorithms are simpler than hierarchical matrix algorithms but can also provide significant computational and memory savings. Just like the hierarchical matrix algorithms, BLR algorithms also have variants that use shared bases or different admissibility conditions \cite{pearce2025randomized,ashcraft2021block,Theo19bridging}.  

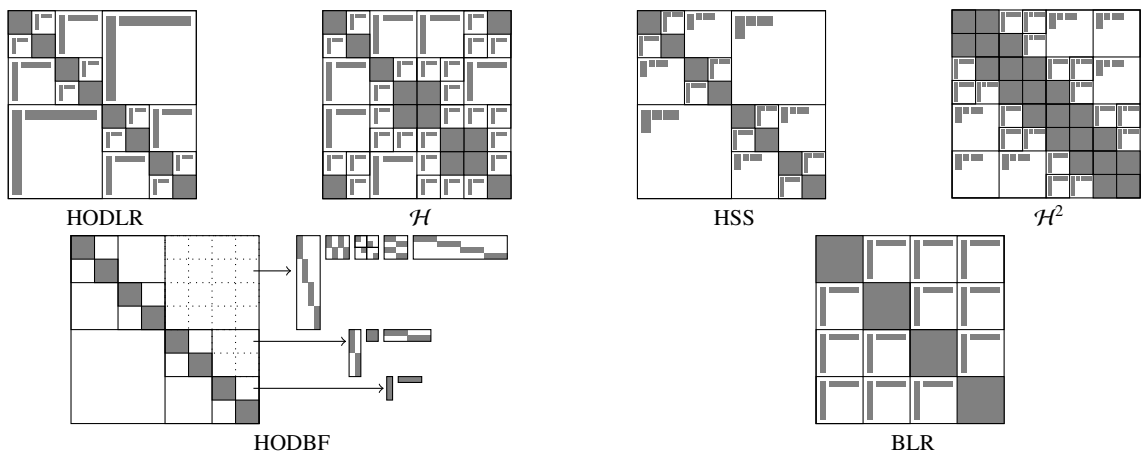
\begin{figure}
	\captionsetup[subfigure]{labelformat=empty}
    \begin{subfigure}[b]{.24\textwidth}
		\begin{center}
			\begin{tikzpicture}[scale=2.5]
			\draw (0,0) rectangle (1,1);
			\draw (0,0) rectangle (0.5,0.5);
			\draw (1,1) rectangle (0.5,0.5);
			\draw (0,1) rectangle (0.25,0.75);
			\draw [fill=gray] (0,1) rectangle (0.25/2,0.75+0.25/2);
			\draw [fill=gray] (0.25/2,0.75+0.25/2) rectangle (0.25,0.75);
			\draw (0.5,0.5) rectangle (0.75,0.25);
			\draw [fill=gray] (0.5,0.5) rectangle (0.75-0.25/2,0.25+0.25/2);
			\draw [fill=gray] (0.75-0.25/2,0.25+0.25/2) rectangle (0.75,0.25);
			\draw (0.25,0.75) rectangle (0.5,0.5);
			\draw [fill=gray] (0.25,0.75) rectangle (0.5-0.25/2,0.5+0.25/2);
			\draw [fill=gray] (0.5-0.25/2,0.5+0.25/2) rectangle (0.5,0.5);
			\draw (0.75,0.25) rectangle (1,0);
			\draw [fill=gray] (0.75,0.25) rectangle (0.75+0.25/2,0.25/2);
			\draw [fill=gray] (0.75+0.25/2,0.25/2) rectangle (1,0);
			
			\path [fill=gray] (0.02,0.02) rectangle (0.07,0.47);
			\path [fill=gray] (0.09,0.42) rectangle (0.47,0.47);
			\path [fill=gray] (0.02+0.5,0.02+0.5) rectangle (0.07+0.5,0.47+0.5);
			\path [fill=gray] (0.09+0.5,0.42+0.5) rectangle (0.47+0.5,0.47+0.5);
			
			\path [fill=gray] (0.52,0.02) rectangle (0.55,0.225);
			\path [fill=gray] (0.57,0.195) rectangle (0.725,0.225);
			\path [fill=gray] (0.52+0.25,0.02+0.25) rectangle (0.55+0.25,0.225+0.25);
			\path [fill=gray] (0.57+0.25,0.195+0.25) rectangle (0.725+0.25,0.225+0.25);
			\path [fill=gray] (0.52-0.5,0.02+0.5) rectangle (0.55-0.5,0.225+0.5);
			\path [fill=gray] (0.57-0.5,0.195+0.5) rectangle (0.725-0.5,0.225+0.5);
			\path [fill=gray] (0.52-0.25,0.02+0.75) rectangle (0.55-0.25,0.225+0.75);
			\path [fill=gray] (0.57-0.25,0.195+0.75) rectangle (0.725-0.25,0.225+0.75);
			
			\path [fill=gray] (0.02,0.02+0.75) rectangle (0.04,0.25/2-0.02+0.75);
			\path [fill=gray] (0.05,0.25/2-0.02-0.02+0.75) rectangle (0.25/2-0.02,0.25/2-0.02+0.75);
			\path [fill=gray] (0.02+0.25/2,0.02+0.75+0.25/2) rectangle (0.04+0.25/2,0.25/2-0.02+0.75+0.25/2);
			\path [fill=gray] (0.05+0.25/2,0.25/2-0.02-0.02+0.75+0.25/2) rectangle (0.25/2-0.02+0.25/2,0.25/2-0.02+0.75+0.25/2);
			\path [fill=gray] (0.02+0.25,0.02+0.5) rectangle (0.04+0.25,0.25/2-0.02+0.5);
			\path [fill=gray] (0.05+0.25,0.25/2-0.02-0.02+0.5) rectangle (0.25/2-0.02+0.25,0.25/2-0.02+0.5);
			\path [fill=gray] (0.02+0.25+0.25/2,0.02+0.5+0.25/2) rectangle (0.04+0.25+0.25/2,0.25/2-0.02+0.5+0.25/2);
			\path [fill=gray] (0.05+0.25+0.25/2,0.25/2-0.02-0.02+0.5+0.25/2) rectangle (0.25/2-0.02+0.25+0.25/2,0.25/2-0.02+0.5+0.25/2);
			
			\path [fill=gray] (0.5+0.02,0.02+0.75-0.5) rectangle (0.5+0.04,0.25/2-0.02+0.75-0.5);
			\path [fill=gray] (0.5+0.05,0.25/2-0.02-0.02+0.75-0.5) rectangle (0.5+0.25/2-0.02,0.25/2-0.02+0.75-0.5);
			\path [fill=gray] (0.5+0.02+0.25/2,0.02+0.75+0.25/2-0.5) rectangle (0.5+0.04+0.25/2,0.25/2-0.02+0.75+0.25/2-0.5);
			\path [fill=gray] (0.5+0.05+0.25/2,0.25/2-0.02-0.02+0.75+0.25/2-0.5) rectangle (0.5+0.25/2-0.02+0.25/2,0.25/2-0.02+0.75+0.25/2-0.5);
			\path [fill=gray] (0.5+0.02+0.25,0.02+0.5-0.5) rectangle (0.5+0.04+0.25,0.25/2-0.02+0.5-0.5);
			\path [fill=gray] (0.5+0.05+0.25,0.25/2-0.02-0.02+0.5-0.5) rectangle (0.5+0.25/2-0.02+0.25,0.25/2-0.02+0.5-0.5);
			\path [fill=gray] (0.5+0.02+0.25+0.25/2,0.02+0.5+0.25/2-0.5) rectangle (0.5+0.04+0.25+0.25/2,0.25/2-0.02+0.5+0.25/2-0.5);
			\path [fill=gray] (0.5+0.05+0.25+0.25/2,0.25/2-0.02-0.02+0.5+0.25/2-0.5) rectangle (0.5+0.25/2-0.02+0.25+0.25/2,0.25/2-0.02+0.5+0.25/2-0.5);
			
			\end{tikzpicture}
			\vspace{-4mm}
		\end{center}
		\caption{HODLR}
		\label{fig:Hformats_HODLR}
	\end{subfigure}
\hfill
	\begin{subfigure}[b]{.24\textwidth}
		\begin{center}
			\begin{tikzpicture}[scale=2.5]
			\draw (0,0) rectangle (1,1);
			\draw (0,0) rectangle (0.5,0.5);
			\draw (1,1) rectangle (0.5,0.5);
			\draw (0,1) rectangle (0.25,0.75);
			\draw (0.5,0.5) rectangle (0.75,0.25);
			\draw (0.25,0.75) rectangle (0.5,0.5);
			\draw (0.75,0.25) rectangle (1,0);
			\draw (0.5,0.5) rectangle (0.75,0.75);
			\draw (0.75,0.75) rectangle (1,1);
			
			\draw [fill=gray] (0.5+0.25/2,0.25/2) rectangle (0.75,0.25);
			\draw (0.5,0) rectangle (0.5+0.25/2,0.25/2);
			
			\draw [fill=gray] (0.5,0.5) rectangle (0.25+0.25/2+0.25,0.25+0.25/2+0.25);
			\draw [fill=gray] (0.5+0.25,0.5-0.25) rectangle (0.25+0.25/2+0.5,0.25+0.25/2);
			
			\draw [fill=gray] (0,1) rectangle (0.25/2,1-0.25/2);
			\draw [fill=gray] (0.25/2,1-0.25/2) rectangle (0.25,0.75);
			\draw [fill=gray] (0.25,0.75) rectangle (0.25+0.25/2,0.75-0.25/2);
			\draw [fill=gray] (0.25+0.25/2,0.75-0.25/2) rectangle (0.5,0.5);
			\draw [fill=gray] (0.5,0.5) rectangle (0.5+0.25/2,0.5-0.25/2);
			\draw [fill=gray] (0.5+0.25/2,0.5-0.25/2) rectangle (0.75,0.25);
			\draw [fill=gray] (0.75,0.25) rectangle (0.75+0.25/2,0.25/2);
			\draw [fill=gray] (0.75+0.25/2,0.25/2) rectangle (1,0);
			
			\draw (0.75,0.75) rectangle (0.75-0.25/2,0.75-0.25/2);
			\draw (1,0.5) rectangle (1-0.25/2,0.5-0.25/2);
			
			
			\draw [fill=gray] (1,1) rectangle (0.75+0.25/2,0.75+0.25/2);
			\draw (0.75,0.75) rectangle (0.75+0.25/2,0.75+0.25/2);
			
			\path [fill=gray] (0.02,0.02+0.25) rectangle (0.05,0.225+0.25);
			\path [fill=gray] (0.07,0.195+0.25) rectangle (0.225,0.225+0.25);
			\path [fill=gray] (0.02+0.25,0.02) rectangle (0.05+0.25,0.225);
			\path [fill=gray] (0.07+0.25,0.195) rectangle (0.225+0.25,0.225);
			
			\draw (0.0,0.0) rectangle (0.25,0.25);
			\draw (0.25,0.25) rectangle (0.5,0.5);
			\draw (0.25/2,0.25/2) rectangle (0.25,0.25);
			\draw (0.25,0.25) rectangle (0.25+0.25/2,0.25+0.25/2);
			\draw [fill=gray] (0.25+0.25/2,0.25+0.25/2) rectangle (0.5,0.5);
			\draw [fill=gray] (0.0,0.0) rectangle (0.25/2,0.25/2);
			
			\path [fill=gray] (0.02,0.02+0.5-0.25/2-0.25) rectangle (0.04,0.25/2-0.02+0.5-0.25/2-0.25);
			\path [fill=gray] (0.05,0.25/2-0.02-0.02+0.5-0.25/2-0.25) rectangle (0.25/2-0.02,0.25/2-0.02+0.5-0.25/2-0.25);
			\path [fill=gray] (0.02+0.25/2,0.02+0.5-0.25/2-0.25) rectangle (0.04+0.25/2,0.25/2-0.02+0.5-0.25/2-0.25);
			\path [fill=gray] (0.05+0.25/2,0.25/2-0.02-0.02+0.5-0.25/2-0.25) rectangle (0.25/2-0.02+0.25/2,0.25/2-0.02+0.5-0.25/2-0.25);
			\path [fill=gray] (0.02+0.25/2,0.02+0.5-0.25/2-0.25/2-0.25) rectangle (0.04+0.25/2,0.25/2-0.02+0.5-0.25/2-0.25/2-0.25);
			\path [fill=gray] (0.05+0.25/2,0.25/2-0.02-0.02+0.5-0.25/2-0.25/2-0.25) rectangle (0.25/2-0.02+0.25/2,0.25/2-0.02+0.5-0.25/2-0.25/2-0.25);
			\path [fill=gray] (0.25+0.02,0.25+0.02+0.5-0.25/2-0.25) rectangle (0.25+0.04,0.25+0.25/2-0.02+0.5-0.25/2-0.25);
			\path [fill=gray] (0.25+0.05,0.25+0.25/2-0.02-0.02+0.5-0.25/2-0.25) rectangle (0.25+0.25/2-0.02,0.25+0.25/2-0.02+0.5-0.25/2-0.25);
			\path [fill=gray] (0.25+0.02,0.25/2+0.02+0.5-0.25/2-0.25) rectangle (0.25+0.04,0.25-0.02+0.5-0.25/2-0.25);
			\path [fill=gray] (0.25+0.05,0.25-0.02-0.02+0.5-0.25/2-0.25) rectangle (0.25-0.02+0.25/2,0.25-0.02+0.5-0.25/2-0.25);
			\path [fill=gray] (0.25+0.02+0.25/2,0.25+0.02+0.5-0.25/2-0.25/2-0.25) rectangle (0.25+0.04+0.25/2,0.25+0.25/2-0.02+0.5-0.25/2-0.25/2-0.25);
			\path [fill=gray] (0.25+0.05+0.25/2,0.25+0.25/2-0.02-0.02+0.5-0.25/2-0.25/2-0.25) rectangle (0.25+0.25/2-0.02+0.25/2,0.25+0.25/2-0.02+0.5-0.25/2-0.25/2-0.25);

			\path [fill=gray] (0.02,0.02+0.5) rectangle (0.05,0.225+0.5);
			\path [fill=gray] (0.07,0.195+0.5) rectangle (0.225,0.225+0.5);
			\path [fill=gray] (0.02+0.25,0.02+0.75) rectangle (0.05+0.25,0.225+0.75);
			\path [fill=gray] (0.07+0.25,0.195+0.75) rectangle (0.225+0.25,0.225+0.75);
			
			\path [fill=gray] (0.02+0.5,0.02+0.75) rectangle (0.05+0.5,0.225+0.75);
			\path [fill=gray] (0.07+0.5,0.195+0.75) rectangle (0.225+0.5,0.225+0.75);
			\path [fill=gray] (0.02+0.75,0.02+0.5) rectangle (0.05+0.75,0.225+0.5);
			\path [fill=gray] (0.07+0.75,0.195+0.5) rectangle (0.225+0.75,0.225+0.5);
			
			\path [fill=gray] (0.02+0.75,0.02+0) rectangle (0.04+0.75,0.25/2-0.02+0);
			\path [fill=gray] (0.05+0.75,0.25/2-0.02-0.02+0) rectangle (0.25/2-0.02+0.75,0.25/2-0.02+0);
			\path [fill=gray] (0.02+0.75+0.25/2,0.02+0+0.25/2) rectangle (0.04+0.75+0.25/2,0.25/2-0.02+0+0.25/2);
			\path [fill=gray] (0.05+0.75+0.25/2,0.25/2-0.02-0.02+0+0.25/2) rectangle (0.25/2-0.02+0.75+0.25/2,0.25/2-0.02+0+0.25/2);
			\path [fill=gray] (0.02+0.5,0.02+0+0.25) rectangle (0.04+0.5,0.25/2-0.02+0+0.25);
			\path [fill=gray] (0.05+0.5,0.25/2-0.02-0.02+0+0.25) rectangle (0.25/2-0.02+0.5,0.25/2-0.02+0+0.25);
			\path [fill=gray] (0.02+0.5+0.25/2,0.02+0+0.25+0.25/2) rectangle (0.04+0.5+0.25/2,0.25/2-0.02+0+0.25+0.25/2);
			\path [fill=gray] (0.05+0.5+0.25/2,0.25/2-0.02-0.02+0+0.25+0.25/2) rectangle (0.25/2-0.02+0.5+0.25/2,0.25/2-0.02+0+0.25+0.25/2);
			
			\path [fill=gray] (0.02,0.02+0.75) rectangle (0.04,0.25/2-0.02+0.75);
			\path [fill=gray] (0.05,0.25/2-0.02-0.02+0.75) rectangle (0.25/2-0.02,0.25/2-0.02+0.75);
			\path [fill=gray] (0.02+0.25/2,0.02+0.75+0.25/2) rectangle (0.04+0.25/2,0.25/2-0.02+0.75+0.25/2);
			\path [fill=gray] (0.05+0.25/2,0.25/2-0.02-0.02+0.75+0.25/2) rectangle (0.25/2-0.02+0.25/2,0.25/2-0.02+0.75+0.25/2);
			\path [fill=gray] (0.02+0.25,0.02+0.5) rectangle (0.04+0.25,0.25/2-0.02+0.5);
			\path [fill=gray] (0.05+0.25,0.25/2-0.02-0.02+0.5) rectangle (0.25/2-0.02+0.25,0.25/2-0.02+0.5);
			\path [fill=gray] (0.02+0.25+0.25/2,0.02+0.5+0.25/2) rectangle (0.04+0.25+0.25/2,0.25/2-0.02+0.5+0.25/2);
			\path [fill=gray] (0.05+0.25+0.25/2,0.25/2-0.02-0.02+0.5+0.25/2) rectangle (0.25/2-0.02+0.25+0.25/2,0.25/2-0.02+0.5+0.25/2);
			
			\path [fill=gray] (0.02+0.75,0.02+0.75) rectangle (0.75+0.04,0.25/2-0.02+0.75);
			\path [fill=gray] (0.05+0.75,0.25/2-0.02-0.02+0.75) rectangle (0.75+0.25/2-0.02,0.25/2-0.02+0.75);
			\path [fill=gray] (0.02+0.75+0.25/2,0.02+0.75) rectangle (0.75+0.04+0.25/2,0.25/2-0.02+0.75);
			\path [fill=gray] (0.05+0.75+0.25/2,0.25/2-0.02-0.02+0.75) rectangle (0.75+0.25/2-0.02+0.25/2,0.25/2-0.02+0.75);
			\path [fill=gray] (0.02+0.75,0.02+0.75+0.25/2) rectangle (0.75+0.04,0.25/2-0.02+0.75+0.25/2);
			\path [fill=gray] (0.05+0.75,0.25/2-0.02-0.02+0.75+0.25/2) rectangle (0.75+0.25/2-0.02,0.25/2-0.02+0.75+0.25/2);

			\path [fill=gray] (0.02+0.5,0.02+0) rectangle (0.04+0.5,0.25/2-0.02+0);
			\path [fill=gray] (0.05+0.5,0.25/2-0.02-0.02+0) rectangle (0.25/2-0.02+0.5,0.25/2-0.02+0);
			\path [fill=gray] (0.02+0.5+0.25/2,0.02+0) rectangle (0.04+0.5+0.25/2,0.25/2-0.02+0);
			\path [fill=gray] (0.05+0.5+0.25/2,0.25/2-0.02-0.02+0) rectangle (0.25/2-0.02+0.5+0.25/2,0.25/2-0.02+0);
			\path [fill=gray] (0.02+0.5,0.02+0+0.25/2) rectangle (0.04+0.5,0.25/2-0.02+0+0.25/2);
			\path [fill=gray] (0.05+0.5,0.25/2-0.02-0.02+0+0.25/2) rectangle (0.25/2-0.02+0.5,0.25/2-0.02+0+0.25/2);
			
			\path [fill=gray] (0.25+0.02+0.25,0.02+1-0.25/2-0.25) rectangle (0.25+0.04+0.25,0.25/2-0.02+1-0.25/2-0.25);
			\path [fill=gray] (0.25+0.05+0.25,0.25/2-0.02-0.02+1-0.25/2-0.25) rectangle (0.25+0.25/2-0.02+0.25,0.25/2-0.02+1-0.25/2-0.25);
			\path [fill=gray] (0.25+0.02+0.25+0.25/2,0.02+1-0.25/2-0.25) rectangle (0.25+0.04+0.25+0.25/2,0.25/2-0.02+1-0.25/2-0.25);
			\path [fill=gray] (0.25+0.05+0.25+0.25/2,0.25/2-0.02-0.02+1-0.25/2-0.25) rectangle (0.25+0.25/2-0.02+0.25+0.25/2,0.25/2-0.02+1-0.25/2-0.25);
			\path [fill=gray] (0.25+0.02+0.25+0.25/2,0.02+1-0.25/2-0.25/2-0.25) rectangle (0.25+0.04+0.25+0.25/2,0.25/2-0.02+1-0.25/2-0.25/2-0.25);
			\path [fill=gray] (0.25+0.05+0.25+0.25/2,0.25/2-0.02-0.02+1-0.25/2-0.25/2-0.25) rectangle (0.25+0.25/2-0.02+0.25+0.25/2,0.25/2-0.02+1-0.25/2-0.25/2-0.25);
			
			\path [fill=gray] (0.5+0.02+0.25,0.02+1-0.25/2-0.5) rectangle (0.5+0.04+0.25,0.25/2-0.02+1-0.25/2-0.5);
			\path [fill=gray] (0.5+0.05+0.25,0.25/2-0.02-0.02+1-0.25/2-0.5) rectangle (0.5+0.25/2-0.02+0.25,0.25/2-0.02+1-0.25/2-0.5);
			\path [fill=gray] (0.5+0.02+0.25+0.25/2,0.02+1-0.25/2-0.5) rectangle (0.5+0.04+0.25+0.25/2,0.25/2-0.02+1-0.25/2-0.5);
			\path [fill=gray] (0.5+0.05+0.25+0.25/2,0.25/2-0.02-0.02+1-0.25/2-0.5) rectangle (0.5+0.25/2-0.02+0.25+0.25/2,0.25/2-0.02+1-0.25/2-0.5);
			\path [fill=gray] (0.5+0.02+0.25+0.25/2,0.02+1-0.25/2-0.25/2-0.5) rectangle (0.5+0.04+0.25+0.25/2,0.25/2-0.02+1-0.25/2-0.25/2-0.5);
			\path [fill=gray] (0.5+0.05+0.25+0.25/2,0.25/2-0.02-0.02+1-0.25/2-0.25/2-0.5) rectangle (0.5+0.25/2-0.02+0.25+0.25/2,0.25/2-0.02+1-0.25/2-0.25/2-0.5);
	
			\end{tikzpicture}
			\vspace{-4mm}
		\end{center}
		\caption{$\mathcal{H}$}
		\label{fig:Hformats_H}
	\end{subfigure}
	\hfill
	\begin{subfigure}[b]{.24\textwidth}
		\begin{center}
			\begin{tikzpicture}[scale=2.5]
			\draw (0,0) rectangle (1,1);
			\draw (0,0) rectangle (0.5,0.5);
			\draw (1,1) rectangle (0.5,0.5);
			\draw (0,1) rectangle (0.25,0.75);
			\draw [fill=gray] (0,1) rectangle (0.25/2,0.75+0.25/2);
			\draw [fill=gray] (0.25/2,0.75+0.25/2) rectangle (0.25,0.75);
			\draw (0.5,0.5) rectangle (0.75,0.25);
			\draw [fill=gray] (0.5,0.5) rectangle (0.75-0.25/2,0.25+0.25/2);
			\draw [fill=gray] (0.75-0.25/2,0.25+0.25/2) rectangle (0.75,0.25);
			\draw (0.25,0.75) rectangle (0.5,0.5);
			\draw [fill=gray] (0.25,0.75) rectangle (0.5-0.25/2,0.5+0.25/2);
			\draw [fill=gray] (0.5-0.25/2,0.5+0.25/2) rectangle (0.5,0.5);
			\draw (0.75,0.25) rectangle (1,0);
			\draw [fill=gray] (0.75,0.25) rectangle (0.75+0.25/2,0.25/2);
			\draw [fill=gray] (0.75+0.25/2,0.25/2) rectangle (1,0);
			
			\path [fill=gray] (0.02,0.35) rectangle (0.07,0.47);
			\path [fill=gray] (0.08,0.42) rectangle (0.13,0.47);
			\path [fill=gray] (0.14,0.42) rectangle (0.22,0.47);
			
			\path [fill=gray] (0.02+0.5,0.35+0.5) rectangle (0.07+0.5,0.47+0.5);
			\path [fill=gray] (0.08+0.5,0.42+0.5) rectangle (0.13+0.5,0.47+0.5);
			\path [fill=gray] (0.14+0.5,0.42+0.5) rectangle (0.22+0.5,0.47+0.5);
			
			
			
			
			
			\path [fill=gray] (0.25+0.01+0.5,0.01) rectangle (0.25+0.03+0.5,0.25/2-0.01);
			\path [fill=gray] (0.25+0.035+0.5,0.25/2-0.025-0.01) rectangle (0.25+0.055+0.5,0.25/2-0.01);
			\path [fill=gray] (0.25+0.06+0.5,0.25/2-0.025-0.01) rectangle (0.25+0.25/2-0.01+0.5,0.25/2-0.01);
			\path [fill=gray] (0.25/2+0.25+0.01+0.5,0.25/2+0.01) rectangle (0.25/2+0.25+0.03+0.5,0.25/2+0.25/2-0.01);
			\path [fill=gray] (0.25/2+0.25+0.035+0.5,0.25/2+0.25/2-0.025-0.01) rectangle (0.25/2+0.25+0.055+0.5,0.25/2+0.25/2-0.01);
			\path [fill=gray] (0.25/2+0.25+0.06+0.5,0.25/2+0.25/2-0.025-0.01) rectangle (0.25/2+0.25+0.25/2-0.01+0.5,0.25/2+0.25/2-0.01);
			\path [fill=gray] (0.01+0.5,0.01+0.25) rectangle (0.03+0.5,0.25/2-0.01+0.25);
			\path [fill=gray] (0.035+0.5,0.25/2-0.025-0.01+0.25) rectangle (0.055+0.5,0.25/2-0.01+0.25);
			\path [fill=gray] (0.06+0.5,0.25/2-0.025-0.01+0.25) rectangle (0.25/2-0.01+0.5,0.25/2-0.01+0.25);
			\path [fill=gray] (0.25/2+0.01+0.5,0.25/2+0.01+0.25) rectangle (0.25/2+0.03+0.5,0.25/2+0.25/2-0.01+0.25);
			\path [fill=gray] (0.25/2+0.035+0.5,0.25/2+0.25/2-0.025-0.01+0.25) rectangle (0.25/2+0.055+0.5,0.25/2+0.25/2-0.01+0.25);
			\path [fill=gray] (0.25/2+0.06+0.5,0.25/2+0.25/2-0.025-0.01+0.25) rectangle (0.25/2+0.25/2-0.01+0.5,0.25/2+0.25/2-0.01+0.25);
			
			\path [fill=gray] (-0.5+0.25+0.01+0.5,0.5+0.01) rectangle (-0.5+0.25+0.03+0.5,0.5+0.25/2-0.01);
			\path [fill=gray] (-0.5+0.25+0.035+0.5,0.5+0.25/2-0.025-0.01) rectangle (-0.5+0.25+0.055+0.5,0.5+0.25/2-0.01);
			\path [fill=gray] (-0.5+0.25+0.06+0.5,0.5+0.25/2-0.025-0.01) rectangle (-0.5+0.25+0.25/2-0.01+0.5,0.5+0.25/2-0.01);
			\path [fill=gray] (-0.5+0.25/2+0.25+0.01+0.5,0.5+0.25/2+0.01) rectangle (-0.5+0.25/2+0.25+0.03+0.5,0.5+0.25/2+0.25/2-0.01);
			\path [fill=gray] (-0.5+0.25/2+0.25+0.035+0.5,0.5+0.25/2+0.25/2-0.025-0.01) rectangle (-0.5+0.25/2+0.25+0.055+0.5,0.5+0.25/2+0.25/2-0.01);
			\path [fill=gray] (-0.5+0.25/2+0.25+0.06+0.5,0.5+0.25/2+0.25/2-0.025-0.01) rectangle (-0.5+0.25/2+0.25+0.25/2-0.01+0.5,0.5+0.25/2+0.25/2-0.01);
			\path [fill=gray] (-0.5+0.01+0.5,0.5+0.01+0.25) rectangle (-0.5+0.03+0.5,0.5+0.25/2-0.01+0.25);
			\path [fill=gray] (-0.5+0.035+0.5,0.5+0.25/2-0.025-0.01+0.25) rectangle (-0.5+0.055+0.5,0.5+0.25/2-0.01+0.25);
			\path [fill=gray] (-0.5+0.06+0.5,0.5+0.25/2-0.025-0.01+0.25) rectangle (-0.5+0.25/2-0.01+0.5,0.5+0.25/2-0.01+0.25);
			\path [fill=gray] (-0.5+0.25/2+0.01+0.5,0.5+0.25/2+0.01+0.25) rectangle (-0.5+0.25/2+0.03+0.5,0.5+0.25/2+0.25/2-0.01+0.25);
			\path [fill=gray] (-0.5+0.25/2+0.035+0.5,0.5+0.25/2+0.25/2-0.025-0.01+0.25) rectangle (-0.5+0.25/2+0.055+0.5,0.5+0.25/2+0.25/2-0.01+0.25);
			\path [fill=gray] (-0.5+0.25/2+0.06+0.5,0.5+0.25/2+0.25/2-0.025-0.01+0.25) rectangle (-0.5+0.25/2+0.25/2-0.01+0.5,0.5+0.25/2+0.25/2-0.01+0.25);
			
			\path [fill=gray] (0.015,0.15+0.5) rectangle (0.05,0.235+0.5);
			\path [fill=gray] (0.06,0.2+0.5) rectangle (0.09,0.235+0.5);
			\path [fill=gray] (0.1,0.2+0.5) rectangle (0.16,0.235+0.5);
			\path [fill=gray] (0.015+0.25,0.15+0.75) rectangle (0.05+0.25,0.235+0.5+0.25);
			\path [fill=gray] (0.06+0.25,0.2+0.75) rectangle (0.09+0.25,0.235+0.5+0.25);
			\path [fill=gray] (0.1+0.25,0.2+0.75) rectangle (0.16+0.25,0.235+0.5+0.25);
			
			\path [fill=gray] (0.5+0.015,0.15+0.5-0.5) rectangle (0.5+0.05,0.235+0.5-0.5);
			\path [fill=gray] (0.5+0.06,0.2+0.5-0.5) rectangle (0.5+0.09,0.235+0.5-0.5);
			\path [fill=gray] (0.5+0.1,0.2+0.5-0.5) rectangle (0.5+0.16,0.235+0.5-0.5);
			\path [fill=gray] (0.5+0.015+0.25,0.15+0.75-0.5) rectangle (0.5+0.05+0.25,0.235+0.5+0.25-0.5);
			\path [fill=gray] (0.5+0.06+0.25,0.2+0.75-0.5) rectangle (0.5+0.09+0.25,0.235+0.5+0.25-0.5);
			\path [fill=gray] (0.5+0.1+0.25,0.2+0.75-0.5) rectangle (0.5+0.16+0.25,0.235+0.5+0.25-0.5);
			
			\end{tikzpicture}
			\vspace{-4mm}
		\end{center}
		\caption{HSS}
		\label{fig:Hformats_HSS}
	\end{subfigure}
\hfill
\begin{subfigure}[b]{.24\textwidth}
	\begin{center}
		\begin{tikzpicture}[scale=2.5]
		\draw (0,0) rectangle (1,1);
		\draw (0,0) rectangle (0.5,0.5);
		\draw (1,1) rectangle (0.5,0.5);
		\draw [fill=gray] (0,1) rectangle (0.25,0.75);
		\draw [fill=gray] (0.5,0.5) rectangle (0.75,0.25);
		\draw [fill=gray] (0.25,0.75) rectangle (0.5,0.5);
		\draw [fill=gray] (0.75,0.25) rectangle (1,0);
		\draw (0,0) rectangle (0.25,0.25);
		\draw (0.25,0.25) rectangle (0.5,0.5);
		\draw (0.5,0.5) rectangle (0.75,0.75);
		\draw (0.75,0.75) rectangle (1,1);
		
		\draw [fill=gray] (0.25/2,0.5+0.25/2) rectangle (0.25,0.75);
		\draw [fill=gray] (0.25,0.75) rectangle (0.25+0.25/2,0.75+0.25/2);
		
		\draw [fill=gray] (0.25+0.25/2,0.25+0.25/2) rectangle (0.5,0.5);
		\draw [fill=gray] (0.5,0.5) rectangle (0.25+0.25/2+0.25,0.25+0.25/2+0.25);
		
		\draw [fill=gray] (0.25+0.25/2+0.25,0.25+0.25/2-0.25) rectangle (0.5+0.25,0.5-0.25);
		\draw [fill=gray] (0.5+0.25,0.5-0.25) rectangle (0.25+0.25/2+0.5,0.25+0.25/2);
		
		\draw (0,1) rectangle (0.25/2,1-0.25/2);
		\draw (0.25/2,1-0.25/2) rectangle (0.25,0.75);
		\draw (0.25,0.75) rectangle (0.25+0.25/2,0.75-0.25/2);
		\draw (0.25+0.25/2,0.75-0.25/2) rectangle (0.5,0.5);
		\draw (0.5,0.5) rectangle (0.5+0.25/2,0.5-0.25/2);
		\draw (0.5+0.25/2,0.5-0.25/2) rectangle (0.75,0.25);
		\draw (0.75,0.25) rectangle (0.75+0.25/2,0.25/2);
		\draw (0.75+0.25/2,0.25/2) rectangle (1,0);
		
		\draw (0,0.5) rectangle (0.25/2,0.5+0.25/2);
		\draw (0.5,1) rectangle (0.5-0.25/2,1-0.25/2);
		\draw (0.25,0.25) rectangle (0.25+0.25/2,0.25+0.25/2);
		\draw (0.5,0) rectangle (0.5+0.25/2,0.25/2);
		\draw (0.75,0.75) rectangle (0.75-0.25/2,0.75-0.25/2);
		\draw (1,0.5) rectangle (1-0.25/2,0.5-0.25/2);
		
		\path [fill=gray] (0.01,0.01+0.5) rectangle (0.03,0.25/2-0.01+0.5);
		\path [fill=gray] (0.035,0.25/2-0.025-0.01+0.5) rectangle (0.055,0.25/2-0.01+0.5);
		\path [fill=gray] (0.06,0.25/2-0.025-0.01+0.5) rectangle (0.25/2-0.01,0.25/2-0.01+0.5);
		\path [fill=gray] (0.01,0.01+0.5+0.25/2) rectangle (0.03,0.25/2-0.01+0.5+0.25/2);
		\path [fill=gray] (0.035,0.25/2-0.025-0.01+0.5+0.25/2) rectangle (0.055,0.25/2-0.01+0.5+0.25/2);
		\path [fill=gray] (0.06,0.25/2-0.025-0.01+0.5+0.25/2) rectangle (0.25/2-0.01,0.25/2-0.01+0.5+0.25/2);
		\path [fill=gray] (0.01+0.25/2,0.01+0.5) rectangle (0.03+0.25/2,0.25/2-0.01+0.5);
		\path [fill=gray] (0.035+0.25/2,0.25/2-0.025-0.01+0.5) rectangle (0.055+0.25/2,0.25/2-0.01+0.5);
		\path [fill=gray] (0.06+0.25/2,0.25/2-0.025-0.01+0.5) rectangle (0.25/2-0.01+0.25/2,0.25/2-0.01+0.5);
		
		\path [fill=gray] (0.01+0.25,0.01+0.5-0.25) rectangle (0.03+0.25,0.25/2-0.01+0.5-0.25);
		\path [fill=gray] (0.035+0.25,0.25/2-0.025-0.01+0.5-0.25) rectangle (0.055+0.25,0.25/2-0.01+0.5-0.25);
		\path [fill=gray] (0.06+0.25,0.25/2-0.025-0.01+0.5-0.25) rectangle (0.25/2-0.01+0.25,0.25/2-0.01+0.5-0.25);
		\path [fill=gray] (0.01+0.25+0.25/2,0.01+0.5-0.25) rectangle (0.03+0.25+0.25/2,0.25/2-0.01+0.5-0.25);
		\path [fill=gray] (0.035+0.25+0.25/2,0.25/2-0.025-0.01+0.5-0.25) rectangle (0.055+0.25+0.25/2,0.25/2-0.01+0.5-0.25);
		\path [fill=gray] (0.06+0.25+0.25/2,0.25/2-0.025-0.01+0.5-0.25) rectangle (0.25/2-0.01+0.25+0.25/2,0.25/2-0.01+0.5-0.25);
		\path [fill=gray] (0.01+0.25,0.01+0.5-0.25+0.25/2) rectangle (0.03+0.25,0.25/2-0.01+0.5-0.25+0.25/2);
		\path [fill=gray] (0.035+0.25,0.25/2-0.025-0.01+0.5-0.25+0.25/2) rectangle (0.055+0.25,0.25/2-0.01+0.5-0.25+0.25/2);
		\path [fill=gray] (0.06+0.25,0.25/2-0.025-0.01+0.5-0.25+0.25/2) rectangle (0.25/2-0.01+0.25,0.25/2-0.01+0.5-0.25+0.25/2);
		
		\path [fill=gray] (0.01+0.5,0.01) rectangle (0.03+0.5,0.25/2-0.01);
		\path [fill=gray] (0.035+0.5,0.25/2-0.025-0.01) rectangle (0.055+0.5,0.25/2-0.01);
		\path [fill=gray] (0.06+0.5,0.25/2-0.025-0.01) rectangle (0.25/2-0.01+0.5,0.25/2-0.01);
		\path [fill=gray] (0.01+0.5+0.25/2,0.01) rectangle (0.03+0.5+0.25/2,0.25/2-0.01);
		\path [fill=gray] (0.035+0.5+0.25/2,0.25/2-0.025-0.01) rectangle (0.055+0.5+0.25/2,0.25/2-0.01);
		\path [fill=gray] (0.06+0.5+0.25/2,0.25/2-0.025-0.01) rectangle (0.25/2-0.01+0.5+0.25/2,0.25/2-0.01);
		\path [fill=gray] (0.01+0.5,0.01+0.25/2) rectangle (0.03+0.5,0.25/2-0.01+0.25/2);
		\path [fill=gray] (0.035+0.5,0.25/2-0.025-0.01+0.25/2) rectangle (0.055+0.5,0.25/2-0.01+0.25/2);
		\path [fill=gray] (0.06+0.5,0.25/2-0.025-0.01+0.25/2) rectangle (0.25/2-0.01+0.5,0.25/2-0.01+0.25/2);
		
		\path [fill=gray] (0.01+0.5-0.25/2,0.01+1-0.25/2) rectangle (0.03+0.5-0.25/2,0.25/2-0.01+1-0.25/2);
		\path [fill=gray] (0.035+0.5-0.25/2,0.25/2-0.025-0.01+1-0.25/2) rectangle (0.055+0.5-0.25/2,0.25/2-0.01+1-0.25/2);
		\path [fill=gray] (0.06+0.5-0.25/2,0.25/2-0.025-0.01+1-0.25/2) rectangle (0.25/2-0.01+0.5-0.25/2,0.25/2-0.01+1-0.25/2);
		\path [fill=gray] (0.01+0.5-0.25,0.01+1-0.25/2) rectangle (0.03+0.5-0.25,0.25/2-0.01+1-0.25/2);
		\path [fill=gray] (0.035+0.5-0.25,0.25/2-0.025-0.01+1-0.25/2) rectangle (0.055+0.5-0.25,0.25/2-0.01+1-0.25/2);
		\path [fill=gray] (0.06+0.5-0.25,0.25/2-0.025-0.01+1-0.25/2) rectangle (0.25/2-0.01+0.5-0.25,0.25/2-0.01+1-0.25/2);
		\path [fill=gray] (0.01+0.5-0.25/2,0.01+1-0.25) rectangle (0.03+0.5-0.25/2,0.25/2-0.01+1-0.25);
		\path [fill=gray] (0.035+0.5-0.25/2,0.25/2-0.025-0.01+1-0.25) rectangle (0.055+0.5-0.25/2,0.25/2-0.01+1-0.25);
		\path [fill=gray] (0.06+0.5-0.25/2,0.25/2-0.025-0.01+1-0.25) rectangle (0.25/2-0.01+0.5-0.25/2,0.25/2-0.01+1-0.25);
		
		\path [fill=gray] (0.25+0.01+0.5-0.25/2,0.01+1-0.25/2-0.25) rectangle (0.25+0.03+0.5-0.25/2,0.25/2-0.01+1-0.25/2-0.25);
		\path [fill=gray] (0.25+0.035+0.5-0.25/2,0.25/2-0.025-0.01+1-0.25/2-0.25) rectangle (0.25+0.055+0.5-0.25/2,0.25/2-0.01+1-0.25/2-0.25);
		\path [fill=gray] (0.25+0.06+0.5-0.25/2,0.25/2-0.025-0.01+1-0.25/2-0.25) rectangle (0.25+0.25/2-0.01+0.5-0.25/2,0.25/2-0.01+1-0.25/2-0.25);
		\path [fill=gray] (0.25+0.01+0.5-0.25,0.01+1-0.25/2-0.25) rectangle (0.25+0.03+0.5-0.25,0.25/2-0.01+1-0.25/2-0.25);
		\path [fill=gray] (0.25+0.035+0.5-0.25,0.25/2-0.025-0.01+1-0.25/2-0.25) rectangle (0.25+0.055+0.5-0.25,0.25/2-0.01+1-0.25/2-0.25);
		\path [fill=gray] (0.25+0.06+0.5-0.25,0.25/2-0.025-0.01+1-0.25/2-0.25) rectangle (0.25+0.25/2-0.01+0.5-0.25,0.25/2-0.01+1-0.25/2-0.25);
		\path [fill=gray] (0.25+0.01+0.5-0.25/2,0.01+1-0.25-0.25) rectangle (0.25+0.03+0.5-0.25/2,0.25/2-0.01+1-0.25-0.25);
		\path [fill=gray] (0.25+0.035+0.5-0.25/2,0.25/2-0.025-0.01+1-0.25-0.25) rectangle (0.25+0.055+0.5-0.25/2,0.25/2-0.01+1-0.25-0.25);
		\path [fill=gray] (0.25+0.06+0.5-0.25/2,0.25/2-0.025-0.01+1-0.25-0.25) rectangle (0.25+0.25/2-0.01+0.5-0.25/2,0.25/2-0.01+1-0.25-0.25);
		
		\path [fill=gray] (0.5+0.01+0.5-0.25/2,0.01+1-0.25/2-0.5) rectangle (0.5+0.03+0.5-0.25/2,0.25/2-0.01+1-0.25/2-0.5);
		\path [fill=gray] (0.5+0.035+0.5-0.25/2,0.25/2-0.025-0.01+1-0.25/2-0.5) rectangle (0.5+0.055+0.5-0.25/2,0.25/2-0.01+1-0.25/2-0.5);
		\path [fill=gray] (0.5+0.06+0.5-0.25/2,0.25/2-0.025-0.01+1-0.25/2-0.5) rectangle (0.5+0.25/2-0.01+0.5-0.25/2,0.25/2-0.01+1-0.25/2-0.5);
		\path [fill=gray] (0.5+0.01+0.5-0.25,0.01+1-0.25/2-0.5) rectangle (0.5+0.03+0.5-0.25,0.25/2-0.01+1-0.25/2-0.5);
		\path [fill=gray] (0.5+0.035+0.5-0.25,0.25/2-0.025-0.01+1-0.25/2-0.5) rectangle (0.5+0.055+0.5-0.25,0.25/2-0.01+1-0.25/2-0.5);
		\path [fill=gray] (0.5+0.06+0.5-0.25,0.25/2-0.025-0.01+1-0.25/2-0.5) rectangle (0.5+0.25/2-0.01+0.5-0.25,0.25/2-0.01+1-0.25/2-0.5);
		\path [fill=gray] (0.5+0.01+0.5-0.25/2,0.01+1-0.25-0.5) rectangle (0.5+0.03+0.5-0.25/2,0.25/2-0.01+1-0.25-0.5);
		\path [fill=gray] (0.5+0.035+0.5-0.25/2,0.25/2-0.025-0.01+1-0.25-0.5) rectangle (0.5+0.055+0.5-0.25/2,0.25/2-0.01+1-0.25-0.5);
		\path [fill=gray] (0.5+0.06+0.5-0.25/2,0.25/2-0.025-0.01+1-0.25-0.5) rectangle (0.5+0.25/2-0.01+0.5-0.25/2,0.25/2-0.01+1-0.25-0.5);
		
		\path [fill=gray] (0.015,0.15) rectangle (0.05,0.235);
		\path [fill=gray] (0.06,0.2) rectangle (0.09,0.235);
		\path [fill=gray] (0.1,0.2) rectangle (0.16,0.235);
		\path [fill=gray] (0.25+0.015,0.15) rectangle (0.25+0.05,0.235);
		\path [fill=gray] (0.25+0.06,0.2) rectangle (0.25+0.09,0.235);
		\path [fill=gray] (0.25+0.1,0.2) rectangle (0.25+0.16,0.235);
		\path [fill=gray] (0.015,0.25+0.15) rectangle (0.05,0.25+0.235);
		\path [fill=gray] (0.06,0.25+0.2) rectangle (0.09,0.25+0.235);
		\path [fill=gray] (0.1,0.25+0.2) rectangle (0.16,0.25+0.235);
		
		\path [fill=gray] (0.015+0.5,0.15+0.75) rectangle (0.05+0.5,0.235+0.75);
		\path [fill=gray] (0.06+0.5,0.2+0.75) rectangle (0.09+0.5,0.235+0.75);
		\path [fill=gray] (0.1+0.5,0.2+0.75) rectangle (0.16+0.5,0.235+0.75);
		\path [fill=gray] (0.015+0.75,0.15+0.75) rectangle (0.05+0.75,0.235+0.75);
		\path [fill=gray] (0.06+0.75,0.2+0.75) rectangle (0.09+0.75,0.235+0.75);
		\path [fill=gray] (0.1+0.75,0.2+0.75) rectangle (0.16+0.75,0.235+0.75);
		\path [fill=gray] (0.015+0.75,0.15+0.5) rectangle (0.05+0.75,0.235+0.5);
		\path [fill=gray] (0.06+0.75,0.2+0.5) rectangle (0.09+0.75,0.235+0.5);
		\path [fill=gray] (0.1+0.75,0.2+0.5) rectangle (0.16+0.75,0.235+0.5);
		\end{tikzpicture}
		\vspace{-4mm}
	\end{center}
	\caption{$\mathcal{H}^2$}
	\label{fig:Hformats_H2}
\end{subfigure}

\hfill
\begin{subfigure}[b]{.5\textwidth}
	\begin{center}
		\begin{tikzpicture}[scale=2.5]
		\def\myshift{.03}
		\def\r{1/32}
		\def\rr{1/16}
		\def\h{1/2}
		\def\q{1/4}
		\def\o{1/8}
		
		\draw (0,0) rectangle (1,1);
		\foreach \x in {0,1,...,7} {
			\draw [fill=gray] (\x*\o,1-\x*\o) rectangle ++(\o,-\o);
		}
		\foreach \x in {0,1,...,3} {
			\draw (\x*\q,1-\x*\q) rectangle ++(\q,-\q);
		}
		\foreach \x in {0,1} {
			\draw (\x*\h,1-\x*\h) rectangle ++(\h,-\h);
		}
		\foreach \x in {0,2} {
			\draw[dotted] (\h+\x*\o,\h) rectangle ++(\o,\h);
		}
		\foreach \x in {0,2} {
			\draw[dotted] (\h,1-\x*\o) rectangle ++(\h,-\o);
		}

		
		\draw[dotted] (\h+\q,\h) rectangle ++(\o,-\o);
		\draw[dotted] (1-\o,\q) rectangle ++(\o,\o);
		
		\draw [->] (1-\myshift,1-\q+\o / 2) -- ++(.2,0);
		\draw [->] (1-\myshift,\q+3./2.*\o) -- ++(.4+2*\r+\myshift,0);
		\draw [->] (1-\myshift,\o+\o/2) -- ++(.6+2*\r+\myshift,0);
		
		\tikzset{shift={(1.2,0)}}
		\foreach \x in {0,1,...,3} {
			\fill [gray] (\x*\r,1-\x*\o) rectangle ++(\r,-\o);
		};
		\draw (0,1) rectangle (\r*4,\h);
		\tikzset{shift={(\r*4+\myshift,0)}}
		\foreach \x in {0,1} {
			\fill [gray] (\x*\r,1-\x*\rr) rectangle ++(\r,-\rr);
			\fill [gray] (\o/2+\x*\r,1-\x*\rr) rectangle ++(\r,-\rr);
		};
		\draw (0,1) rectangle (\o,1-\o);
		\tikzset{shift={(\r*4+\myshift,0)}}
		\foreach \x in {0,1} {
			\foreach \y in {0,1} {
				\fill [gray] (\x*\o/2+\y*\r,1-\y*\o/2-\x*\r) rectangle ++(\r,-\r);
				\draw (\x*\o/2,1-\y*\o/2) rectangle ++(\o/2,-\o/2);
			}
		}
		\draw (0,1) rectangle (\o/2,1-\o/2);
		\tikzset{shift={(\r*4+\myshift,0)}}
		\foreach \x in {0,1} {
			\fill [gray] (\x*\rr,1-\x*\r) rectangle ++(\rr,-\r);
			\fill [gray] (\x*\rr,1-\x*\r-\o/2) rectangle ++(\rr,-\r);
		};
		\draw (0,1) rectangle (\o,1-\o);
		\tikzset{shift={(\r*4+\myshift,0)}}
		\foreach \x in {0,1,...,3} {
			\fill [gray] (\x*\o,1-\x*\r) rectangle ++(\o,-\r);
		};
		\draw (0,1) rectangle (\h,1-\o);

		\tikzset{shift={(-9*\r-2*\myshift,0)}}
		\foreach \x in {0,1} {
			\fill [gray] (\x*\r,\h-\x*\o) rectangle ++(\r,-\o);
		};
		\draw (0,\q) rectangle (\r*2,\h);
		
		\tikzset{shift={(\r*2+\myshift,0)}}
		\fill [gray] (0,\h) rectangle (\o/2,\h-\o/2);
		\draw (0,\h) rectangle (\o/2,\h-\o/2);
		
		\tikzset{shift={(\r*2+\myshift,0)}}
		\foreach \x in {0,1} {
			\fill [gray] (\x*\o,\h-\x*\r) rectangle ++(\o,-\r);
		};
		\draw (0,\h) rectangle (\q,\h-\o/2);
		
		\tikzset{shift={(-4*\r-2*\myshift+.2,0)}}
		\fill [gray] (0,\q) rectangle (\r,\q-\o);
		\draw (0,\q) rectangle (\r,\q-\o);
		
		\tikzset{shift={(\r+\myshift,0)}}
		\fill [gray] (0,\q) rectangle (\o,\q-\r);
		\draw (0,\q) rectangle (\o,\q-\r);
		\end{tikzpicture}
		\vspace{-4mm}
	\end{center}
	\caption{HODBF}
	\label{fig:Hformats_HODBF}
\end{subfigure}
\hfill
	\begin{subfigure}[b]{.45\textwidth}
		\begin{center}
			\begin{tikzpicture}[scale=2.5]
			\draw (0,0) rectangle (1,1);
			\draw (0,0) rectangle (0.5,0.5);
			\draw (1,1) rectangle (0.5,0.5);
			\draw [fill=gray] (0,1) rectangle (0.25,0.75);
			\draw [fill=gray] (0.5,0.5) rectangle (0.75,0.25);
			\draw [fill=gray] (0.25,0.75) rectangle (0.5,0.5);
			\draw [fill=gray] (0.75,0.25) rectangle (1,0);
			\draw (0,0) rectangle (0.25,0.25);
			\draw (0.25,0.25) rectangle (0.5,0.5);
			\draw (0.5,0.5) rectangle (0.75,0.75);
			\draw (0.75,0.75) rectangle (1,1);
			
			\path [fill=gray] (0.02,0.02) rectangle (0.05,0.225);
			\path [fill=gray] (0.07,0.195) rectangle (0.225,0.225);
			\path [fill=gray] (0.02+0.25,0.02) rectangle (0.05+0.25,0.225);
			\path [fill=gray] (0.07+0.25,0.195) rectangle (0.225+0.25,0.225);
			\path [fill=gray] (0.02+0.5,0.02) rectangle (0.05+0.5,0.225);
			\path [fill=gray] (0.07+0.5,0.195) rectangle (0.225+0.5,0.225);
			\path [fill=gray] (0.02,0.02+0.25) rectangle (0.05,0.225+0.25);
			\path [fill=gray] (0.07,0.195+0.25) rectangle (0.225,0.225+0.25);
			\path [fill=gray] (0.02,0.02+0.5) rectangle (0.05,0.225+0.5);
			\path [fill=gray] (0.07,0.195+0.5) rectangle (0.225,0.225+0.5);
			\path [fill=gray] (0.02+0.25,0.02+0.25) rectangle (0.05+0.25,0.225+0.25);
			\path [fill=gray] (0.07+0.25,0.195+0.25) rectangle (0.225+0.25,0.225+0.25);
			\path [fill=gray] (0.02+0.75,0.02+0.25) rectangle (0.05+0.75,0.225+0.25);
			\path [fill=gray] (0.07+0.75,0.195+0.25) rectangle (0.225+0.75,0.225+0.25);
			\path [fill=gray] (0.02+0.5,0.02+0.5) rectangle (0.05+0.5,0.225+0.5);
			\path [fill=gray] (0.07+0.5,0.195+0.5) rectangle (0.225+0.5,0.225+0.5);
			\path [fill=gray] (0.02+0.75,0.02+0.5) rectangle (0.05+0.75,0.225+0.5);
			\path [fill=gray] (0.07+0.75,0.195+0.5) rectangle (0.225+0.75,0.225+0.5);
			\path [fill=gray] (0.02+0.25,0.02+0.75) rectangle (0.05+0.25,0.225+0.75);
			\path [fill=gray] (0.07+0.25,0.195+0.75) rectangle (0.225+0.25,0.225+0.75);
			\path [fill=gray] (0.02+0.5,0.02+0.75) rectangle (0.05+0.5,0.225+0.75);
			\path [fill=gray] (0.07+0.5,0.195+0.75) rectangle (0.225+0.5,0.225+0.75);
			\path [fill=gray] (0.02+0.75,0.02+0.75) rectangle (0.05+0.75,0.225+0.75);
			\path [fill=gray] (0.07+0.75,0.195+0.75) rectangle (0.225+0.75,0.225+0.75);
			\end{tikzpicture}
			\vspace{-4mm}
		\end{center}
		\caption{BLR}
		\label{fig:Hformats_BLR}
	\end{subfigure}
\hfill
\caption{Illustration of several types of hierarchical matrices: HODLR, $\mathcal{H}$, HSS and $\mathcal{H}^2$, HODBF (the butterfly extension of HODLR), and BLR (single-level partitioning-based). \rev{In HODLR, $\mathcal{H}$ and BLR, each admissible block $A^\ell_{st}$ is represented as $A^\ell_{st}\approx U_{st}V^T_{st}$ where $U_{st}$ and $V^T_{st}$ are plotted. In HSS and $\mathcal{H}^2$, each admissible block $A^\ell_{st}$ is represented as $A^\ell_{st}\approx U_{s}B_{st}V^T_{t}$ where $U_{s}$ and $V^T_{t}$ are not explicitly formed and can be constructed from the transfer matrices via \cref{eqn:nested_basis}. At non-leaf levels, the transfer matrices $E_{s}$, $E^T_{t}$ and the coupling matrices $B_{st}$ are plotted.}}\label{fig::Hformats}.   
\end{figure}

\begin{table}[h!]
\centering
\begin{tabular}{lcccccccc}
\hline
\textbf{Format} 
& \textbf{Storage} 
& \textbf{Construction} 
& \textbf{Factorization}
& \textbf{Apply}
& \textbf{Adm}
& \textbf{Shared Basis}
& \textbf{Dim} \\
\hline

HODLR 
& $\bigo{n r \log n}$ 
& $\bigo{n r^{\alpha} \log^{\beta} n}$ 
& $\bigo{n r^{2} \log^{2} n}$ 
& $\bigo{n r \log n}$ 
& Weak
& No
& $d\leq 2$ \\[4pt]

$\mathcal{H}$ 
& $\bigo{n r \log n}$ 
& $\bigo{n r^{\alpha} \log^{\beta} n}$ 
& $\bigo{n r^{2} \log^{2} n}$ 
& $\bigo{n r \log n}$ 
& Strong
& No
& $d\geq 3$ \\[4pt]

HSS 
& $\bigo{n r}$ 
& $\bigo{n r^\alpha}$ 
& $\bigo{n r^{2}}$ 
& $\bigo{n r}$ 
& Weak
& Yes
& $d\leq 2$ \\[4pt]

$\mathcal{H}^2$ 
& $\bigo{n r}$ 
& $\bigo{n r^\alpha}$ 
& $\bigo{n r^{2}}$ 
& $\bigo{n r}$ 
& Strong
& Yes
& $d\geq 3$ \\[4pt]

\hline
\end{tabular}
\caption{Summary of four major classes of hierarchical matrix representation of an $n\times n$ matrix. $r$ denotes the maximum numerical rank in the off-diagonal blocks, $d$ denotes the dimensionality of the underlying scientific or machine learning kernels. $\alpha=1,2$, $\beta=1,2$. ``Adm" means "admissibility", ``Dim" means ``dimensionality", and ``Apply" represents application of the matrix or its inverse to a single vector. Note that $\mathcal{H}$ and $\mathcal{H}^2$ matrices also support weak admissibility, but their major advantage over other hierarchical matrix algorithms is attributed to their support for strong admissibility for high-dimensional problems.}\label{tab:hierarchical_matrix}
\end{table}

\subsection{Existence of hierarchical matrix structure in practical applications}\label{sec:app_example}
The efficiency of the rank-structured matrix algorithms heavily relies on, after certain preprocessing, the existence of the low-rank structure for the matrix operators appearing in many scientific computing, physics, engineering, and machine learning applications. Here we list a few application examples where rank-structured matrix algorithms are known to be effective. 

\subsubsection{Integral equations}
For the simulation of wave phenomena (e.g., in acoustics, elastics, electromagnetics and optics \cite{manic2017efficient,Sushnikova2023,shaeffer2008direct,Sadeed2022VIE,martinsson2005fast,corona2015n,hackbusch2002h2,Han_2013_butterflyLU,liu2023detecting}), fluid flow (e.g. Stokes flow \cite{barnett2015spectrally}) and plasma dynamics (e.g., MHD \cite{khalichi2025taylor}, Poisson solvers \cite{birdsall2018plasma}), one oftentimes solves a linear system with matrix elements $A_{ij}$ in the form of 
\begin{align}
\rev{A_{ij}=\int_{D_i}\int_{D_j}\psi_i(y)\,G(x,y)\,\phi_j(y)\,dx\,dy}. \label{eq:IE}
\end{align}
Here $\psi_i$ and $\phi_j$ are local basis functions (often $\psi_i=\phi_i$), \rev{$D_i$ and $D_j$} form the domain of integration, and $G$ is the Green's function for Helmholtz, Laplace, Yukawa or Stokes kernels, etc. For example, the Green's function of the integral equation (IE) for the 3D Laplace kernel is 
\begin{align}
G(x,y) = \frac{1}{4\pi|x-y|}. \label{eq:laplace}
\end{align}
For off-diagonal blocks representing Green's function interaction between well-separated spatial groups, one can prove the algebraic compressibility for analytical \cite{ying2012pedestrian,engquist2018approximate} and semi-analytical \cite{liu2023fast} Green's function for both smooth and oscillatory kernels. On the other hand, the compressibility of off-diagonal blocks of $A^{-1}$, essential for preserving the computational advantage of fast direct solvers, remains an active research area \cite{Angleitner2021,Faustmann2015ExistenceO,faustmann2016existence}. 

\subsubsection{Kernel matrices}
Many machine learning and statistics algorithms, e.g., kernel ridge regression and Gaussian processes, require the inversion of kernel matrices in the form of a Gaussian kernel with elements
\begin{align}
A_{ij}=\exp{-\frac{\|x_i-x_j\|^2}{2\sigma^2}} \label{eq:kernelmatrix}
\end{align}
with hyperparameter $\sigma$, or many other kernels. Due to smoothness or rapid decay (with respect to $\|x_i-x_j\|$) of the kernels, the kernel matrix $A$ exhibits low numerical ranks in the off-diagonal blocks and rank-structured matrix algorithms have been widely used \cite{rebrova2018study,chavez2020scalable,chenhan2016inv,chenhan2017n,ambikasaran2015fast,litvinenko2020hlibcov,minden2017fast,ambikasaran2013n}. Moreover, when optimizing hyperparameters in Gaussian processes with gradient-based methods, one often constructs a hierarchical matrix representation for $\partial_{\sigma}A$ as well \cite{geoga2020scalable}.

\subsubsection{PDEs and Schur Complements}\label{sec:pde}
Discretization of many elliptic PDEs (e.g., Poisson, Helmholtz, Yukawa, convection-diffusion, convection-reaction, and incompressible Stokes, etc.) leads to a sparse linear system with matrix $A$. In supernodal or multifrontal sparse direct solvers, block elimination of the properly permuted system leads to Schur complements in the form of 
\begin{align}
S=A_{ss} - A_{sr}A_{rr}^{-1}A_{rs},~~~~ A=\begin{bmatrix} A_{rr} & A_{rs} \\ A_{sr} & A_{ss} \end{bmatrix}. \label{eq:schur}
\end{align} 
The Schur complement can be viewed as the numerical Green's function for the underlying PDE \cite{engquist2018approximate, chandrasekaran2010numerical,faustmann2013new,FaustmannHFEM} and hence exhibits low-rank structures. Here we assume $A$ is $2\times2$ block partitioned for simplicity. Note that in reality, the block elimination process often leads to Schur complements in the form of 
\begin{align}
S=F_{ss} - F_{sr}F_{rr}^{-1}F_{rs}.\label{eq:schur_frontal}
\end{align}
Here $F$ are the dense frontal matrices where $F_{rr}$ and $F_{ss}$ are represented as hierarchical matrices, and $F_{rs}$ and $F_{sr}$ are presented as either hierarchical matrices or single low-rank compression \cite{claus2023sparse}, depending on the admissibility condition.

\subsubsection{PDE-constrained inverse problem and optimization}

In PDE-constrained optimization and inverse problems \cite{feliu2021approximate,alger2019data,ambartsumyan2020hierarchical}, one optimizes some misfit term $f(u(m))$ with respect to model parameter $m$ subject to a PDE constraint $A(m)u-b=0$, which leads to a Hessian matrix in the form of 
\begin{align}
G=A^{-T}\partial^2_{mm}fA^{-1}+S.\label{eq:hessian}
\end{align}
Here $A$ is the discretized PDE operator, $A^{-1}$ can be computed via hierarchical matrix algorithms, $\partial^2_{mm}f$ and $S$ are local operators. One can show that $G$ also inherits the hierarchical matrix structure of $A^{-1}$ \cite{grasedyck2003construction,ambartsumyan2020hierarchical,bebendorf2003existence} and can be efficiently inverted by hierarchical matrix algorithms.   

In addition to the above-mentioned classes of matrices, many other special matrices such as Toeplitz matrix, Hankel matrix, Fourier integral operator, and spherical harmonic transform also exhibit hierarchical rank structures.



\subsection{Algorithm description with an $\mathcal{H}^2$ matrix example}
\subsubsection{Preprocessing}\label{sec:preprocess_H}
The preprocessing phase is typically required for all existing rank-structured matrix algorithms to reveal the proper hierarchical structure to achieve optimal compressibility and computational efficiency. The preprocessing phase determines (a) clustering: permutation information that block partitions the matrix into, typically, multiple levels and (b) admissibility assignment: which blocks at each level of the permuted matrix can be algebraically compressed. When geometrical information is available for each row/column, e.g., Cartesian coordinates $x_i$ in \cref{eq:kernelmatrix}, or any geometrical point in the support of local basis functions $\psi_i,\phi_j$ in \cref{eq:IE}, clustering algorithms such as KD tree, cobblestone-like \cite{shaeffer2008direct} or random projection tree \cite{chavez2020scalable} are commonly used. For dense frontal matrices arising from sparse multifrontal solvers with no geometrical information, the frontal matrices are clustered using graph partitioning algorithms such as recursive bisection \cite{liu2021sparse,ghysels2016efficient}.    

Similarly, admissibility assignment needs to estimate how far away a pair of block rows and block columns are from each other. When geometrical information is available, the distance and diameters of two geometrical groups can be easily computed via geometrical distance (Note that even frontal matrices for PDEs can inherit the geometrical information \cite{zhu2025recursive} when the PDE is discretized on e.g., a 3D Cartesian grid). However, for (SPD) kernel matrices algebraically constructed without geometrical information, kernel distances have been exploited to determine the admissibility \cite{zhu2025recursive}. Similarly, for frontal matrices without geometrical information, one can rely on graph distance \cite{liu2021sparse}, or adaptive algorithms (i.e., attempting compressing each block but further partitioning it if the rank is too high) \cite{Claus25}. In addition to admissibility, sometimes it suffices to compute a set of nearest neighbors for each row/column. This can be computed with k-nearest neighbor search \cite{bhatia2010surveynearestneighbortechniques}, approximate nearest neighbor search \cite{chavez2020scalable} or graph-based algorithms \cite{liu2021sparse}. 

As an example, consider an $\mathcal{H}^2$ representation of the kernel matrix in~\cref{eq:kernelmatrix} with geometrical information $x_i$. A KD tree-based clustering will generate a cluster tree $I$ of $L=\bigo{\log n}$ levels (see \cref{fig:h2algo}(b)). Each pair of clusters $(I_s,I_t)$ (or simply $(s,t)$) at each level $\ell$ is deemed admissible; 
\begin{equation}
\texttt{adm}(s,t)=1,~\texttt{if}~\frac{\texttt{Diam}(s) + \texttt{Diam}(t)}{2} \leq \eta \: \texttt{Dist}(s, t) ,
\end{equation}
where $\texttt{Diam}$ and $\texttt{Dist}$ are the diameters and distance of the two clusters. The block $A_{st}$ for the admissible pair, i.e., admissible block, is algebraically compressed and the inadmissible block is kept as dense. \Cref{fig:h2algo}(a) shows the matrix structure with a set of $n=2^{15}$ 3D geometry points using the admissibility parameter $\eta=0.7$ (green and red indicate admissible and inadmissible blocks, respectively). The set of clusters that form inadmissible pairs with a cluster $t$ are denoted as $\mathcal{N}_t$. The set of clusters (a) that form admissible pairs with cluster $t$ and (b) whose parents form inadmissible pairs with the parent of $t$, are denoted as $\mathcal{F}_t$. In $\mathcal{H}^2$ matrix format, each admissible block is compressed as $A_{st} = U_{s} B_{st} V_{t}^T$ where $B_{st}$ is an $r\times r$ coupling matrix, and $U_{s}$ and $V_{t}^T$ are column and row bases. The basis for leaf nodes in the cluster tree are stored explicitly, and the basis for an inner node $t$ of the cluster tree is defined in terms of the basis of its children $t_1$ and $t_2$ using transfer matrices $E_{t_1}$ and $E_{t_2}$, resulting in a nested and shared basis:
\begin{equation}
\label{eqn:nested_basis}
U_t = \begin{bmatrix}
U_{t_1} & \\
& U_{t_2}
\end{bmatrix} \begin{bmatrix}
E_{t_1} \\
E_{t_2}
\end{bmatrix}.
\end{equation}
See the $\mathcal{H}^2$ illustration in  \cref{fig::Hformats} where the leaf-level inadmissible blocks, coupling matrices, leaf-level basis matrices, and higher-level transfer matrices are shown. 

In addition to $\mathcal{H}^2$ matrices, different choices of admissibility parameter ($\eta\geq1$ (weak admissibility) or $\eta\leq0.5$ (strong admissibility)) and shared basis property lead to all the other hierarchical matrix formats in~\cref{tab:hierarchical_matrix} and~\cref{fig::Hformats}. It is worth noting that in independent basis-based formats such as $\mathcal{H}$ or HODLR matrices, the low-rank compression of each admissible block becomes $A_{st} \approx U_{st} V_{st}^T$ of \cref{eq:low-rank} instead of $A_{st} \approx U_{s} B_{st} V_{t}^T$ (again, see~\cref{fig::Hformats}). 

\subsubsection{Construction}\label{sec:construct_H}
The construction phase essentially computes the low-rank compression $A_{st} \approx U_{st} V_{st}^T$ of \cref{eq:low-rank} for all admissible blocks to form a computationally efficient approximation $H\approx A$. In order to achieve lower than $\bigo{n^2}$ computational complexity for large matrices, one needs to use alternative algorithms to truncated SVD or rank-revealing QR that forms a full $n\times n$ admissible block.
Instead, optimal-complexity algorithms rely on the assumption of either (a) any entry of the block $A_{st}$ can be efficiently computed on the fly, or (b) the block and its transpose can be rapidly applied to any random vector $\Omega$ as $A_{st}\Omega$ and $A_{st}^T\Omega$. The former leads to purely algebraic algorithms such as adaptive cross approximation (ACA) or Nystr\"{o}m method \cite{Bebendorf2000ACA,Bebendorf2006BEMACA,Zhao2005ACAEM,liu2019parallelACA,xia2024nystrom}, and proxy point or Chebyshev interpolation-based methods \cite{ye2020analytical,xing2020interpolative} requiring geometrical information. These algorithms have been extended for butterfly compression as well \cite{Candes_butterfly_2009,li_butterfly_2015,Pang2020IDBF}. On the other hand, the latter leads to random sketching-type algorithms such as randomized SVD  \cite{low-rankLiberty}, QR \cite{MingGu2017qr,MingGu2018SRQR}, UTV \cite{Martinsson2017utv} and butterfly \cite{liu_butterfly:2020,li_butterfly_2015} algorithms. These faster algorithms can achieve up to an optimal $\bigo{nr}$ or $\bigo{nr^2}$ computational complexity.    

When it comes to efficient construction of rank structured matrix representations, the above assumptions become (a) entry-evaluation: any entry of $A$ can be efficiently computed, (b) fully sketching: $A\Omega$ and $A^T\Omega$ can be rapidly computed, or (c) partially sketching: a hybrid version of (a) and (b). In practical applications, entry-evaluation applies to integral equations \cref{eq:IE} and kernel matrices \cref{eq:kernelmatrix}, etc.; randomly sketching applies to frontal matrices \cref{eq:schur_frontal} and Hessian matrices \cref{eq:hessian}, etc; partially sketching is often applied to frontal matrices (when the update term $F_{sr}F_{rr}^{-1}F_{rs}$ in \cref{eq:schur_frontal} is formed as a low-rank compression first).   

In the context of $\mathcal{H}^2$ matrices, the adaptation of fast low-rank compression algorithms becomes more sophisticated due to the shared-basis property. In other words, one needs to compute a single low-rank basis $U_s$ of $A_{st}$, $t \notin \mathcal{N}_s$ for each cluster $s$ and a single low-rank basis $V_t$ of $A_{st}$, $s \notin \mathcal{N}_t$ for each cluster $t$ via entry evaluation or sketching. Fast entry evaluation-based $\mathcal{H}^2$ algorithms include proxy surface-based \cite{huang2020h2pack,minden2017fast,martinsson2005fast,xing2020interpolative}, polynomial interpolation-based \cite{hackbusch2002h2,bendoraityte2008distributed,zampini2022h2opus}, and ACA-type \cite{bebendorf2012constructing,zhao2019fast} algorithms. On the other hand, fast sketching-based $\mathcal{H}^2$ algorithms include fully sketching-based top-down algorithms \cite{lin2011fast,levitt2022randomized,zampini2022h2opus}, fully sketching-based bottom-up algorithm \cite{Yesypenko}, and partially sketching-based bottom-up algorithm \cite{boukaram2025}.  

\begin{figure*}[thp!]
	\centering
	\includegraphics[width=0.9\textwidth]{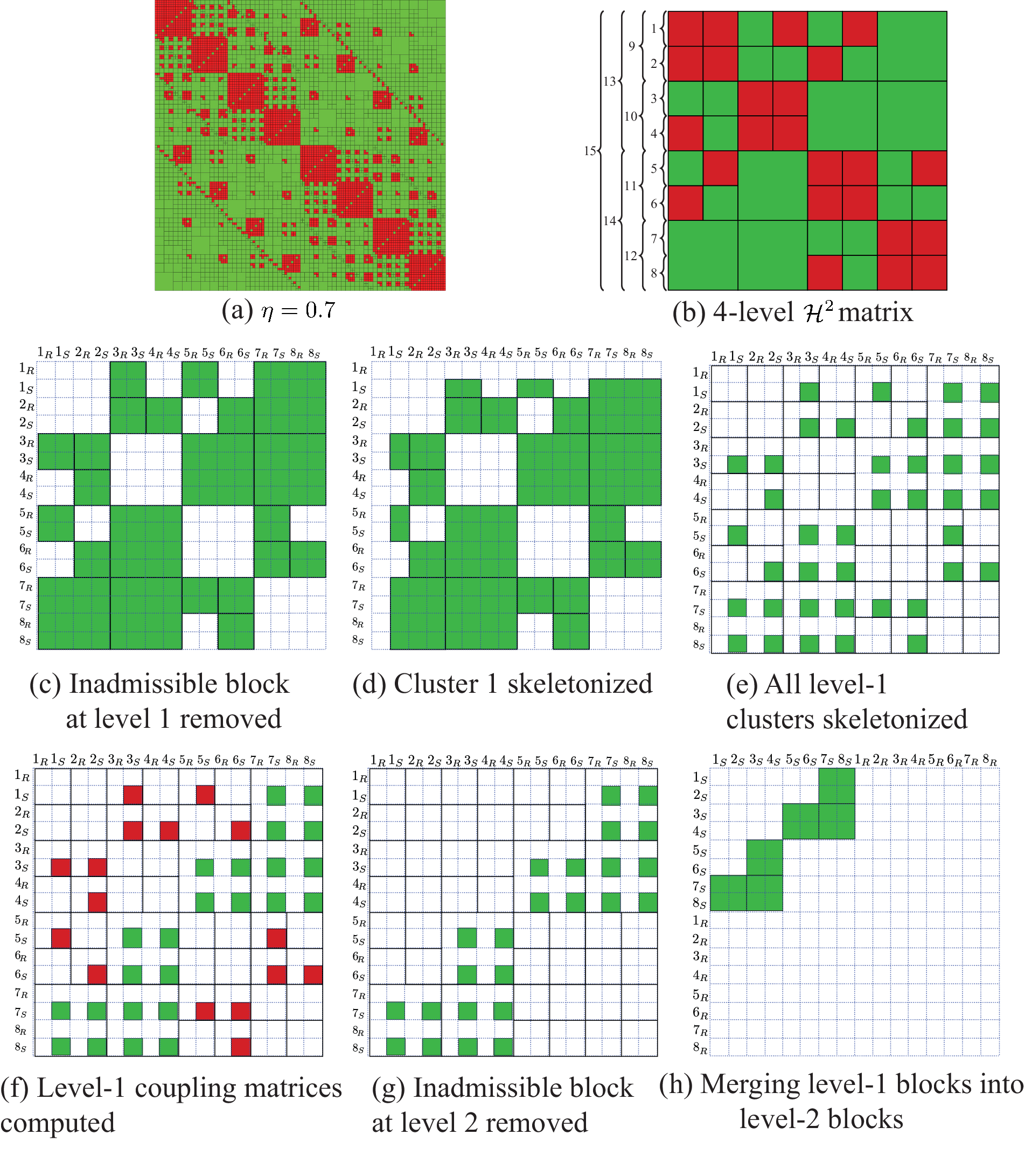}
	\caption{\cite[figs.~1 and 4]{boukaram2025} (a) Block partitioning of a hierarchical matrix for a 3D problem of size $N=2^{15}$ with different $\eta$. (b) Illustration of a 4-level $\mathcal{H}^2$ matrix. (c)-(h) First two levels of the partially sketching-based algorithm in \cite{boukaram2025}. }
	\label{fig:h2algo}
\end{figure*}

\begin{figure*}[th!]
	\centering
	\includegraphics[width=0.9\textwidth]{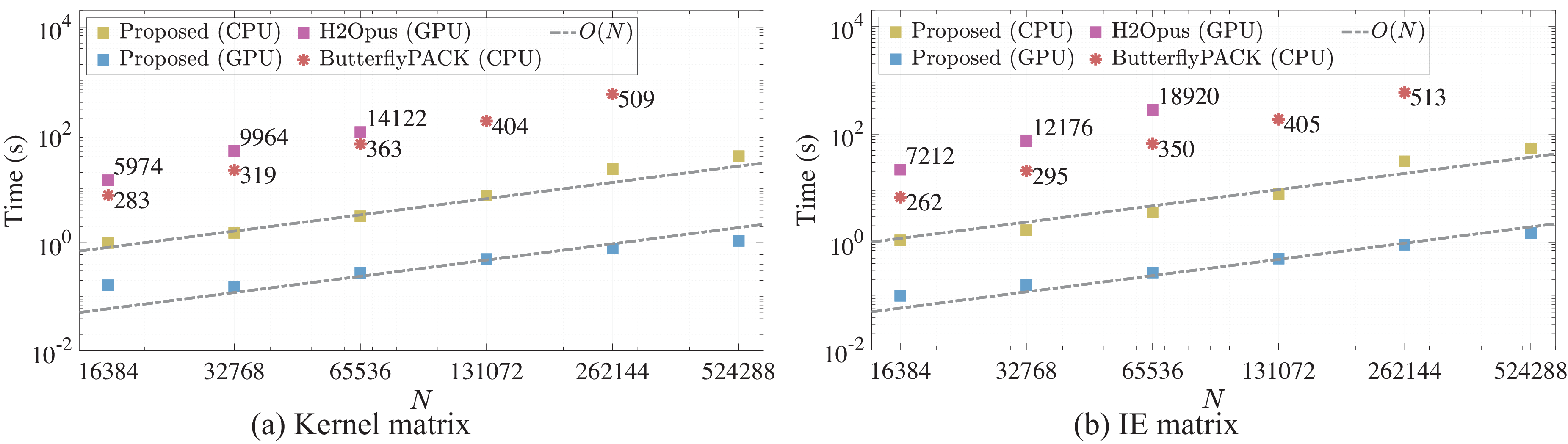}
	\caption{\cite[fig. 5]{boukaram2025} The execution time for the CPU and GPU implementations of the construction algorithm \cite{boukaram2025} for the kernel and IE matrices. Also shown is the top-down construction algorithm  implemented in ButterflyPACK \cite{levitt2022randomized} and H2Opus \cite{zampini2022h2opus}, with its data points labeled with the total number of random samples $d$. In comparison, algorithm \cite{boukaram2025} requires only $d=256$.}\label{fig:construction_H2}
\end{figure*}

Here we give a brief description of the partially sketching-based bottom-up algorithm \cite{boukaram2025} using the example of \cref{fig:h2algo}(b). Note that admissible blocks only exist at the lower two levels. Assuming $A$ is real and symmetric for simplicity of explanation, the algorithm uses interpolative decomposition (ID) for admissible blocks such that $A_{st}\approx U_sB_{st}U_t^T$, and $B_{st}=A({\widetilde{I}_s,\widetilde{I}_t})$ for some skeleton sets $\widetilde{I}_s \subseteq I_s$ and $\widetilde{I}_t \subseteq I_t$. 
From the sketching data $Y=A\Omega$ with some random matrix $\Omega \in \mathbb{R}^{n\times d}$, $d>r$, the algorithm first computes $Y_s=Y(I_s,:)-\sum_{b\in \mathcal{N}_s}D_{sb}\Omega_b$ with $\Omega_b=\Omega(I_b,:)$ for each leaf-level cluster $s$. Note that $D_{sb}$ denotes leaf-level inadmissible blocks and is explicitly computed via entry evaluation. $Y_s$ effectively sketches the column space of all admissible blocks at block row $s$ (see~\cref{fig:h2algo}(c)), and $U_s$ and $\widetilde{I}_s$ (also denoted as $I_{s_S}$) can be computed from $Y_s$ via ID. Note that $A$ in~\cref{fig:h2algo}(c) has been reordered such that the redundant indices $I_{s_R}=I_s\setminus I_{s_S}$ appear before skeleton indices $I_{s_S}$ for each cluster $s$. After applying a transformation associated with $U_s$, only the skeletonized rows $A(\widetilde{I}_s,:)$ \rev{(i.e., rows of $A$ corresponding to skeleton sets $\widetilde{I}_s$)} and columns $A(:,\widetilde{I}_s)$ remain (see \cref{fig:h2algo}(d)(e)). At the next level, one computes $Y_s=\begin{bmatrix}
U_{s_1}Y_{s_1}\\
U_{s_2}Y_{s_2}
\end{bmatrix}-\begin{bmatrix}
\sum_{b\in\mathcal{F}_{s_1}}B_{s_1b}U_{b}^T\Omega_{b} \\
\sum_{b\in\mathcal{F}_{s_2}}B_{s_2b}U_{b}^T\Omega_{b}
\end{bmatrix}$ for each level-2 cluster $s$ with $s_1,s_2$ being its children. Note that the coupling matrices $B_{s_1b}$ and $B_{s_2b}$ are explicitly computed via entry-evaluation (see the blocks in red in \cref{fig:h2algo}(f)) and $\Omega_{b}$ and $T_b$ for a level-1 cluster $b$ have been scaled by $U_b$. $Y_s$ effectively sketches the column space of all admissible blocks shown in green in~\cref{fig:h2algo}(h) from which transfer matrices $E_{s_1}$ and $E_{s_2}$ can be computed via ID. Finally, the coupling matrices $B_{st}$ at level 2 are explicitly formed via entry evaluation. As can be seen, this algorithm leverages both sketching and entry evaluation, and can achieve the optimal $\bigo{nr^2}$ computational complexity. Numerical results in \cite{boukaram2025} report up to a $1000\times$ speedup over an existing GPU implementation of the top-down sketching-based algorithm \cite{zampini2022h2opus} and a $660\times$ speedup over a CPU implementation of the algorithm \cite{levitt2022randomized} (see~\cref{fig:construction_H2}).

In addition to $\mathcal{H}^2$ matrices, fast construction algorithms have also been applied in many other rank-structured matrix algorithms, including entry-evaluation-based algorithms for $\mathcal{H}$ \cite{grasedyck2003construction}, HODLR \cite{ambikasaran2015fast}, HODBF \cite{Sadeed2022VIE}, HSS \cite{chavez2020scalable} and BLR \cite{shaeffer2008direct}, as well as sketching-based algorithms for $\mathcal{H}$, HODLR \cite{lin2011fast,levitt2022randomized}, HSS \cite{gorman2019robust,martinsson2011fast,levitt2024linear} and uniform BLR \cite{pearce2025randomized}, etc.

\subsubsection{Factorization and Solve}\label{sec:factor_H}
The factorization phase efficiently computes the inverse of the rank-structured matrix $H^{-1}$ or its equivalent, by leveraging arithmetic operations in the compressed formats such as block addition and multiplication \cite{grasedyck2003construction}, Sherman-Morrison-Woodbury formula \cite{ambikasaran2013n}, and skeletonization \cite{martinsson2005fast}. Once $H^{-1}$ has been computed, it can be used in the solve phase to compute $X=H^{-1}B$ for any RHS $B$ leveraging fast arithmetic operations as well. It is worth noting that the factorization (and solve) algorithms can significantly differ from each other based on different rank-structured matrix formats. These include recursive block LU \cite{hackbusch2003introduction,grasedyck2003construction,Han_2017_butterflyLUPEC,heldring2025fast} for $\mathcal{H}$ or $\mathcal{H}^2$ matrices, single-level block LU for BLR matrices \cite{shaeffer2008direct}, block ULV or Cholesky for HSS \cite{chandrasekaran2006fast,chandrasekaran2007fast,xia2010fast,gillman2012direct}, multiplicative inversion for HODLR \cite{ambikasaran2014fast,ambikasaran2015fast,ambikasaran2013n} and HODBF \cite{Liu_2017_HODBF,Sadeed2022VIE}, recursive skeletonization for HSS and $\mathcal{H}^2$ \cite{martinsson2005fast,minden2017recursive}. These algorithms can achieve an up to $\bigo{nr^2}$ optimal computational complexity as shown in~\cref{tab:hierarchical_matrix}. While many factorization algorithms for weak-admissible formats such as HODLR \cite{ambikasaran2013n} or HSS \cite{martinsson2005fast} can guarantee a constant growth factor in $r$ during factorization, these formats are less preferable for high-dimensional problems as $r$ does not stay independent of $n$ due to weak admissibility. Although there exist specialized algorithms \cite{corona2015n,l2016hierarchical} for PDEs on structured grids, strong-admissible formats such as $\mathcal{H}$ or $\mathcal{H}^2$ matrices exhibit much smaller ranks.

As efficient and parallel $\mathcal{H}^2$ factorization algorithms represent one state-of-the-art rank-structured matrix algorithm research area, we summarize existing $\mathcal{H}^2$ algorithms in the following. The algorithms of B{\"o}rm et al. \cite{borm13_lr} and Ma et al. \cite{ma2017accuracy} represent the early $\mathcal{H}^2$ factorization algorithms based on recursive block addition/multiplication and local update, which can attain $\bigo{nr^2}$ computational complexity, albeit with high constant prefactors. Later, inverse fast multipole (IFMM) methods \cite{coulier2017inverse,ambikasaran2014inverse,Pouransari2017IFMM,chandrasekaran2007fast} were developed which rewrite the dense system as an extended sparse system using FMM bases and perform block LU elimination. More recently, the recursive skeletonization with strong admissibility (RS-S) \cite{minden2017recursive} was proposed leveraging both interpolative decomposition and block LU elimination. RS-S exhibits much smaller prefactors and is more amenable to efficient parallelization. A sequence of follow-up works exists~\cite{liang2024on,ma2022scalable,Ma2019,Yesypenko,Sushnikova2023,boukaram2025linear}.   

\begin{figure}
	\centering
	\includegraphics[width=0.9\doublecolumnimgwidth]{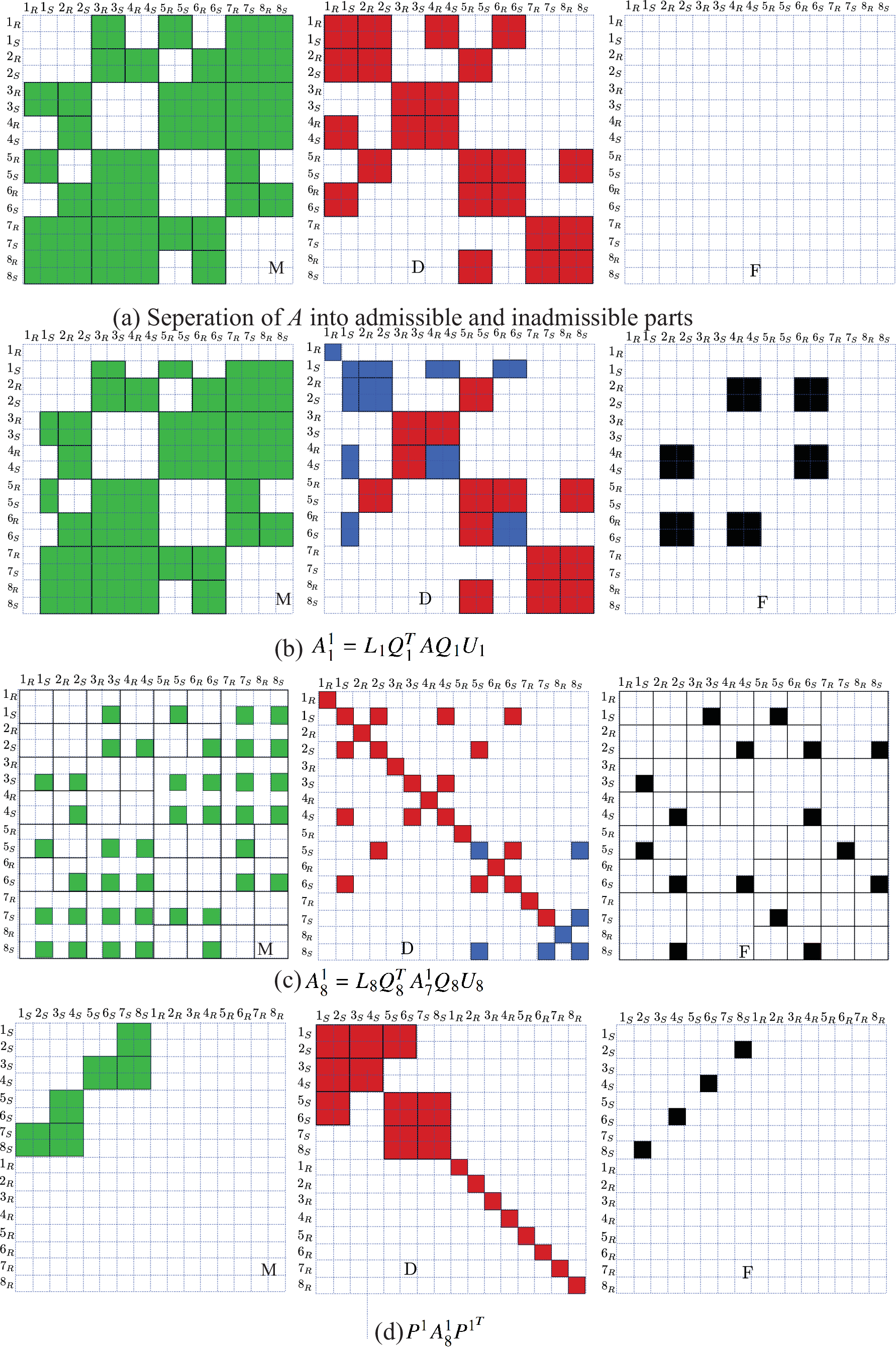}  
	\caption{\small \cite[figs. 4-12]{boukaram2025linear} \rev{Illustration of the factorization algorithm of \cite{boukaram2025linear} on the $\mathcal{H}^2$ example of \cref{fig:h2algo}(b). (a) The splitting of $A=M+D+F$ where $M$ and $D$ contain respectively all admissible and inadmissible blocks at level 1 and $F$ contains the fill-ins to be generated. Note that $A$ has been reordered such that for each cluster $t$, its index set  $I_t=[t_R,t_S]$ consists of the redundant indices $t_R$ and skeletonization indices $t_S$, respectively. (b) Skeletonization using $Q_1$ and elimination using $L_1, U_1$ for cluster 1. After the skeletonization, the elimination of block row/column $1_R$ introduces modifications to existing inadmissible blocks (shown in blue) and new fill-ins which are admissible (shown in black). (c) Skeletonization using $Q_8$ and elimination using $L_8, U_8$ for cluster 8. (d) Permutation $P^1$ is applied to permute the diagonal blocks of redundant indices to the lower right corner, which is then LU factorized as $L_r^1U_r^1$. Note that certain level-1 admissible blocks (green and black) now become level-2 inadmissible blocks.} }\label{fig:h2_factor}
\end{figure}

Here we give a high-level overview of the $\mathcal{H}^2$ factorization algorithm in \cite{boukaram2025linear} using the 4-level $\mathcal{H}^2$ matrix of \cref{fig:h2algo}(b). For simplicity of explanation, we write $H=M+D+F$ where $M$ and $D$ contain respectively all admissible and inadmissible blocks at level 1 and $F$ contains the fill-ins to be generated (see~\cref{fig:h2_factor}(a)). 
For cluster 1, let $Q_1 = \texttt{diag}(\widetilde{Q}_1, I, \cdots, I)$ where $\widetilde{Q}_1=[U_1^\bot,U_1]$, $\bar{H}=Q_1^THQ_1$ skeletonizes cluster 1. Next, LU elimination of $\bar{H}_{1_R1_R}$ by $L_1$ and $U_1$ modifies blocks in $D$ (shown in blue in~\cref{fig:h2_factor}(b)) and introduces fill-ins in $F$ (shown in black in~\cref{fig:h2_factor}(b)). Let $A^\ell_s$ denote the matrix after LU elimination of $s_R$ at level $\ell$ by $L_s$ and $U_s$, \rev{we have $A_1^1=L_1Q^T_1H Q_1U_1$. With a slight abuse of notation, we still split $A_1^1$ (and the results of all subsequent $A_s^\ell$) into $A_1^1=M+D+F$.} We repeat these operations for cluster 2 with $A_2^1=L_2Q^T_2A_1^1 Q_2 U_2$ (see \cref{fig:h2_factor}(c)) and for all the remaining level-1 clusters until $A_8^1 = L_8Q^T_8A_7^1 Q_8 U_8 $ (see \cref{fig:h2_factor}(d)). 
Note that the fill-in blocks (in black) appearing at the row block of cluster $s$ are used to augment the original $U_s$. The blocks in $M$ and $F$ corresponding to inadmissible blocks at level 2, marked in red in \cref{fig:h2_factor}(e), are then moved into $D$. Finally, a permutation $P^1$ is applied to permute all the diagonal blocks of redundant indices to the lower right corner (see \cref{fig:h2_factor}(f)), followed by an LU factorization $L_r^1U_r^1$ that eliminates these redundant blocks. The above procedure is repeated for level $2,3,\ldots,L$, until at the highest level $L$ there are no more admissible blocks left and a final LU factorization $L_LU_L$ is performed. This LU-like factorization can be summarized as 

\begin{equation}
H = \left( \prod_{\ell=1}^L \left(\prod_{s\mathrm{at} \ell} Q_s L^{-1}_s \right) P^{\ell^T} L_r^\ell\right) L_L U_L \left( \prod_{\ell=L}^1 \left(U_r^\ell P^\ell \prod_{s}U_s^{-1} Q_s^{T} \right) \right).\label{eq:H2-factor}
\end{equation}
  
It is worth noting that unlike weak-admissible formats such as HODLR and HSS, the rank growth of $H^{-1}$ for $\mathcal{H}$ or $\mathcal{H}^2$ matrices is due to the fill-in blocks (in black in \cref{fig:h2_factor}), which has not been fully understood to date \cite{Angleitner2021,Faustmann2015ExistenceO,faustmann2016existence}.    

\begin{figure*}[th!]
	\centering
	\includegraphics[width=\textwidth]{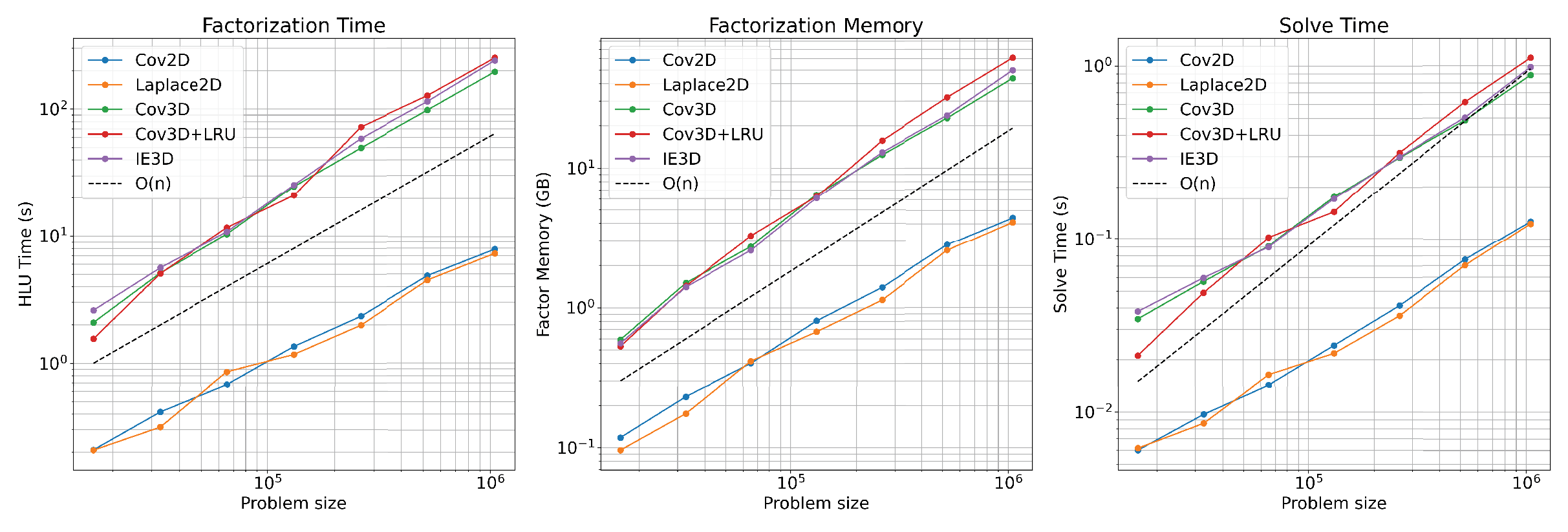}
	\caption{\cite[figs.~13 and~16]{boukaram2025linear} The factorization time, memory requirement and solve time of the shared-memory parallel $\mathcal{H}^2$ algorithm of \cite{boukaram2025linear}.}\label{fig:factor_H2}
\end{figure*}

Once the factorization in \cref{eq:H2-factor} has been computed, $H^{-1}$ can be applied to any RHS vectors $B$ by applying each factor in \cref{eq:H2-factor} one by one. Note that the inverses of $Q_s$, $P^\ell$, $L_r^\ell$ and $U_r^\ell$ are readily available now. As a few examples, \cref{fig:factor_H2} demonstrates the factorization time, factorization memory and solve time when applying the algorithm of \cite{boukaram2025linear} to 2D and 3D, IE and kernel matrices of varying sizes on a shared-memory CPU architecture. As can be clearly seen, the $\bigo{n}$ complexities have been validated.    

Needless to say, depending on the accuracy of $U_s$ and condition number of $A$, the factorization in \cref{eq:H2-factor} can be either used as a direct solution operator or efficient preconditioner.   
  
\subsection{Parallelization of rank-structured matrix algorithm}\label{sec:parallel_H}
Beyond the superior asymptotic complexities of rank-structured matrix algorithms shown in~\cref{tab:hierarchical_matrix}, their efficient implementation on modern parallel computers represents a core enabler for real-world large-scale multi-physics, multi-scale, and high-fidelity simulations. That said, these algorithms can face significant parallel scalability challenges as (a) different parallel data layouts can introduce vastly different communication patterns, (b) many algorithms feature sequences of small and irregular-sized numerical linear algebra operations making them hard to achieve peak flop performance of modern hardware, and (c) many factorization algorithms involve complex data dependencies and recursive computations that are sequential in nature. In what follows, we review existing parallelization strategies for rank-structured matrices to tackle these challenges.   

\subsubsection{Distributed-memory process layout}\label{sec:layout_H}
\begin{figure*}[th!]
	\centering
	\includegraphics[width=\textwidth]{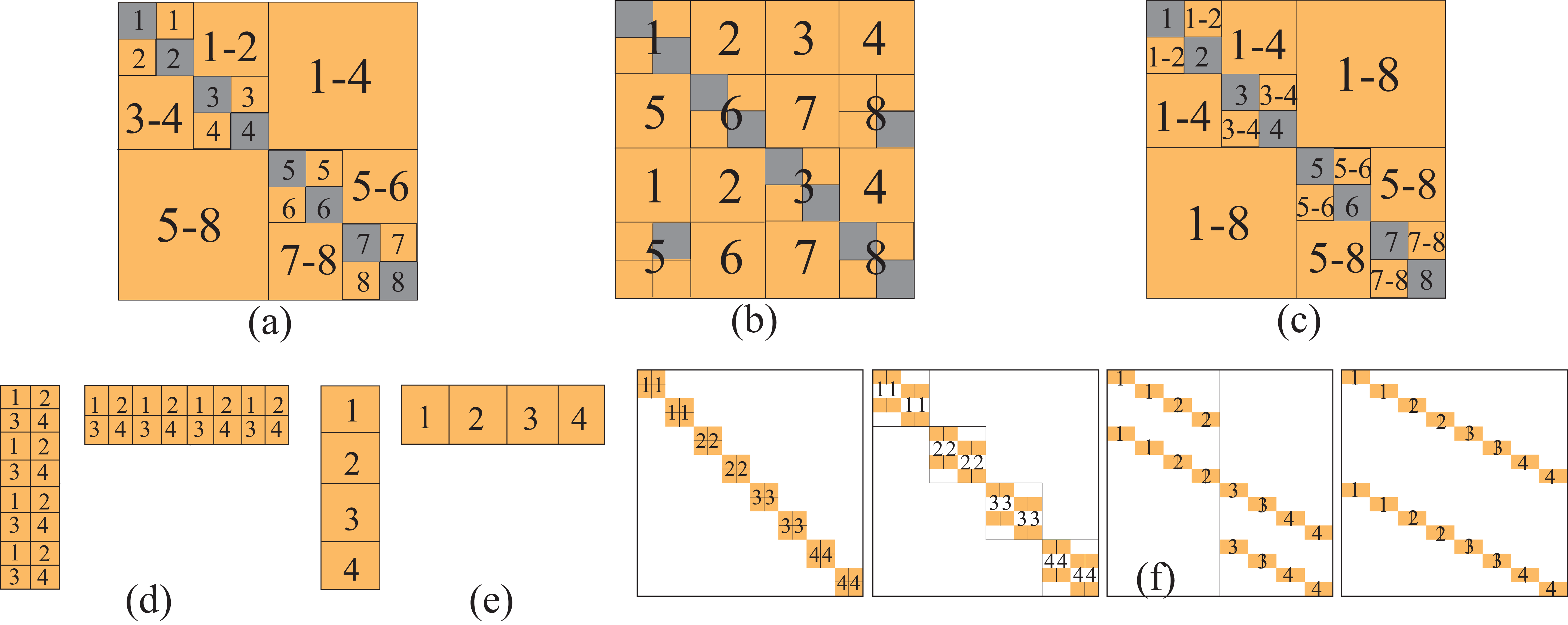}
	\caption{(Top) Parallel process layouts for distributing a hierarchical matrix on 8 MPI processes: (a) 1D block layout for HODLR, HODBF matrices \cite{chenhan2016inv,Sadeed2022VIE}, $\mathcal{H}$ matrices \cite{li2020distributed}, and $\mathcal{H}^2$ matrices \cite{bendoraityte2008distributed,zampini2022h2opus}, (b) 2D block-cyclic layout for lattice-$\mathcal{H}$ matrix algorithm \cite{yamazaki2019distributed} and BLR algorithm of \cite{claus2023sparse}, and (c) tree-based layout for HSS \cite{rouet2016distributed}. (Bottom) Parallel process layouts for distributing a single low-rank or butterfly compression on 4 MPI processes: (d) 2D block-cyclic layout for a low-rank representation used in \cite{rouet2016distributed}, (e) 1D block layout for a low-rank representation used in \cite{li2020distributed}, and (f) A parallel FFT-like layout for a butterfly representation used in \cite{liu_butterfly:2020,Sadeed2022VIE,liu2021sparse}.}\label{fig:parallel_layout}
\end{figure*}

When the matrix size $n$ is large enough such that the entire rank-structured matrix cannot fit onto a single shared-memory machine, distributed-memory parallelism becomes essential to fully unlock the potential of these algorithms. The optimal parallel data layout should be able to simultaneously achieve (a) good load balance for matrices with general sizes and non-uniform ranks on a non-power-of-two number of processes, and (b) minimal communication during construction, factorization and solve. Here we summarize a few existing parallel data layouts using examples of~\cref{fig:parallel_layout}.  

\begin{figure*}[th!]
	\centering
	\includegraphics[width=0.9\textwidth]{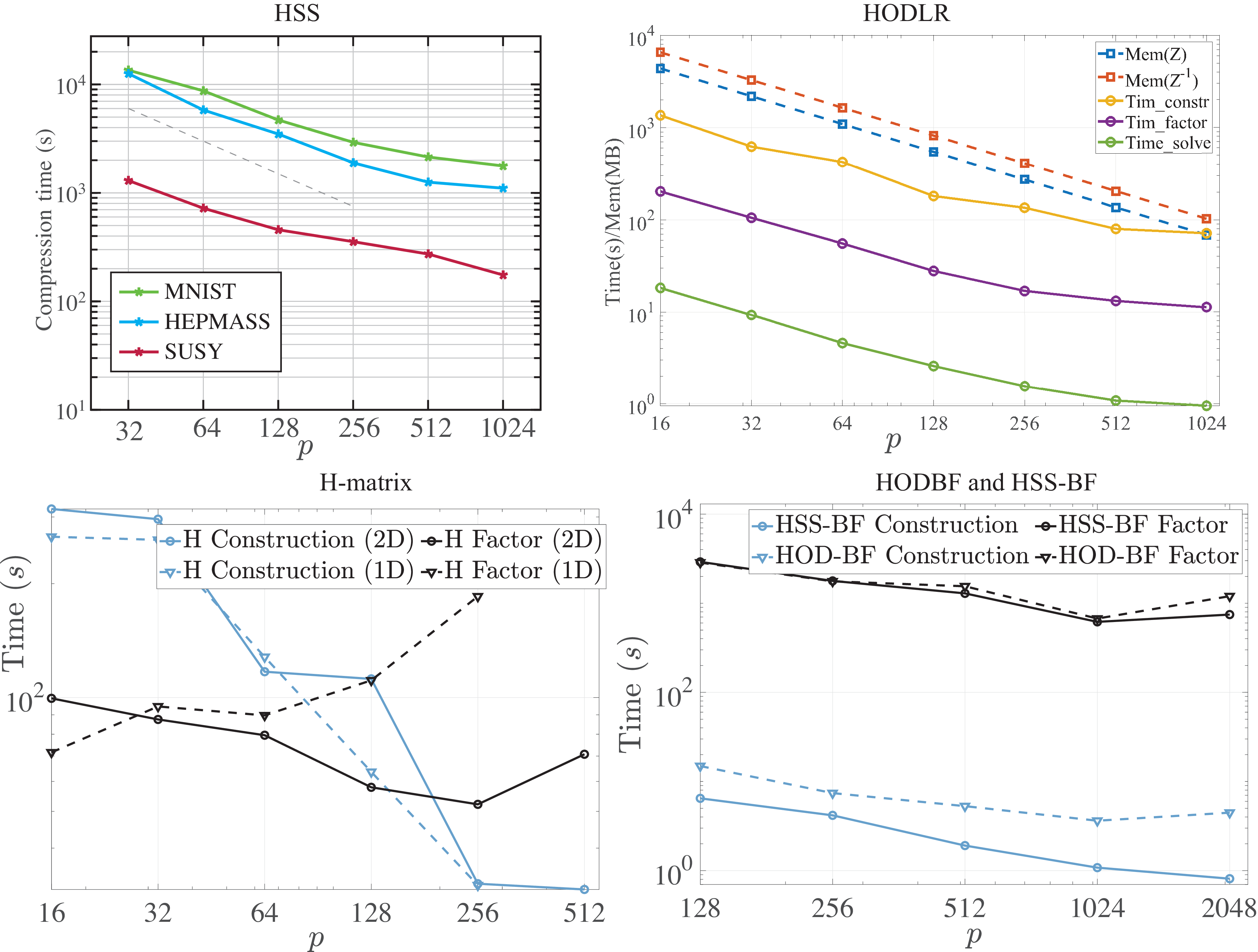}
	\caption{Strong scaling (on distributed-memory systems) of various phases of several hierarchical matrix algorithms: HSS (\cite[fig. 7]{chavez2020scalable}), HODLR, $\mathcal{H}$ matrix, HODBF and HSSBF. \rev{In general, weak-admissible algorithms can attain much better parallel efficiencies than strong-admissible algorithms (e.g., $\mathcal{H}$ and $\mathcal{H}^2$). For $\mathcal{H}$ matrix algorithms, the 2D block-cyclic process layout can achieve a higher parallel efficiency than the 1D block process layout.}}\label{fig:parallel_bpack}
\end{figure*}

For a hierarchical matrix, the most common parallel layout is perhaps the 1D block layout of  \cref{fig:parallel_layout}(a), which has been used for HODLR and HODBF matrices \cite{chenhan2016inv,Sadeed2022VIE}, $\mathcal{H}$ matrices \cite{li2020distributed,guo2015mpi}, and $\mathcal{H}^2$ matrices \cite{bendoraityte2008distributed,zampini2022h2opus} for the construction phase and matrix-vector multiplication. This layout exhibits good load balance, but may require extra data redistribution in matrix-vector multiplication when the input vectors also follow the 1D block layout. Moreover, although this layout leads to manageable communication in the factorization phase for weak-admissible formats such as HODLR and HODBF matrices \cite{Sadeed2022VIE} (see~\cref{fig:parallel_bpack} for the strong scaling performance of HODLR, HODBF and HSS-BF algorithms in the ButterflyPACK package \cite{butterfly_web}), they post significant challenges for recursive computation-based $\mathcal{H}$ and $\mathcal{H}^2$ factorization algorithms. In \cite{yamazaki2019distributed}, a so-called lattice-$\mathcal{H}$ matrix algorithm representing a hybrid algorithm between BLR and $\mathcal{H}$ uses a 2D block-cyclic layout as shown in  \cref{fig:parallel_layout}(b), which improves parallel efficiency of the factorization phase at the cost of introducing more flop operations. See  \cref{fig:parallel_bpack} for the performance difference between 1D block and 2D block-cyclic layouts for the lattice-$\mathcal{H}$ algorithm in ButterflyPACK \cite{butterfly_web}. In addition, the HSS algorithm of \cite{rouet2016distributed} uses a tree-based parallel layout that simplifies the communication pattern during the HSS-ULV factorization and some strong scaling results are shown in  \cref{fig:parallel_bpack} for HSS construction. Other parallel layouts have been used in parallel HIF \cite{li2017distributed}, $\mathcal{H}^2$ \cite{liang2024on} and BLR \cite{claus2023sparse} factorization algorithms, etc. 

When it comes down to the parallel layout of each individual off-diagonal block (e.g., the top-level off-diagonal block of \cref{fig:parallel_layout}(a)), one can use the 2D block-cyclic \cite{rouet2016distributed} or 1D block \cite{li2020distributed} layout for each low-rank factor, and a parallel FFT-like layout for each butterfly representation \cite{poulson2014parallel,liu_butterfly:2020}. See  \cref{fig:parallel_layout}(d)-(e) for an illustration.    

\subsubsection{Fine-grained parallelism on CPU and GPU}\label{sec:fine_H}
Efficient parallel implementation of an algorithm (or its local workload of each MPI process) on shared-memory CPU and GPU architectures must take into consideration the irregular data structures in each phase of a rank-structured matrix algorithm. For example, the matrix-vector multiplication of an $\mathcal{H}$ matrix involves an $\bigo{n/2^\ell}$ number of fine-grained GEMV operations of non-uniform sizes at each tree level $\ell$. As another example, the factorization of a HODLR matrix involves an $\bigo{n/2^\ell}$ number of fine-grained LU decompositions of variable-size diagonal blocks at each level $\ell$. 
Common implementation strategies are to flatten and marshal the tree metadata so that pertinent operations can be batched on GPUs or CPUs, and the CPU-GPU data transfer can be minimized. This is particularly true for shared-basis algorithms such as $HSS$ and $\mathcal{H}^2$ where even at higher levels of the tree, operations only involve small-sized blocks. 
These strategies have been leveraged in GPU implementation of BLR factorization~\cite{Claus25,keyes20_haha}, HODLR factorization \cite{chen2022solving}, $\mathcal{H}^2$ multiplication \cite{boukaram19_gpu,keyes20_haha,zampini2022h2opus,yesypenko2025simplified}, sketching-based $\mathcal{H}^2$ construction \cite{boukaram2025}, and shared-memory CPU-based $\mathcal{H}^2$ factorization \cite{boukaram2025linear}. Needless to say, the same strategies implemented on GPUs can behave much better than those implemented on CPUs, see the example of $\mathcal{H}^2$ construction \cite{boukaram2025} in \cref{fig:construction_H2}.  

\subsubsection{Strategies to tackle data dependency}

The other parallelization challenge is the data dependency, particularly in the factorization phase, of many rank-structured matrix algorithms, e.g., the recursive computation-based $\mathcal{H}$ and $\mathcal{H}^2$ factorization algorithms \cite{hackbusch2003introduction,grasedyck2003construction}, the skeletonization-based $\mathcal{H}^2$ factorization algorithms \cite{minden2017recursive}, and the sketching-based $\mathcal{H}$ matrix construction algorithms \cite{lin2011fast}. Common practices to exploit parallelism within data dependency are threefold: (a) One can leverage graph-based algorithms such as graph coloring to explore operations that can be simultaneously executed. This has been demonstrated for $\mathcal{H}^2$ factorization \cite{liang2024on} and construction \cite{levitt2022randomized}. (b) Runtime scheduling systems such as StarPU \cite{augonnet:inria-00550877}, PARSEC \cite{cao2019performance} and OmpSs can be leveraged based on the DAG information of e.g., BLR \cite{al2020solving,cao2019performance} and $\mathcal{H}$ matrix \cite{yamazaki2019distributed,kriemann2013h,aliaga2017task} algorithms. (c) One can potentially modify the original factorization algorithm to remove, to some extent, the data dependency. Examples include the lattice-$\mathcal{H}$ \cite{yamazaki2019distributed} algorithm that behaves as BLR (which is easier to parallelize) at the higher tree levels, and the $\mathcal{H}^2$/BLR/HSS algorithms of \cite{ma2022scalable,spendlhofer2025interplay,Ma_2024} which precompute the augmented basis matrices due to fill-ins associated with inadmissible blocks.

\section{Combining Structure-sparse and Data-sparse Formats}\label{sec:data-and-structure}
Many sparse matrices arising from PDE discretization on structured and unstructured space and time grids (see a few examples in~\cref{sec:pde}) exhibit rank structures in their inverse or equivalent factorization \cite{bebendorf2003existence,chandrasekaran2010numerical,FaustmannHFEM,faustmann2013new,engquist2018approximate}. Hence, the rank-structured matrix algorithms introduced in~\cref{sec:data-sparse} can be combined with sparse supernodal or multifrontal solvers to construct fast direct solvers or efficient preconditioners. Moreover, other data-sparse compression algorithms such as lossy floating-point compression can also be combined with rank-structured matrix algorithms and sparse direct solvers. In what follows, we focus on reviewing multifrontal sparse solvers with data-sparse compression. That said, there are also attempts to augment supernodal sparse solvers with algebraic compression, such as the integration of HODLR \cite{faverge:hal-01187882} and BLR \cite{nies2019testing} in the PaStiX package \cite{henon2002pastix}, and the use of other low-rank compression techniques in a supernodal-based Cholesky solver \cite{chadwick2015efficient}. 


\begin{algorithm}
	\begin{flushleft}
		\textbf{Input:} $A \in \mathbb{R}^{n \times n}$, $b \in \mathbb{R}^{n}$ \\
		\textbf{Output:} $x \approx A^{-1} b$
	\end{flushleft}
	\begin{algorithmic}[1]
		\State $A \leftarrow P (D_r A D_c Q_c) P^\top$  \hfill \Comment{Scaling, and permutation for stability and fill reduction}
		\State \label{alg_prec_line:sepreorder} $A \leftarrow \widehat{P} A \widehat{P}^\top$ \hfill \Comment{Rank-reducing separator reordering}
		\State Build assembly tree: define $I^{\text{s}}_\tau$ and $I_\tau^{\text{u}}$ for every frontal matrix $F_\tau$
		\For{nodes $\tau$ in assembly tree in topological order}
		\If{no compression}
		\State construct $F_{\tau}$ and perform arithmetic as a dense matrix  
		\ElsIf{ZFP compression}
		\State construct $F_{\tau}$ and perform arithmetic as a dense matrix  
		\State compress as zfp matrix
		\ElsIf{BLR compression}
		\State \label{alg_prec_line:extadd} $\scriptsize F_{\tau} \leftarrow \begin{bmatrix} A(I_{\tau}^{\text{s}},I_{\tau}^{\text{s}}) & A(I_{\tau}^{\text{s}},I_{\tau}^{\text{u}}) \\
		A(I_{\tau}^{\text{u}},I_{\tau}^{\text{s}}) & 0 \end{bmatrix} \extadd T_{\tau_1} \extadd T_{\tau_2}$
		\State \label{alg_prec_line:LU} $P_\tau L_\tau U_\tau \leftarrow F_{11}$ \hfill \Comment{LU with partial pivoting} 
		\State \label{alg_prec_line:comp} $F_{11}, F_{12}, F_{21} \leftarrow \text{BLR compress} (F_{11}, F_{12}, F_{21})$ 
		\State \label{alg_prec_line:upd1} $F_{12} \leftarrow L_{\tau}^{-1} P_{\tau}^\top F_{12}$ 
		\State \label{alg_prec_line:upd2} $F_{21} \leftarrow F_{21} U_{\tau}^{-1}$ 
		\State \label{alg_prec_line:CB} $T_{\tau} \leftarrow F_{22} - F_{21} F_{12}$ \hfill \Comment{Schur update} 
		\ElsIf{HODBF compression} 
		\State $F_{11} \leftarrow \text{HODBF compress}
		\left(A(I_{\tau}^{\text{s}},I_{\tau}^{\text{s}})
		\extadd T_{\tau_1} \extadd T_{\tau_2} \right)$ \label{alg_prec_line:F11_HODBF}
		\State \label{alg_prec_line:F11_inv} $F_{11}^{-1} \leftarrow \text{HODBF invert}\left(F_{11}\right)$\hfill  
		\State \label{alg_prec_line:F12_construction} $F_{12} \leftarrow \text{butterfly compress}\left(
		A(I_{\tau}^{\text{s}},I_{\tau}^{\text{u}})
		\extadd T_{\tau_1} \extadd T_{\tau_2} \right)$  
		\State \label{alg_prec_line:F21_construction} $F_{21} \leftarrow \text{butterfly compress}\left(
		A(I_{\tau}^{\text{u}},I_{\tau}^{\text{s}})
		\extadd T_{\tau_1} \extadd T_{\tau_2} \right)$  
		\State \label{alg_prec_line:S_construction} $S \leftarrow \text{butterfly compress}\left( F_{21} F_{11}^{-1} F_{12}
		\right)$\Comment{Sketching-based construction}  
		\State \label{alg_prec_line:CB_construction} $T_{\tau} \leftarrow \text{HODBF compress}
		\left(T_{\tau_1} \extadd T_{\tau_2} - S\right)$ \hfill 
		\ElsIf{HSS compression}
		\State \label{alg_prec_line:extadd1} $F_{\tau} \leftarrow \text{HSS compress}\left( \begin{bmatrix} A(I_{\tau}^{\text{s}},I_{\tau}^{\text{s}}) & A(I_{\tau}^{\text{s}},I_{\tau}^{\text{u}}) \\
		A(I_{\tau}^{\text{u}},I_{\tau}^{\text{s}}) & 0 \end{bmatrix} \extadd T_{\tau_1} \extadd T_{\tau_2}\right)$\Comment{Sketching-based construction} 	
		\State \label{alg_prec_line:LU1} $U_\tau L_\tau V_\tau \leftarrow F_{11}$ \hfill \Comment{HSS ULV} 
		\State \label{alg_prec_line:upd11} $F_{12} \leftarrow L_{\tau}^{-1} U_{\tau}^T F_{12}$ 
		\State \label{alg_prec_line:upd21} $F_{21} \leftarrow F_{21} V_{\tau}^{T}$ 
		\State \label{alg_prec_line:CB1} $T_{\tau} \leftarrow F_{22} - F_{21} F_{12}$ \hfill \Comment{Low-rank update} 		
		\EndIf
		\EndFor
		\State $x \leftarrow \text{GMRES}(A, b, M: u \leftarrow D_c Q_c P^\top \widehat{P}^\top \,\, \text{bwd-solve}(\text{fwd-solve}(\widehat{P} P D_r v ) ))$ \label{alg_prec_line:precGMRES}
	\end{algorithmic}
	\caption{Sparse rank-structured multifrontal factorization using ZFP, BLR, HSS and HODBF matrix compression, followed by a GMRES iterative solve using the multifrontal factorization as an efficient preconditioner.}
	\label{alg:preconditioner}
\end{algorithm}

\subsection{Overview of data-sparse multifrontal sparse solvers and preconditioners}

Data-sparse multifrontal solvers have a similar algorithmic workflow to exact multifrontal solvers, except that the dense frontal matrices and their arithmetic are compressed using various compression techniques depending on their sizes and underlying PDE characteristics. 

\begin{figure*}[bh!]
	\centering
	\includegraphics[width=\textwidth]{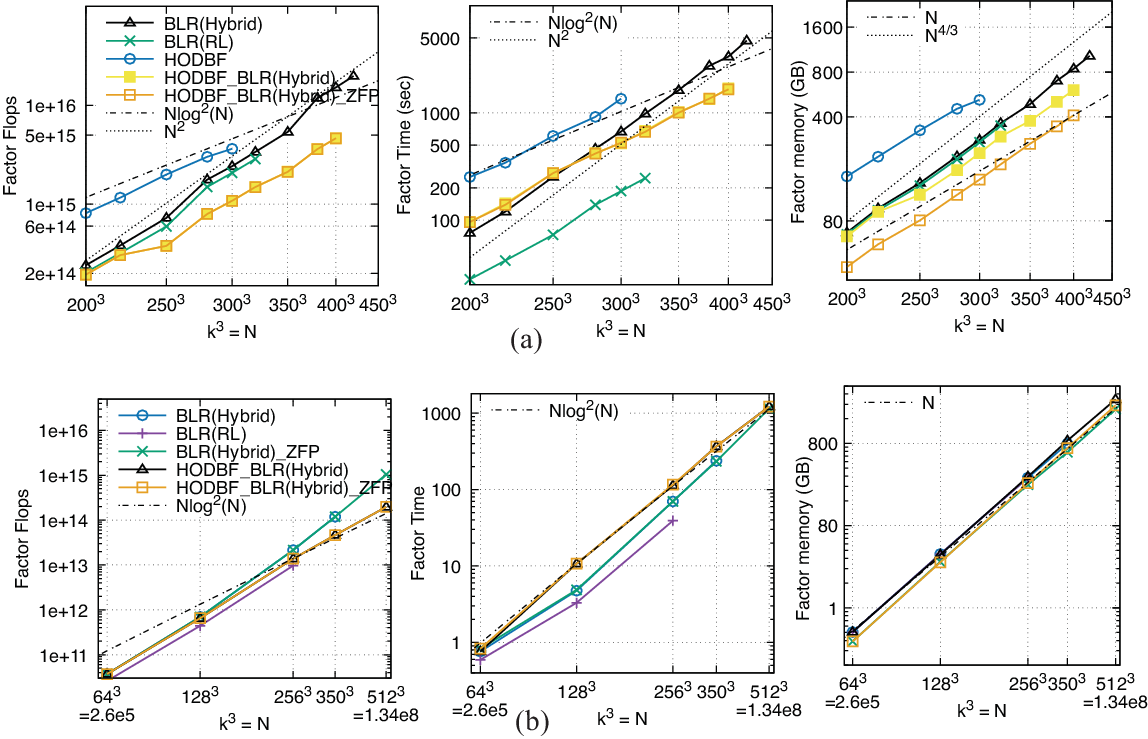}
	\caption{\cite[figs. 8 and 9]{claus2023sparse} Characteristics of the factorization phase including flops, CPU time and memory for a multifrontal sparse solver with composite compression algorithms using 64 NERSC Cori CPU nodes.}\label{fig:hybrid_compression}
\end{figure*}

The overall workflow is described in \Cref{alg:preconditioner}. Just like exact multifrontal and supernodal solvers, the matrix $A$ is first preprocessed as~\cref{eq:factorization} with diagonal scaling matrices $D_r$ and $D_c$, and permutation matrices $P_r$ and $P_c$. For example, a typical choice for the fill-in reduction permutation $P_c$ is nested dissection, from which the frontal matrices can be viewed as graphs between the separators consisting of subsets of vertices of the original graph $A$. As explained in~\cref{sec:preprocess_H}, the separators need to be further clustered to reveal rank structures inside each frontal matrix, via graph partitioning algorithms such as recursive bisection \cite{liu2021sparse,ghysels2016efficient} or geometrical information if available \cite{zhu2025recursive}. This permutation is denoted $\widehat{P}$ in Algorithm \ref{alg:preconditioner}. Based on $P_c$, a structure called the assembly tree is constructed consisting of dense frontal matrices as the nodes (see~\cref{fig:hybrid_compression} for an assembly tree example). For each non-root node, the frontal matrix has a $2\times2$ block structure $F_\tau=\begin{bmatrix}F_{11}&F_{12}\\F_{21}&F_{22}\end{bmatrix}$, where $F_{11}$ corresponds to fully summed variables $I_\tau^s$, and $F_{22}$ with indices $I_\tau^u$ corresponds to the trailing submatrix to be Schur complement updated by $F_{11}$ and will contribute to the parent frontal matrix via extend-add. The data-sparse multifrontal solvers will then traverse the assembly tree in the topological order, and for each (non-leaf) frontal matrix perform factorization of $F_{11}$, update panels $F_{12}/F_{21}$, Schur complement update $F_{22}$ as $T$, and propagate $T$ into its parent using extend-add. For the root frontal matrix, only $F_{11}=F_\tau$ is factorized. 
Each frontal matrix can be compressed using any existing data-sparse compression algorithm such as lossy floating-point compression including ZFP \cite{lindstrom2025zfp} and SZ \cite{di2016fast}, or rank-structured matrix algorithms including $\mathcal{H}$, $\mathcal{H}^2$, HSS, BLR, HODLR, HODBF and their other variants. Most existing data-sparse multifrontal solvers use a single compression algorithm, but there also exist solvers that leverage more than one compression algorithm (i.e., hybrid or composite compression algorithms). 
Since lossy floating-point compression only works well for very small frontal matrices or the inadmissible blocks inside each rank-structured matrix algorithm, we refer to data-sparse multifrontal sparse solvers simply as rank-structured multifrontal sparse solvers in the rest of this section. Note that as the frontal matrix arithmetic is approximated by the compression methods using a prescribed tolerance, the multifrontal solver only computes $M\approx A^{-1}$, which can be used either as a direct solver or a preconditioner (in e.g., GMRES) depending on the tolerance and application needs. In what follows, we review existing rank-structured multifrontal sparse solvers in literature. 

\begin{figure*}[bh!]
	\centering
	\includegraphics[width=0.9\textwidth]{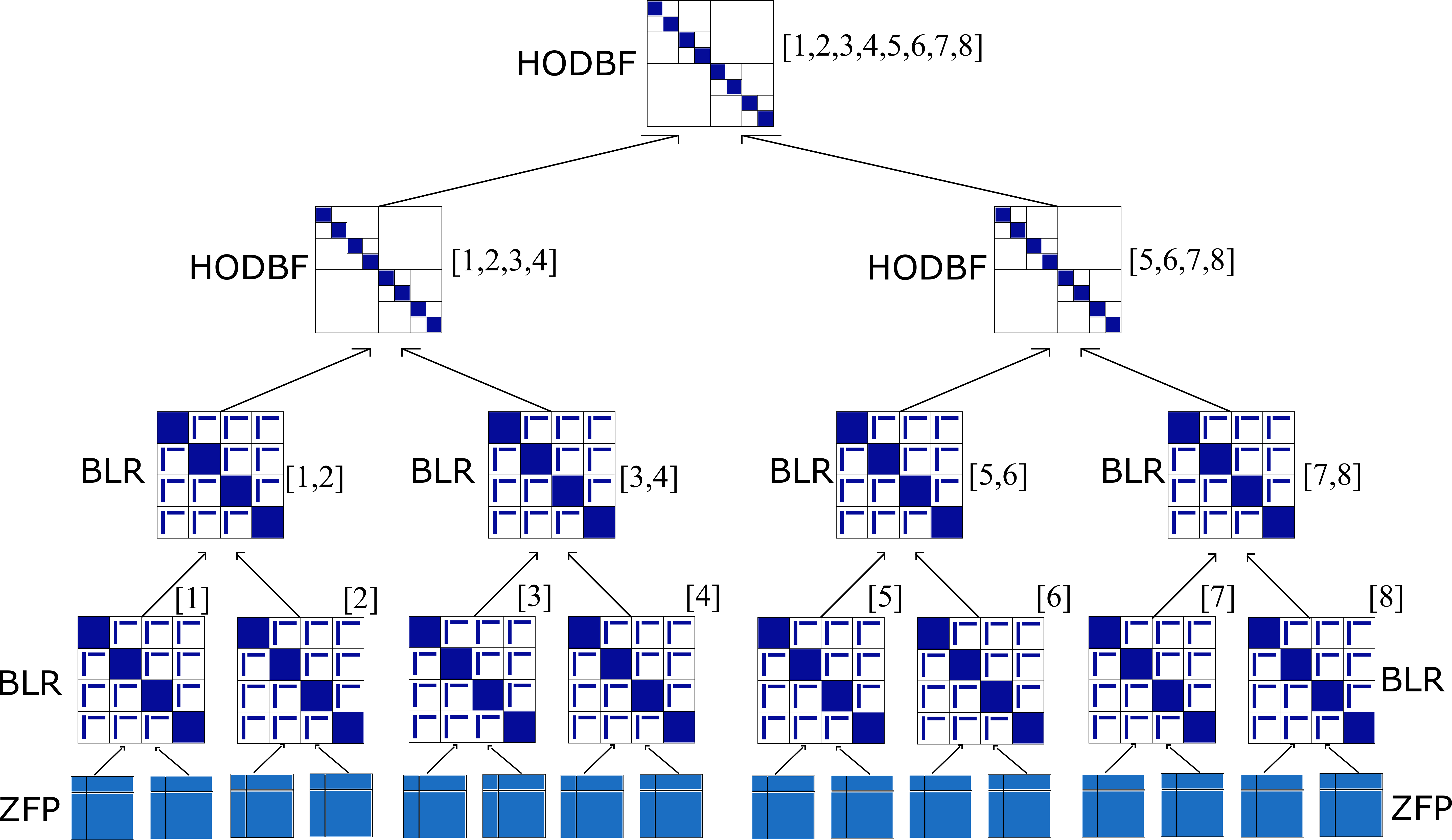}
	\caption{\cite[fig. 5]{claus2023sparse} The assembly tree of rank-structured multifrontal solvers using different compression algorithms. The solver is parallelized using 8 MPI processes.}\label{fig:multifrontal}
\end{figure*}

$\mathcal{H}$ and $\mathcal{H}^2$ represent the first hierarchical matrix algorithms applied to frontal matrices. Earlier works consider recursive computation-based factorization algorithms \cite{grasedyck2008parallel,zhou2015Hfem,Bebendorf2000ACA} which exhibit large computational overheads and difficulty in efficient parallelization \cite{grasedyck2008parallel}, albeit with promising complexity claims. Modern $\mathcal{H}^2$ algorithms and variants such as IFMM and skeletonization-based algorithms have been recently introduced to multifrontal solvers \cite{yesypenko2024slablu,Yesypenko,Pouransari2017IFMM,zhu2025recursive}, representing an active research area. In addition to these strong-admissible algorithms, weak-admissible algorithms such as HSS \cite{ghysels2017robust,ghysels2016efficient,xia2010superfast,xia2013randomized,Wang2016HSS,wang20113d,xia2017effective} represent one of the foundational algorithms in rank-structured multifrontal solver package STRUMPACK \cite{strumpack_web}. As shown in Algorithm \ref{alg:preconditioner}, HSS is used to compress $F_\tau$, typically using sketching-based construction algorithms to ensure compression accuracy in fully algebraic settings. Next, $F_{11}$ (which is also a HSS matrix) is factorized using HSS-ULV, and $F_{22}$ can be low-rank updated into $T_\tau$ in an efficient manner. As HSS turns out to be less efficient for high-dimensional problems (e.g. 3D PDEs), its high-dimensional variants such as HSS2D \cite{xia2014n} and HIF \cite{ho2016hierarchical,feliu2020recursively,cambier2020algebraic} have been integrated into multifrontal solvers as well. Other weak-admissible formats such as HODLR and HODBF have also been integrated into multifrontal solvers \cite{liu2021sparse} where the HODBF algorithm is well-suited to compress large frontal matrices arising from high-frequency wave equations. As shown in Algorithm \ref{alg:preconditioner}, HODBF is used to compress $F_{11}$, $F_{12}$, $F_{21}$ and $T_{\tau}$ separately, and the Schur complement update $S$ is compressed using sketching-based algorithms \cite{liu_butterfly:2020}. Finally, BLR has also been a popular method for rank-structured multifrontal solvers \cite{mary2017block,amestoy2015improving,aminfar2016fast,nies2019testing,Theo19performance}, which is available in both MUMPS \cite{amestoy2000mumps} and STRUMPACK \cite{strumpack_web}. BLR has several variants \cite{amestoy2017complexity,claus2023sparse}, each with different characteristics in flop performance, memory usage and communication complexity. We show the right-looking (RL)-BLR in Algorithm \ref{alg:preconditioner}, which requires forming the frontal matrix in dense format and performs the arithmetic in compressed representations. Other more memory efficient BLR variants such as hybrid BLR \cite{claus2023sparse} have also been integrated into STRUMPACK. It is worth mentioning that in addition to standard finite difference or finite element discretizations, rank-structured multifrontal solvers have also been developed for higher order discretizations such as the hierarchical merging of Poincar{\'e}--Steklov (HPS) solvers \cite{martinsson2013direct,hao2016direct,gillman2014n} and spectral methods \cite{Shen2016Spectral}. 
 
\emph{Remark on numerical stability.} When one frontal matrix is dense without low-rank compression, $F_{11}$ can be factorized as LU with partial pivoting $P_\tau$ which is numerically stable for ill-conditioned frontal matrices. However, rank-structured matrix algorithms can only allow partial pivoting on each leaf-level diagonal block, yielding these algorithms more prone to numerical instability. One exception is the BLR algorithm \cite{mary2017block}, which allows permutation of the blocks during factorization. In practice, a frontal matrix may need to be scaled or preconditioned before performing the hierarchical matrix arithmetic as demonstrated in \cite{xia2017effective,feliu2020recursively,cambier2020algebraic}.

\begin{table*} [!tp]
	\footnotesize
	\centering	
\begin{tabular}{|c|c|cccc|cccc|cccc|c|}
    \hline
    \rule{0pt}{2.3ex}
    & & \multicolumn{4}{c}{rank $r(m)$}  & \multicolumn{4}{|c}{factor flops} & \multicolumn{4}{|c|}{solve flops/factor memory} \\
    problem & dim & BF & HSS & BLR & $\mathcal{H}^2$ & BF & HSS & BLR & $\mathcal{H}^2$ & BF & HSS & BLR & $\mathcal{H}^2$  \\
    \hline
    \hline
    \rule{0pt}{2.3ex}
    \multirow{2}{*}{Helmholtz} & $2$ & $\log m$ & $m$ & $m$ & $m$ & $n$ & $n^{3/2}$ & $n^{3/2}$ & $n^{3/2}$ & $n$ & $n\log n$ & $n\log n$ & $n\log n$ \\
    \cline{3-14}
    \cline{2-6}
    \rule{0pt}{2.3ex}
    & $3$ & $m^{1/4}$ & $m$ & $m$ & $m$ & $n\log^2 n$ & $n^{2}$ & $n^{2}$ & $n^{2}$ & $n$ & $n^{4/3}$ & $n^{4/3}$ & $n^{4/3}$ \\
    \hline
    \hline
    \rule{0pt}{2.3ex}
    \multirow{2}{*}{Poisson} & $2$ & $\log m$ & $\log m$ & $1$ & $1$ & $n$ & $n$ & $n\log n$ & $n$ & $n$ & $n$ & $n$ & $n$ \\
    \cline{2-14}
    \rule{0pt}{2.3ex}
    & $3$ & $m^{1/4}$ & $m^{1/2}$ & $1$ & $1$ & $n\log^2 n$ & $n^{4/3}$ & $n^{4/3}$ & $n$ & $n$ & $n$ & $n\log n$ & $n$ \\
    \hline
\end{tabular}
	\vspace{-4pt}
	\caption{Asymptotic complexity of rank-structured matrix (HODBF (denoted as ``BF" in the table), HSS, BLR and $\mathcal{H}^2$) enhanced multifrontal solvers for 2D and 3D, Helmholtz and Poisson equations. Other variants like HIF \cite{ho2016hierarchical}, HSS2D \cite{xia2014n}, and multilevel BLR \cite{Theo19bridging} are not shown here. The ``big O" notation $\bigo{\cdot}$ has been dropped. Here $m$ denotes the size of a front and $n$ is the size of the global sparse system.}\label{tab:cc}
	\vspace{-10pt}
\end{table*} 

\emph{Remark on computational complexity.} The asymptotic complexity of rank-structured multifrontal solvers is closely related to the numerical rank of each frontal matrix compressed by a specific rank-structured matrix algorithm. As the frontal matrix represents numerical Green's functions of the governing PDEs, acting on separators that are orthogonal or parallel to each other, the rank behavior can be estimated to derive the overall sparse solver complexity. 
In \cref{tab:cc} we summarize the asymptotic complexities for the factor flops, solve flops and factor memory of different rank-structured multifrontal solvers using two classes of 2D and 3D PDEs, namely Poisson and high-frequency Helmholtz. We list the rank of each frontal matrix of size $m\times m$, $r(m)$, for HODBF, HSS, BLR and $\mathcal{H}^2$ algorithms as well as the overall solver complexity in terms of the sparse matrix size $n$. It is worth noting that for 3D high-frequency Helmholtz, the theoretical rank estimate is $r(m)=\bigo{m}$ for HSS, BLR, and $\mathcal{H}$/$\mathcal{H}^2$, etc. \cite{liu2021sparse,engquist2018approximate}, which leads to no theoretical asymptotic complexity reduction compared to $\bigo{n^2}$. However, one often empirically observes $r(m)=\bigo{m^{0.5}}$ when $n\leq 400^3$ \cite{wang20113d,Wang2016HSS,wang2012massively}, leading to an empirically reduced solver complexity. While the optimal asymptotic complexity depends on the PDE type, dimensionality, problem size and compression algorithm, we remark that in practice there does not necessarily exist a single compression algorithm that yields the best compression performance for all frontal matrices in a multifrontal solver. As such, hybrid or composite compression-based \cite{claus2023sparse} multifrontal solvers may become necessary. As an example, we illustrate a composite multifrontal solver \cite{claus2023sparse} in \cref{fig:multifrontal} where small-sized frontal matrices are compressed using ZFP, small-to-medium-sized ones are compressed using BLR (hybrid BLR or RL BLR), and large ones are compressed using HODBF. \Cref{fig:hybrid_compression} demonstrates that the fully composite solver (labeled as ``HODBF\_BLR(Hybrid)\_ZFP") is capable of handling $n=420^3=74,088,000$ for 3D high-frequency wave equations and $n=512^3=134,217,728$ for 3D singularly perturbed reaction-diffusion equations, using 2048 NERSC Cori CPU cores. 

\begin{figure*}[th!]
	\centering
	\includegraphics[width=0.9\textwidth]{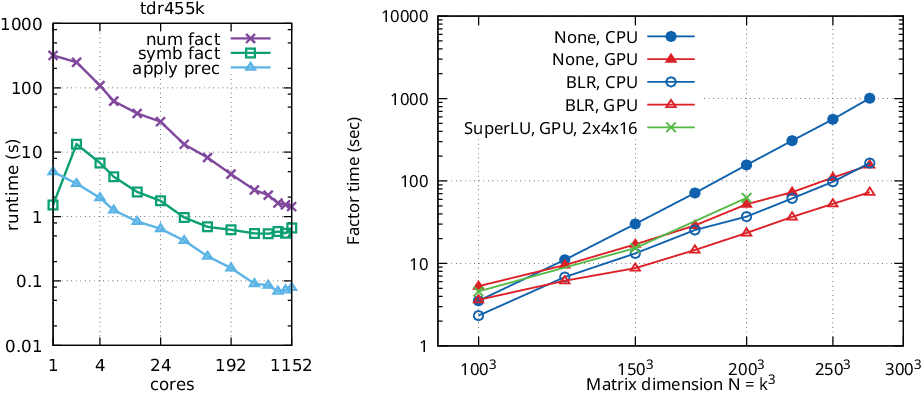}
	\caption{(a) \cite[fig. 6]{ghysels2017robust} Strong scaling of the parallel HSS-multifrontal solver of STRUMPACK. (b) \cite[fig. 1]{Claus25} Performance comparison of the exact multifrontal solver and BLR-multifrontal solver in STRUMPACK and the supernodal solver SuperLU\_DIST on multiple CPU and GPU nodes.  }\label{fig:parallel_multifrontal}
\end{figure*}

\emph{Remark on distributed-memory and fine-grained parallelization.} Distributed-memory parallelization of rank-structured multifrontal solvers represents a core technique to fully unlock the potential of these solvers for large-scale PDE simulations and machine learning computation. Most distributed-memory parallel multifrontal solvers follow a nested parallelization design pattern, which is briefly explained as follows: At the outer layer, the assembly tree is distributed using a tree parallelism approach \cite{amestoy2001asynchronous}. 
The root node frontal matrix is computed on all processes in the root MPI communicator; the root MPI communicator is split into two smaller ones, each handling one of the children frontal matrices at the next level; this procedure continues until a frontal matrix (and the corresponding subtree) is fully occupied by one MPI process. \Cref{fig:multifrontal} shows the distribution of the assembly tree with 8 MPI processes. Note that each parent MPI communicator does not  need to be split into two approximately equal-sized communicators. One can estimate the workload of its child (and the associated subtree) and use the so-called proportional mapping to assign communicator sizes proportional to the estimated workloads \cite{amestoy2001asynchronous,Wang2016HSS}. At the inner layer of the nested parallelization, each rank-structured frontal matrix occupied by more than one MPI process can be parallelized using the data layouts explained in~\cref{sec:layout_H}, and each dense frontal matrix can be parallelized using e.g., a 2D block-cyclic layout. The above-described nested parallelization strategy has been utilized in several rank-structured multifrontal solvers, including BLR-based \cite{Claus25,mary2017block}, HSS-based \cite{ghysels2017robust,Wang2016HSS,wang20113d}, HODBF-based \cite{liu2021sparse} and composite compression-based \cite{claus2023sparse} solvers.
\Cref{fig:parallel_multifrontal}(a) shows the strong scaling performance of different phases of a HSS-based multifrontal solver in STRUMPACK \cite{ghysels2017robust} for a sparse matrix tdr455k from accelerator simulations using up to 1152 CPU cores.    
Fine-grained parallelism inside each frontal matrix can naturally leverage the techniques summarized in~\cref{sec:fine_H}. Beyond that, this kind of parallelism can also be exploited across dense \cite{ghysels2022high} or rank-structured frontal matrices, particularly at lower levels of the assembly tree consisting of large numbers of small-sized frontal matrices. Examples include shared-memory HSS-multifrontal solvers \cite{ghysels2016efficient} and GPU BLR-multifrontal solvers \cite{Claus25}. In addition, GPUs have been exploited for several two-level multifrontal solvers such as \cite{yesypenko2024slablu,kump2025two}. For large-sized frontal matrices, it is worth mentioning that distributed-memory multi-GPU parallelization of a single frontal matrix has also been recently developed leveraging batched vendor GPU BLAS libraries and GPU-aware MPI libraries \cite{Claus25}.  \Cref{fig:parallel_multifrontal}(b) demonstrates the performance of the BLR-multifrontal solver \cite{Claus25} compared with exact supernodal and multifrontal solvers on 2048 CPU cores or 128 GPUs on the NERSC Perlmutter.  

\section{Available Software for Distributed Memory Parallel Computers}
\label{sec:software}
In this section, we list in \cref{tab:software} several direct solver packages (open-source for most of them) leveraging structure-sparse and/or data-sparse algorithms. Here we only mention a few representative packages designed with distributed-memory parallelism, and implemented with HPC programming languages such as C, C++, Fortran or Julia. Note that several of them also provide support for Python or Matlab interfaces. A more complete list can be found at {\url{https://portal.nersc.gov/project/sparse/superlu/SparseDirectSurvey.pdf}} for structure-sparse solvers and {\url{https://github.com/gchavez2/awesome_hierarchical_matrices} for data-sparse solvers.

\begin{table}[hb!]
\centering
\scriptsize
\begin{tabular}{lcccccc}
\hline
\textbf{Software} 
& \textbf{GPU} 
& \textbf{Matrix Type}
& \textbf{Data-sparse}
& \textbf{Programming Language}
& \textbf{Typical Application}\\
\hline

ASKIT \cite{askit-web,chenhan2016inv}
& No
& Dense 
& Yes
& C++ 
& Kernel\\

AHMED \cite{ahmed-web}
& No
& Dense/Sparse 
& Yes
& C++ 
& IE/FEM\\

ButterflyPACK \cite{butterfly_web} 
& No
& Dense 
& Yes
& F/C++/PY 
& IE/Kernel/Optimization\\

DINFMM \cite{dinfmm-web,liang2024on}
& No
& Dense 
& Yes
& Julia 
& IE\\


h2opus \cite{h2opus-web,zampini2022h2opus} 
& Yes
& Dense 
& Yes
& C++ 
& IE/Kernel/Optimization\\

HiCMA \cite{hicma-web}
& Yes
& Dense 
& Yes
& C/C++ 
& IE/Kernel\\

HLibPro \cite{hlibpro-web}
& No
& Dense/Sparse 
& Yes
& C++ 
& IE/Kernel/PDE\\

MUMPS \cite{mumps-web,amestoy2000mumps}
& Yes
& Sparse 
& Yes
& C/F
& Any sparse\\

mxc \cite{mxc-web,Ma_2024}
& Yes
& Dense 
& Yes
& C++ 
& IE/Kernel\\

PaStiX \cite{pastix-web,henon2002pastix}
& Yes
& Sparse 
& Yes
& C/F/PY/Julia 
& Any sparse\\

PARDISO \cite{pardiso-web}
& No
& Sparse 
& No
& C/C++/F/PY/Matlab 
& Any sparse\\

STRUMPACK \cite{strumpack_web}
& Yes
& Dense/Sparse 
& Yes
& C++/F/PY
& IE/Kernel/Optimization/Any sparse\\

SuperLU\_DIST \cite{superlu_web,TOMS_release_2023}
& Yes
& Sparse 
& No
& C/C++/F/PY/Julia/Matlab 
& Any sparse\\

\hline
\end{tabular}
\caption{List (in alphabetical order) of some distributed-memory direct solver packages leveraging structure sparsity and/or data sparsity.``PY" stands for ``Python" and ``F" stands for ``Fortran".}\label{tab:software}
\end{table}

Among the packages in \cref{tab:software}, SuperLU\_DIST and PARDISO are exact direct solvers that leverage only data-sparse algorithms and can be applied to any sparse matrix. 
Other such packages not listed here include PanguLU \cite{pangulu-web}, symPACK \cite{sympack-web} and WSMP \cite{wsmp-web}, etc. On the other hand, the other packages in \cref{tab:software} leveraging data-sparse algorithms for dense matrices implement a variety of rank-structured matrix algorithms described in \cref{sec:overview_H} and usually work as efficient preconditioners. Some of these packages are designed for a particular application such as kernel methods in machine learning and statistics, integral equation (IE) or finite-element method (FEM) matrices, and thus provide additional support at the application layer (e.g., PDE discretization, quadrature, hyperparameter optimization, etc.). Note that there are several dense direct solver packages that only support construction, rather than factorization, of a rank-structured matrix, which we do not list here. Finally, MUMPS, PaStiX, STRUMPACK, AHMED and HLibPro in the table combine structure-sparse and data-sparse algorithms for sparse matrices exhibiting low-rank structures, often arising from PDE or optimization problems. Note that MUMPS, PaStiX and STRUMPACK can behave as exact direct solvers when their data-sparse algorithms are turned off.

\section{Open Problems}\label{sec:open-problems}



Despite tremendous progress in parallel algorithms for both structure-sparse and data-sparse direct solvers, there remain many open problems, both theoretical and practical. We list some of them in the following.

\begin{itemize}
\item
\emph{GPU-resident solvers.} GPU memory capacity today is orders of magnitude larger than a decade ago. As such, many application codes are moved entirely to GPUs, avoiding the data transfer cost between CPU and GPU. This requires the linear solvers to be running on GPUs as well. The open problems are mostly in the preprocessing steps that are largely on CPUs. These are all combinatorial algorithms and much less progress is achieved on GPUs.

\item
\emph{Distributed memory symmetric indefinite sparse direct solvers.} This capability is sorely needed  by the optimization community. 
The symmetric $LDL^T$ factorization is highly desirable in numerical optimization methods that use second-order information, as it exposes matrix inertia needed to assess local convexity and KKT optimality. 
As of now, a robust and widely used algorithm is MA57~\cite{ma57}, but there is no reliable distributed memory code. 

\item
\emph{Communication analysis for data-sparse parallel algorithms.} While communication analysis is well-established for exact dense and structure-sparse direct solvers, there is a dearth of research concerning data-sparse direct solvers. In fact, there exists only a limited body of latency–bandwidth model-based analyses for the construction phase \cite{li2020distributed,liu_butterfly:2020}, with none available for the factorization phase. Accordingly, communication avoiding algorithms remain unexplored for rank-structured matrix algorithms. Moreover, distributed-memory parallel algorithms and dynamical runtime scheduling techniques have not been reported for sketching-based construction and factorization algorithms for strong-admissible rank-structured matrix algorithms.   
\item
\emph{Data-sparse in tensor computations.} The efficiency of rank-structured matrix algorithms for high-dimensional applications—such as 3D integral equations, kernel matrices, and quantum chemistry—remains a significant challenge with substantial potential for optimization. Tensor-based extensions of these algorithms represent a promising frontier for reducing both the constant prefactors and the asymptotic complexities of existing fast direct solvers. Recent efforts include the development of tensor-structured $\mathcal{H}$-matrices \cite{li2025hierarchical} and butterfly algorithms \cite{liu2025block,kielstra2025linear}, tensor-train-based solvers for integral equations \cite{corona2017tensor,nguyen2025tensor}, as well as tensor-based approaches for wave equations with varying frequencies or coefficients. However, these methods primarily target structured data, such as PDEs discretized on regular grids; their applicability to unstructured data remains an open problem. Furthermore, while several parallel algorithms for individual tensor compressions exist \cite{ballard2025tensor}, fully parallelized tensor-based direct solvers have yet to be developed.

\item
\emph{Data-sparse rank analysis for matrix inverse.} While the rank behavior of frontal matrices in sparse solvers for PDEs is relatively well understood, the rank behavior of inverses for integral equation matrices, particularly using strong-admissible-based rank structured algorithms, has been addressed primarily through empirical studies. Theoretical results have only been established for Laplacian kernels with quasi-uniform discretization \cite{Faustmann2015ExistenceO,faustmann2016existence}. Comparing with the rank structure of the integral equation matrices, the ranks in the inverse often exhibit mild growth. This phenomenon is attributed to both the accumulation of numerical errors and the increasingly complex physics (e.g., scattering matrices in wave problems) encoded within the Schur complements. Consequently, rigorous analytical tools for characterizing rank growth in the inverses of integral equation matrices remain largely unavailable.

\item
\emph{Error and spectral analysis for data-sparse matrices.} Error analysis for rank-structured matrices represents an indispensable tool for understanding the reliability of these algorithms and providing practical guidance on algorithm parameter choices. Backward stability analysis exists for certain weak-admissible formats such as HSS \cite{martinsson2011fast,xi2016stability}, but remains largely unavailable for most rank-structured matrix algorithms. Moreover, the effectiveness of rank-structured preconditioners for ill-conditioned systems, such as recent $\mathcal{H}^2$ preconditioners \cite{Ma_2024,spendlhofer2025interplay}, requires rigorous spectral analysis tools.     

\end{itemize}

\begin{acknowledgement}
This work was supported by the U.S. Department of Energy, Office of Science, Office of Advanced Scientific Computing Research's Applied Mathematics Competitive Portfolios program and  Scientific Discovery through Advanced Computing (SciDAC) program through the FASTMath Institute, under contract number DE-AC02-05CH11231.
\end{acknowledgement}

\ignore{
\ethics{Competing Interests}{Please declare any competing interests
in the context of your chapter. The following sentences can be regarded as examples.\newline
This study was funded by [X] [grant number X].\newline
 [Author A] has a received research grant from [Company W].\newline
 [Author B] has received a speaker honorarium from [Company X] and owns stock in [Company~Y].\newline 
 [Author C] is a member of [committee Z].\newline 
The authors have no conflicts of interest to declare that are relevant to the content of this chapter.}
}


\bibliographystyle{spmpsci}
\bibliography{references}

\end{document}